\documentclass[aip,amsmath,amssymb,a4paper,reprint]{revtex4-2}

\usepackage{graphicx}
\usepackage{xcolor}
\usepackage{soul} 
\usepackage{comment}
\usepackage{bm}
\usepackage{wrapfig}
\usepackage{dcolumn}

\newcommand{\e}{\mathrm{e}}
\newcommand{\D}{\mathrm{d}}
\newcommand{\NA}{N_\mathrm{A}}
\newcommand{\kB}{k_\mathrm{B}}
\newcommand{\fa}{f_\mathrm{a}}

\newcommand{\mbE}{\mathbf{E}}
\newcommand{\mbm}{\mathbf{m}}
\newcommand{\mbr}{\mathbf{r}}
\newcommand{\mbx}{\mathbf{x}}
\newcommand{\mby}{\mathbf{y}}
\newcommand{\mbR}{\mathbf{R}}
\newcommand{\mbX}{\mathbf{X}}
\newcommand{\mbalpha}{\boldsymbol{\alpha}}
\newcommand{\mbpi}{\boldsymbol{\pi}}

\newcommand{\epsr}{\varepsilon_\mathrm{r}}
\newcommand{\keff}{\kappa_\mathrm{eff}}
\newcommand{\ur}{u_\mathrm{r}}
\newcommand{\Tref}{T_\mathrm{ref}}
\newcommand{\lamhe}{\lambda_\mathrm{He}}
\newcommand{\etahe}{\eta_\mathrm{He}}
\newcommand{\lamar}{\lambda_\mathrm{Ar}}
\newcommand{\etaar}{\eta_\mathrm{Ar}}
\newcommand{\Pra}{\mathrm{Pr}}
\newcommand{\prar}{\Pra_\mathrm{Ar}}

\newcommand{\CVNIST}{CV${}_\mathrm{NIST90}$}
\newcommand{\CVNBS}{CV${}_\mathrm{NBS76}$}

\newcommand{\betaa}{\beta_\mathrm{a}}
\newcommand{\gammaa}{\gamma_\mathrm{a}}
\newcommand{\RTg}{RT \gamma_\mathrm{a}}

\def\br{\bm{r}}
\def\bA{\bm{A}}

\setcitestyle{numbers,square}

\begin{document}

\title{{\em Ab initio} Calculation of Fluid Properties for Precision Metrology}

\author{Giovanni Garberoglio}
\email{garberoglio@ectstar.eu}
\affiliation{European Centre for Theoretical Studies in Nuclear Physics and
Related Areas (FBK-ECT*), Strada delle Tabarelle 286, 38123 Trento, Italy}
\affiliation{Trento Institute for Fundamental Physics and Applications
  (INFN-TIFPA), via Sommarive 14, 38123 Trento, Italy}

\author{Christof Gaiser}
\affiliation{Physikalisch-Technische Bundesanstalt (PTB), Abbestrasse 2–12,
  10587 Berlin, Germany} 

\author{Roberto M. Gavioso}
\affiliation{Istituto Nazionale di Ricerca Metrologica, Strada delle Cacce
  91, 10135 Torino, Italy}

\author{Allan H. Harvey}
\affiliation{Applied Chemicals and Materials Division, National Institute of
  Standards and Technology, Boulder, CO 80305, United States of America}

\author{Robert Hellmann}
\affiliation{Institut f\"ur Thermodynamik,
  Helmut-Schmidt-Universit\"at/Universit\"at der Bundeswehr Hamburg,
  Holstenhofweg 85, 22043 Hamburg, Germany} 

\author{Bogumi\l{} Jeziorski}
\affiliation{Faculty of Chemistry, University of Warsaw, Pasteura 1, 02-093
  Warsaw, Poland}

\author{Karsten Meier}
\affiliation{Institut f\"ur Thermodynamik,
  Helmut-Schmidt-Universit\"at/Universit\"at der Bundeswehr Hamburg,
  Holstenhofweg 85, 22043 Hamburg, Germany} 

\author{Michael R. Moldover}
\affiliation{Sensor Science Division, National Institute of Standards and
  Technology, Gaithersburg, MD 20899-8360, United States of America}

\author{Laurent Pitre}
\affiliation{LCM-LNE-Cnam, 61 rue du Landy, 93210 La Plaine-Saint Denis,
  France}

\author{Krzysztof Szalewicz}
\affiliation{Department of Physics and Astronomy, University of Delaware,
  Newark, Delaware 19716, United States of America}

\author{Robin Underwood}
\affiliation{National Physical Laboratory (NPL), Teddington TW11 0LW,
  United Kingdom}

\date{\today}

\begin{abstract}
Recent advances regarding the interplay between {\em ab initio} calculations and metrology are
reviewed, with particular emphasis on gas-based techniques used for temperature and pressure
measurements.  Since roughly 2010, several thermophysical quantities -- in particular, virial and
transport coefficients -- can be computed from first principles without uncontrolled approximations
and with rigorously propagated uncertainties. In the case of helium, computational results have
accuracies that exceed the best experimental data by at least one order of magnitude and are
suitable to be used in primary metrology.  The availability of {\em ab initio} virial and transport
coefficients contributed to the recent SI definition of temperature by facilitating measurements of
the Boltzmann constant with unprecedented accuracy. Presently, they enable the development of
primary standards of temperature in the range 2.5--552~K and pressure up to 7~MPa using acoustic
gas thermometry, dielectric constant gas thermometry, and refractive index gas thermometry.  These
approaches will be reviewed, highlighting the effect of first-principles data on their accuracy. The
recent advances in electronic structure calculations that enabled highly accurate solutions for the
many-body interaction potentials and polarizabilities of atoms -- particularly helium -- will be
described, together with the subsequent computational methods, most often based on quantum
statistical mechanics and its path-integral formulation, that provide thermophysical properties and
their uncertainties.  Similar approaches for molecular systems, and their applications, are briefly
discussed.  Current limitations and expected future lines of research are assessed.
\end{abstract}

\maketitle

\tableofcontents

\newpage

\section{Introduction}

On May 20, 2019, the base SI units were redefined by assigning fixed values to the fundamental
constants of nature. By decoupling the base units from specific material artifacts, this
new redefinition is expected to lead to improved scientific instruments, reducing the degradation in
accuracy when measuring quantities at larger or smaller magnitudes than a predefined unit
standard. Additionally, the most accurate experimental technique available at each scale can be used
to implement a primary standard, resulting in easier calibrations, increased accuracies of measuring
devices, and further technological advancements.

Gas-based techniques have been proven to provide unparalleled performance for pressures up to 7 MPa
and for temperatures in the range 2.5~K -- 552~K (with extension to 1000~K or more progressing.~\cite{McEvoy20})

This accomplishment is in large part due to the
availability of thermophysical properties of the working gases (especially helium) calculated {\em
  ab initio} with no uncontrolled approximation and rigorously defined uncertainties, often
resulting in a better accuracy than the best experimental determinations.

These achievements have been facilitated by the increase in supercomputing power and advances in
numerical techniques for electronic structure calculations. For example, state-of-the-art
calculations for up to three He atoms even include relativistic and quantum electrodynamics
effects. In particular, these numerical investigations produce pair and three-body potentials, as
well as single-atom, pair, and three-body polarizabilities, with unprecedented accuracy.

Building on these results, the exact quantum statistical mechanics formulation enabled rigorous
calculations of the coefficients appearing in the density (virial) expansion of the equation of
state, the speed of sound, the dielectric constant, and the refractive index. The path-integral
Monte Carlo (PIMC) method has been shown to provide sufficient accuracy for these quantities. As a
consequence, it has been possible to devise a fully first-principles chain of calculations with
rigorous uncertainty propagation to compute virial coefficients of helium gas.

As a result of these endeavors, since about 2010 thermophysical properties of gaseous helium have been
known from theory with an accuracy that in most cases surpasses that of the most precise
experimental determinations. Currently, the uncertainties of the {\em ab initio} second and third
virial coefficients of helium are at least one order of magnitude smaller than the experimental
ones. The situation is similar for the density dependence of the speed of sound, the dielectric
constant, and the refractive index, where it leads to improved accuracy in Acoustic Gas Thermometry
(AGT), Dielectric Constant Gas Thermometry (DCGT), and Refractive Index Gas Thermometry (RIGT),
respectively.

Section~\ref{sec:expt} describes these gas-based experimental techniques for temperature and
pressure measurement, highlighting how much theoretical knowledge, in the form of virial
coefficients, enters in the uncertainty budget and helps improve the accuracy of measurements. For
each of these approaches, we will describe the operating principles, the range of temperatures and
pressures that can be covered, the most recent technological improvements, and the uncertainty
budget, highlighting the contribution of {\em ab initio} virial coefficients, which has been of
growing importance in the past 25 years.

First-principles calculations of virial coefficients involve two steps: the {\em ab initio}
electronic structure calculation of interatomic potentials and/or polarizabilities, followed by
the solution of the exact quantum statistical equations describing virial coefficients.

We therefore present in Sec.~\ref{sec:abinitio} a critical review of the state of the art of
non-relativistic, relativistic, and quantum electrodynamic electronic structure calculations, with
particular emphasis on the determination of uncertainties.  Our primary focus will be on helium –
which is currently the only substance for which computations can be performed that consistently
exceed the accuracy of the best experiments – but other noble gases will be briefly covered due to
their importance in metrology.  For the sake of completeness, we will recall the hierarchy of
physical theories involved in quantum chemical calculations, with particular emphasis on the Full
Configuration Interaction (FCI) approach, which is exact within a given orbital basis set and is
currently feasible for systems with up to 10 electrons. Relativistic and quantum electrodynamic
effects (expressed as expansions in powers of the fine-structure constant) have been crucial for
achieving the extremely low uncertainty of the latest helium calculations, and are also
progressively important in describing larger atoms (notably, neon and argon).
Additionally, the evaluation of electronic polarizabilities and magnetic susceptibilities will be
discussed.  All of these theoretical advances will be exemplified for the case of helium, where we
will present the current state of the art regarding interaction potentials and many-body
polarizabilities.

Knowledge of interaction potentials and polarizabilities enables the calculation of the coefficients
appearing in the virial expansion of the equation of state, the speed of sound, the dielectric
constant, and the refractive index, which are crucial ingredients in the uncertainty budget of AGT,
DCGT, and RIGT. In the past 15 years, the path-integral approach to quantum statistical mechanics has
been successfully applied in calculating virial coefficients without uncontrolled
approximations. The main features of this method are reviewed in Sec.~\ref{sec:thermo}, with
particular attention to the question of uncertainty propagation from the potentials and the
polarizabilities. In the case of pair properties, an alternative method based on the solution of the
Schr{\"o}dinger equation is available and provides mutual validation of the path-integral results, as
well as enabling the calculation of transport properties.
Most of this review is focused on thermodynamic properties, but {\em ab initio} calculations
also provide viscosity and thermal conductivity. We also briefly review how this leads to
improvements in flow-rate measurements.

Although most efforts have been devoted to noble gases, highly accurate theoretical calculations are
also available for molecular systems and have the potential to enable the same paradigm shift
observed in temperature and pressure metrology also to other types of measurements. We describe in
Sec.~\ref{sec:molecules} the present situation in the first-principles calculation of molecular
properties, and point out a few areas where computational contributions are expected to have an
increasing impact in the near future, namely humidity metrology, measurements of very low pressures,
and atmospheric science.  We will end our review in Sec.~\ref{sec:future}, where future perspectives
and an overview of the status of highly accurate {\em ab initio} calculations will be presented.

\section{Primary Metrology and Thermophysical Properties}
\label{sec:expt}

\subsection{Paradigm reversal in temperature metrology}
\label{sec:paradigm}

\begin{figure}[h]
\center\includegraphics[width=0.9\linewidth]{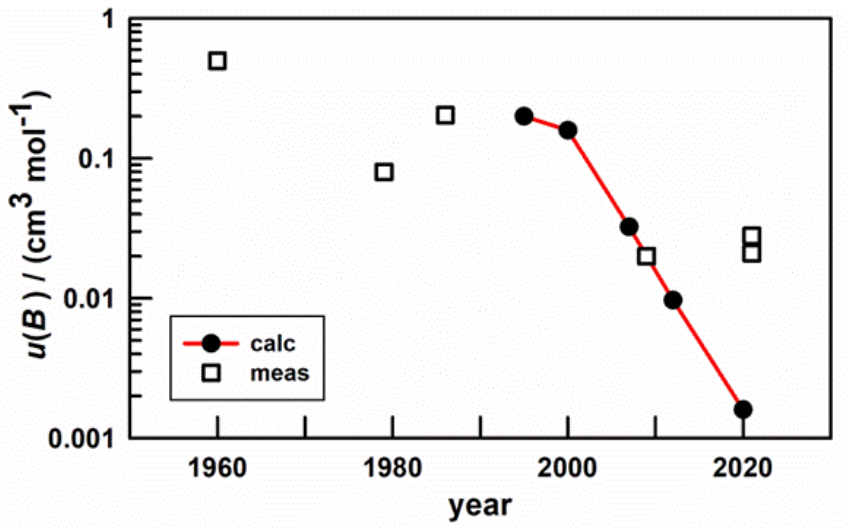} 
\caption{The standard uncertainty of both the measured and the calculated values of the second
  density virial coefficient of ${}^4$He decreased with time. After 2012, $u(B_\mathrm{calc}) <
  u(B_\mathrm{meas})$.  Both $B_\mathrm{calc}$ and $B_\mathrm{meas}$ are evaluated near 24.5561~K,
  which is the defined temperature of the triple point of neon.  Calculated values (circles): Aziz
  {\em et al.} (1995)~\cite{Aziz1995}; Hurly and Moldover (2000)~\cite{Hurly2000}; Hurly and Mehl
  (2007)~\cite{Hurly2007}; Cencek {\em et al.}  (2012)~\cite{Cencek2012}; Czachorowski {\em et al.}
  (2020).~\cite{Czachorowski2020} Measured values (squares): White {\em et al.}
  (1960)~\cite{White1960}; Berry (1979)~\cite{Berry1979}; Kemp {\em et al.} (1986)~\cite{Kemp1986};
  Gaiser and Fellmuth (2009)~\cite{Gaiser2009}; Gaiser and Fellmuth (2021)~\cite{Gaiser2021b};
  Madonna Ripa {\em et al.} (2021)~\cite{MadonnaRipa21}. }
  \label{fig:paradigm-1}
\end{figure}

Traditionally, accurate measurements of temperature-dependent thermophysical properties of gases
[such as: second density virial coefficient $B(T)$, viscosity $\eta(T)$, thermal conductivity
  $\lambda(T)$] have been used to determine parameters in evermore-refined models for interatomic
and intermolecular potentials.  This tradition/paradigm can be traced back to the 18th century when
``\ldots\ Bernoulli had proposed that in Boyle's law the specific volume $v$ be replaced by $(v-b)$,
where $b$ was thought to be the volume of the molecules''.~\cite{Rowlinson1973} During the past 25
years, the accuracy of the calculated thermophysical properties of the noble gases (particularly
helium) has increased dramatically.  An example is shown in Fig.~\ref{fig:paradigm-1}, which shows
how the accuracy of the second virial coefficient $B(T)$ of ${}^4$He improved with time.  The data
plotted are for temperatures near $T_\mathrm{Ne}$, ($T_\mathrm{Ne} \equiv 24.5561$~K is the defined
temperature of the triple point of neon on the international temperature scale, ITS-90.)  Since the
year 2012, the uncertainty of $B(T_\mathrm{Ne})$, as calculated {\em ab initio}, has been smaller
than the uncertainty of the best measurements of $B(T_\mathrm{Ne})$.

The paradigm reversal (replacing measured thermophysical properties of helium with calculated
thermophysical properties) applies to zero-density values of the viscosity $\eta(T)$, thermal
conductivity $\lambda(T)$, ${}^3$He-${}^4$He mutual diffusion coefficient as well as to the density
and acoustic virial coefficients, relative dielectric permittivity (dielectric constant)
$\epsr(p,T)$, relative magnetic permittivity $\mu_\mathrm{r}(p,T)$, and refractive index $n(p,T) =
\sqrt{\epsr \mu_\mathrm{r}}$. For many of these properties, the values calculated for helium are
standards that are used to calibrate apparatus that measures the same properties for other gases.

The paradigm reversals for $\epsr(p,T)$ and $n(p,T)$ have been combined with technical advances in
the measurement of $\epsr(p,T)$ and $n(p,T)$ to develop novel pressure standards.  One standard
operating at optical frequencies and low pressures (100 Pa $\leq p \leq$ 100 kPa) is more accurate
than manometers based on liquid columns.  (See Section~\ref{sec:p_low} and
Ref.~\onlinecite{Egan16}).  Other standards operating at microwave frequencies and higher pressures
(100 kPa $\leq p \leq$ 7 MPa) have enabled exacting tests of mechanical pressure generators based on
the dimensions of a rotating piston in a cylinder.  (See Section ~\ref{sec:p_intermediate} and
Refs.~\onlinecite{Gaiser2020,Gaiser2022}). At still higher pressures (up to 50 MPa), the values of
helium's density calculated from the virial equation of state (VEOS) have
been used to calibrate magnetic suspension densimeters.  A more accurate high-pressure scale may
result.  In Section \ref{sec:flow}, we will comment on {\em ab initio} calculations of transport
properties and their contribution to improved flow metrology.

\begin{figure}[h]
\center\includegraphics[width=0.9\linewidth]{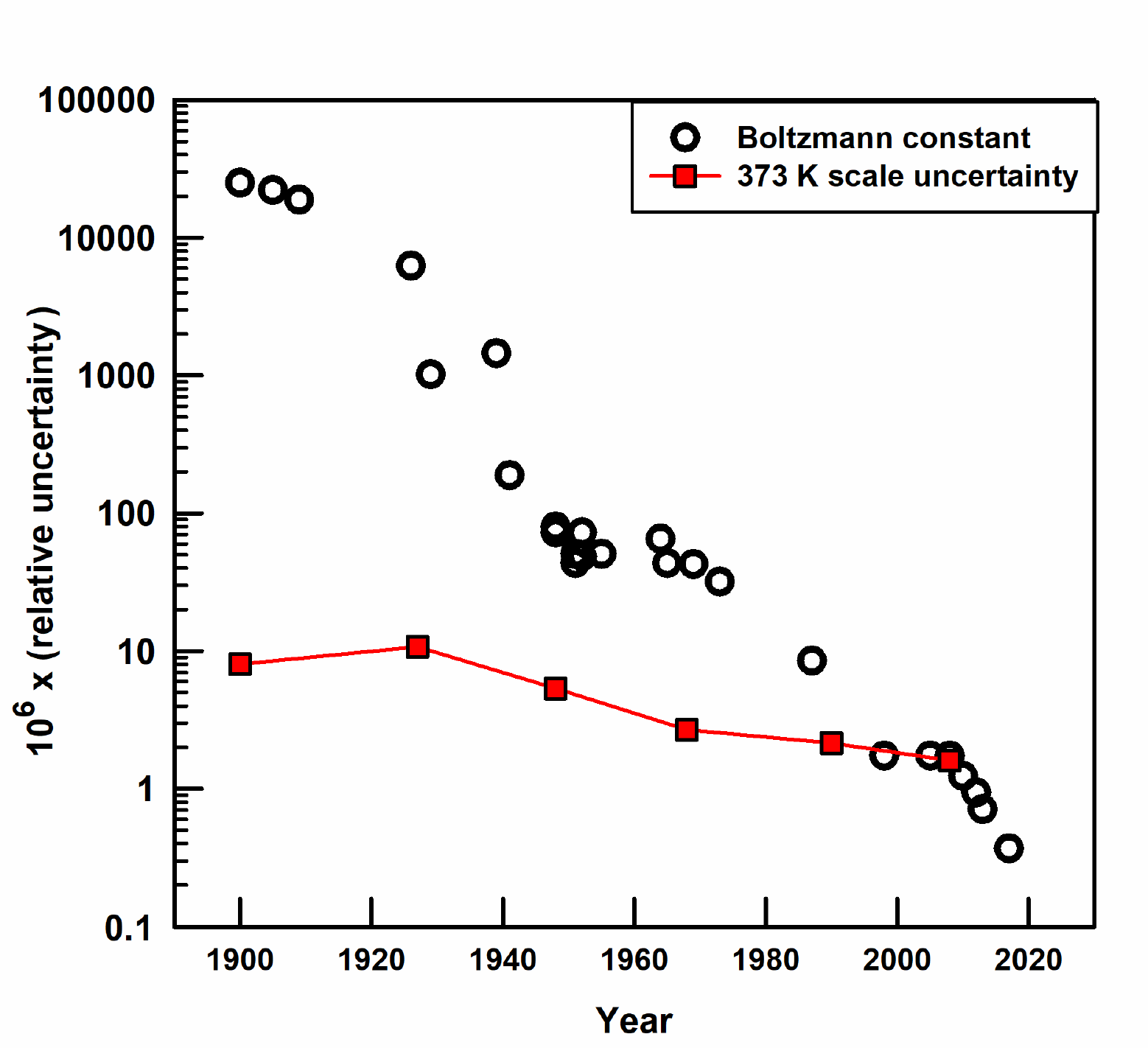} 
\caption{Comparison over time of the standard uncertainty of the Boltzmann constant to the
  reproducibility of the ``practical'' temperature scale. Circles represent the relative uncertainty
  of measurements of the Boltzmann constant, mostly from evaluations by groups such as
  CODATA. Squares represent the relative uncertainty of internationally accepted practical
  temperature scales in the vicinity of the normal boiling point of water.  Adapted from
  Ref.~\onlinecite{White2015}.}
  \label{fig:paradigm-2}
\end{figure}

During the past 25 years, the accurate calculations of the thermophysical properties of the noble
gases have strongly interacted with gas-based measurements of the thermodynamic temperature $T$.  To
put this in context, we compare in Fig.~\ref{fig:paradigm-2} the evolution of “practical”
temperature metrology with “thermodynamic” temperature metrology.~\cite{White2015}

In Fig.~\ref{fig:paradigm-2}, the squares represent estimates of the relative uncertainties
$\ur(T_\mathrm{scale})$ of the “practical” temperature scales disseminated by National Metrology
Institutes (NMIs).  We plot the values of $\ur(T_\mathrm{scale})$ near the boiling point of water at
intervals of roughly 20 years.  Most of the points are at years when the NMIs agreed to disseminate
a new practical scale that was either a better approximation of thermodynamic temperatures and/or an
extension of the practical scale to higher and lower temperatures.  The most recent scale is the
``International Temperature Scale of 1990'' (ITS-90) and temperatures measured using ITS-90 are
denoted $T_{90}$.  The data underlying ITS-90 are constant-volume gas thermometry (CVGT) and
spectral radiation thermometry linked to CVGT.~\cite{Fischer2011} The pre-1990 CVGT was based on the
ideal-gas equation of state, as corrected by virial coefficients either taken from the experimental
literature or measured during the CVGT.  Post-1990 thermometry, together with {\em ab initio}
calculations of virial coefficients, revealed that the authors of ITS-90 underestimated ITS-90’s
uncertainties and were unaware of its biases.  (See Fig.~\ref{fig:AGT} and the discussion at the end
of Sec.~\ref{sec:AGT}).

In Fig.~\ref{fig:paradigm-2}, the circles represent the relative uncertainty of determinations of
the Boltzmann constant $\ur(\kB)$.  To determine $\kB$, one measures the mean energy $\kB T$ per
degree-of-freedom of a system in  thermal equilibrium at the thermodynamic temperature $T$.  During
the interval 1960 to 2019, the thermodynamic temperature of the triple point of water was defined as
$T_\mathrm{TPW} \equiv 273.16$~K, exactly.  Thus, measurements of $\kB T$ that were conducted near
$T_\mathrm{TPW}$ had a negligible uncertainty from $T$ and $\ur(\kB T)$ was an excellent proxy for
$\ur (T)$, the uncertainty of measurements of $T$ under the most favorable conditions.

As displayed in Fig.~\ref{fig:paradigm-2}, $\ur(T_\mathrm{scale})$ decreased from $\sim 10$ ppm to
$\sim 2$ ppm (1 ppm $\equiv$ 1 part in $10^6$) during the 20th century.  Also during the 20th
century, the relative uncertainty $\ur(\kB)$ decreased from $\sim 200,000$~ppm to $\sim 2$~ppm.
Thus, $\ur(\kB) \gg \ur(T_\mathrm{scale})$ for most of the 20th century, even though $\kB$ was a
“fundamental” constant and, therefore, a worthy challenge for metrology.  The 100-fold decrease of
$\ur(\kB)$ from $\sim 40$~ppm in 1973 to $\sim 0.4 $~ppm in 2017 was mostly achieved by refining the
AGT measurements of $\kB$.~\cite{Cohen1973,Newell2018}

In 1995, Aziz {\em et al.}~\cite{Aziz1995} argued that the values of the thermal conductivity
$\lambda(T)$, viscosity $\eta(T)$, and second density virial coefficient $B(T)$ of helium, as
calculated using {\em ab initio} input, were more accurate than the best available measurements of
these quantities.  Subsequently, helium-based AGT measurements of $\kB$ relied on the {\em ab
  initio} values of $\lambda(T)$ to account for the thermoacoustic boundary layer.  Just before the
Boltzmann constant was defined in 2019, the lowest uncertainty measurement of $\kB$ used either the
{\em ab initio} value of thermal conductivity of helium $\lambda_\mathrm{He}(273.16~\mathrm{K})$ or
the value of $\lambda_\mathrm{Ar}(273.16~\mathrm{K})$ that was deduced from ratio measurements using
$\lambda_\mathrm{He}(273.16~\mathrm{K})$ as a standard.~\cite{Pitre2017,Podesta2017}

In 2019, the unit of temperature, the kelvin, was redefined by assigning the fixed numerical value
$1.380649 \times 10^{-23}$ to the Boltzmann constant, $\kB$, when $\kB$ is expressed in the unit
J~K${}^{-1}$.  Thus, the Boltzmann constant can no longer be measured.  However, the
temperature of the triple point of water now has an uncertainty of a few parts in $10^7$, although
the best current value is still 273.16~K.~\cite{Gaiser2022}

As discussed in the next section, the techniques for measuring thermodynamic temperatures are
evolving rapidly.  They are becoming more and more accurate and easier to implement.  We anticipate
NMIs will disseminate thermodynamic temperatures instead of ITS-90 at temperatures below 25 K.  This
would not be possible without the accurate {\em ab initio} values of the thermophysical properties
of helium.

\subsection{Gas thermometry}

\subsubsection{Acoustic gas thermometry}
\label{sec:AGT}

During the past two decades, acoustic gas thermometry (AGT) has emerged as the most accurate primary
thermometry technique over the temperature range $7$~K to $552$~K, achieving uncertainties as low as
$10^{-6} T$. AGT experiments were instrumental in measuring the Boltzmann constant for the
redefinition of the kelvin,~\cite{Fischer2018} and have revealed small, systematic errors in the
ITS-90.~\cite{Fischer2011,Gaiser2022} This section is necessarily brief; for an in-depth review of
AGT, the reader is referred to Ref.~\onlinecite{Moldover2014}.

The underlying principle of AGT is the relationship between thermodynamic temperature, $T$, and the
speed of sound, $w$, in a monatomic gas:
\footnote{Notice that in the literature one might find multiple and inconsistent definitions of the
acoustic virial coefficients, depending on the variable chosen for the expansion of $w^2$ (the
pressure $p$ or the molar density $\rho$) and the powers of $RT$ included in the definition of the
acoustic virials. We used the convention put forward in Ref.~\onlinecite{Gillis1996}; in this case
$\betaa$ has the same dimensions as the second virial coefficient $B$ and $\RTg$ has the same
dimensions as the third virial coefficient $C$.}
\begin{equation}
  w^2 = \frac{\gamma_0 \kB T}{m} \left[
  1 +  \frac{\betaa}{RT} p + \frac{\gammaa}{RT} p^2 + \ldots
  \right],
  \label{eq:agt}
\end{equation}
where $\kB$ is the Boltzmann constant, $R = \kB \NA$ is the gas constant, $m$ is the average
molecular mass of the gas, $\gamma_0$ is the limiting low-pressure value of $c_p / c_v$ where $c_p$
and $c_v$ are the isobaric and isochoric heat capacities, respectively, (this ratio is exactly $5/3$
for a monatomic gas), $p$ is the gas pressure, and $\betaa$ and $\gammaa$ are the
temperature-dependent acoustic virial coefficients. Helium-4 or argon gas is typically used, as
these are considerably less expensive than other noble gases and available in ultra-pure forms,
although xenon has also been used.~\cite{Moldover1999}

Most modern realizations of primary AGT determine the speed of sound from the resonance frequencies
of the acoustic normal modes in a cavity resonator of fixed and stable dimensions. Resonators have
been manufactured from copper, aluminium, and stainless steel, with internal volumes between 0.5
liters and 3 liters. Cavity shapes have either been spherical, quasi-spherical (with smooth,
deliberate deviations from sphericity), or cylindrical. The use of diamond turning to produce
quasi-spherical resonators (QSRs) with extremely accurate forms ($\sim 1~\mu\mathrm{m}$) and
smooth surfaces (average surface roughness on the order of $\approx 3$~nm) has significantly improved
performance.~\cite{Morantz2017} In spherical geometries, the best results are obtained from the
radially symmetric acoustic modes, since these possess high quality factors and are relatively
insensitive to imperfections in the cavity shape. In cylindrical geometries, the longitudinal
plane-wave modes are typically favored.

Two distinct methods of primary AGT exist: absolute and relative. In the absolute method, $T$ is
determined from the limiting low-pressure ($p = 0$) form of Eq.~(\ref{eq:agt}). The terms
$\gamma_0$ and $\kB$ are known exactly; $m$ must be determined by an auxiliary experiment; and $w$
is calculated from the radial acoustic mode frequencies, $\fa$, of the QSR:
\begin{equation}
  w = \frac{\fa - \Delta \fa}{z_\mathrm{a}} (6 \pi^2 V)^{1/3},
  \label{eq:u}
\end{equation}
where $z_\mathrm{a}$ are the acoustic eigenvalues, $\Delta \fa$ is the sum of the acoustic corrections, and $V$ is
the cavity volume. If the longitudinal mode frequencies of a cylindrical cavity are used, the term
proportional to $V^{1/3}$ is replaced with a multiple of the cylinder length. 

Improvements in QSR volume measurements are perhaps the most significant innovation in AGT in the
last two decades, and were driven by efforts to redetermine the Boltzmann constant for the
redefinition of the kelvin. Modern AGT systems measure the frequency $f_\mathrm{m}$ of microwave
resonances in the cavity, which are related to the volume through the equation
\begin{equation}
  \frac{c}{n} = \frac{f_\mathrm{m} - \Delta f_\mathrm{m}}{z_\mathrm{m}} (6 \pi^2 V)^{1/3},
\end{equation}
where $c$ is the speed of light in vacuum, $n$ in the refractive index of the gas in the cavity,
$\Delta f_\mathrm{m}$ is the sum of the electromagnetic corrections, and $z_m$ are the microwave
eigenvalues. The microwave modes do not occur in isolation, being at least 3-fold degenerate in
perfectly spherical cavities. The smooth deformations of the QSR shape lift these degeneracies,
enabling accurate measurement of the individual mode frequencies. A key theoretical result is that
(to first order) the mean frequency of these mode groups is unaffected by volume-preserving shape
deformations.~\cite{Mehl1986}
 
In diamond-turned QSRs, the relative uncertainty in $V$ from the microwave method can be less than
$1 \times 10^{-6}$.~\cite{Underwood2010} This was made possible by improvements in theory,~\cite{Mehl2009}
resonator shape accuracy, and studies of small perturbations due to probes.~\cite{Underwood2010}
Recently, it has been demonstrated that comparable uncertainties can be achieved with 
low-cost microwave equipment.~\cite{Yang2018,Corbellini2013} Accurate microwave
dimensional measurements have also been performed in cylindrical acoustic resonators.~\cite{Zhang2016}

Relative primary AGT measures thermodynamic temperature ratios:
\begin{equation}
  \frac{T}{T_\mathrm{ref}} = \frac{w^2}{w^2_\mathrm{ref}},
\end{equation}
where $w_\mathrm{ref}$ is the measured speed of sound at a known reference temperature
$T_\mathrm{ref}$. Most AGT 
determinations of ($T – T_{90}$) use the relative method. The main advantages are that the molecular mass
term, $m$, cancels in the ratio, and that only the relative volume $V/V_\mathrm{ref}$ need be measured. Also, many
small perturbations to the acoustic and microwave frequencies ({\em e.g.}, due to shape deformations)
either fully or partially cancel in the ratio. As a result, excellent results can be obtained using
resonators with modest form accuracies that would be unsuited to absolute AGT. The disadvantages are
that relative AGT propagates underlying errors and uncertainty in $T_\mathrm{ref}$, and can require the
apparatus to operate over a wide temperature range when no suitable reference points are nearby. 

In both absolute and relative primary AGT, maintaining gas purity is of critical
importance. Impurities will shift the average molecular mass of the gas, and hence the speed of
sound, by an amount that depends on the mass contrast between the bulk gas and impurity. For
example, the speed of sound in helium is approximately 16 times more sensitive to water vapor than
in argon. Impurities can either be present in the gas source or arise from outgassing or leaks in the
apparatus itself.

Relative AGT requires only that $m$ remain unchanged between the measurements at $T$ and
$T_\mathrm{ref}$. Temperature dependence in $m$ can arise through several mechanisms: impurities such
as water, hydrocarbons, or heavy noble gases can be condensed out at low temperatures; higher
temperatures ($>500$~K) cause significant outgassing from the walls of steel
resonators.~\cite{Strouse2003} Gas purity is vastly improved by
maintaining a flow of gas (typically $<50~\mu\mathrm{mol}$/s) through the resonator and supply manifold.

Absolute AGT has more stringent requirements on gas purity than relative AGT. To determine an
accurate value for $m$, both the isotopic abundance of the gas and any residual impurities must be
quantified. Reactive impurities, including water, can be removed from the source gas using gas
purifiers, and noble gas impurities can be removed from helium using a cold trap.~\cite{Pitre2017}
The isotopic ratios ${}^{36}$Ar/${}^{40}$Ar and ${}^{38}$Ar/${}^{40}$Ar in argon, and
${}^3$He/${}^4$He in helium, have been determined by mass spectrometry, and vary significantly from
source to source.~\cite{Yang2015} Alternatively, isotopically pure ${}^{40}$Ar gas can be used,
although this is only available in small quantities and at great expense.~\cite{Moldover1986}

The low uncertainty of the AGT technique arises from the excellent agreement between acoustic theory
and experiment. The simplicity of Eq.~(\ref{eq:u}) hides a number of temperature-, pressure-,
and mode-dependent corrections that constitute the term $\Delta \fa$. The largest of these are the
thermoacoustic boundary layer corrections, which arise from an irreversible heat exchange between
the oscillating gas and resonator walls.~\cite{Moldover1986,Gillis2012} This effect both lowers the
frequency of the acoustic resonances and broadens them; a valuable cross-check of experiment and
theory can be made by comparing the predicted and measured resonance widths. The radial-mode
boundary layer correction in QSRs is approximately proportional to the square root of the gas
thermal conductivity – in cylinders, the gas viscosity also features in the
correction.~\cite{Zhang2010} For most temperature ranges, the uncertainty in these parameters can be
considered negligibly small for both helium and argon due to improved {\em ab initio} calculations (see
section \ref{sec:flow}).

AGT measurements are typically conducted on isotherms in a pressure range between $25$~kPa and
$500$~kPa, with the optimum pressure range depending on several factors such as the type of gas,
temperature, and particular details of the apparatus.~\cite{Moldover2009} At low pressures, the
accuracy in determining $\fa$ is compromised by weak acoustic signals, interference from neighboring
modes due to resonance broadening, and the need to account for details of the interaction of the gas
with the resonator's walls.~\cite{Sharipov2016} At high pressures, higher-order virial terms are
required to account for molecular interactions, and the elastic recoil of the resonator walls
becomes increasingly significant. The shell recoil effect, which shifts $\fa$ in proportion to gas
density,~\cite{Mehl1985} is difficult to predict in real resonators~\cite{Gavioso2010,Truong2011}
because of the complex mechanical properties of the joint(s) formed when the cavity resonator is
assembled.

For this and other reasons, it is not common practice to use Eq.~(\ref{eq:agt}) to determine $T$
from $w$; instead, the measured data are fitted to low-order polynomials that account for the virial
coefficients and perturbations that are proportional to pressure. Isotherm measurements have the
advantage of data redundancy and reduced uncertainty, but are very slow to execute, with each
pressure point taking several hours. Single-state AGT,~\cite{Gavioso2019} which utilizes
low-uncertainty {\em ab initio} calculations of $\betaa$ and $\gammaa$ in helium, offers a much
faster means of primary thermometry.

\begin{figure}[h]
\center\includegraphics[width=0.9\linewidth]{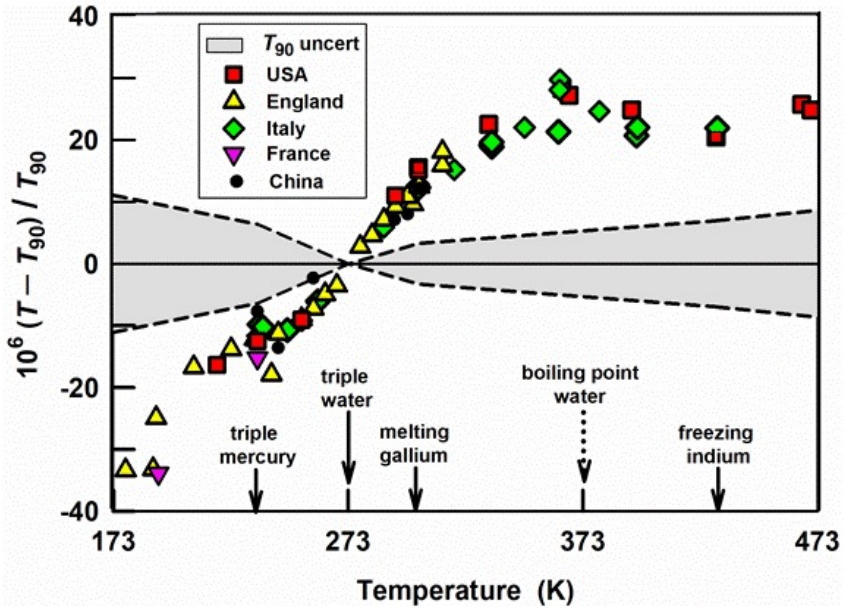} 
\caption{Post-1990 acoustic measurements of $T-T_{90}$.  The shaded area encloses 1990 estimates of
  the fractional uncertainties of ITS-90.  The acoustic measurements indicate that ITS-90 has an
  error of $\sim 25 \times 10^{-6}T$ near water's boiling point and $\sim -35 \times 10^{-6}T$ near
  173~K.
  The solid arrows indicate some ITS-90 fixed points.  (The boiling point of water is not a fixed
  point on ITS-90.)  Data sources: USA: Refs.~\onlinecite{Ripple07,Strouse2003,Moldover1999};
  England: Refs.~\onlinecite{Underwood2017,Ewing2000}; Italy: Ref.~\onlinecite{Benedetto2004};
  France: Ref.~\onlinecite{Pitre2006}; China: Ref.~\onlinecite{Zhang20}. 
  }
  \label{fig:AGT}
\end{figure}

Figure~\ref{fig:AGT} compares AGT measurements from 5 countries with ITS-90.  The AGT data indicate
that ITS-90 has an error of $\sim 25 \times 10^{-6}T$ near water's boiling point and $\sim -35
\times 10^{-6}T$ near 173~K. Near $T_\mathrm{TPW}$, the derivative $\D T_{90}/\D T \approx 1 + 1.0
\times 10^{-4}$.  This implies that heat capacity measurements made using ITS-90 will generate
values of the heat capacity that are 0.010\% larger than the true heat capacity.  However, we are
not aware of heat capacity measurement uncertainties as low as 0.01\%.

Prior to the AGT publications shown in Fig.~\ref{fig:AGT}, Astrov {\em et al.} corrected an estimate
used in their CVGT.  They had used measurements of the linear thermal expansion of a metal sample to
estimate the thermal expansion of the volume of their CVGT ``bulb''.~\cite{Astrov95} Using
additional expansion measurements, Astrov {\em et al.} corrected their $T-T_{90}$ results.  They now
agree, within combined uncertainties, with the AGT data.  (Because AGT uses microwave resonances to
measure the cavity's volume {\em in situ}, it is not subject to errors from auxiliary measurements
of thermal expansion.)

\subsubsection{Dielectric constant gas thermometry}

DCGT, developed in the seventies in the U.K.~\cite{Gugan80,Gugan91} and later improved by
PTB,~\cite{Luther96,Gaiser2015} is now a well-established method of primary thermometry.  The basic
idea of DCGT is to replace the density in the equation of state of a gas by the relative
permittivity (dielectric constant) $\epsr$ and to measure it by the relative capacitance changes at
constant temperature:
\begin{equation}
  \frac{\Delta C_\mathrm{c}}{C_\mathrm{c}} \equiv
  \frac{C_\mathrm{c}(p)-C_\mathrm{c}(0)}{C_\mathrm{c}(0)}
  =  \epsr - 1+ \epsr \keff p.
  \label{eq:DCGT1}
\end{equation}

In Eq.~(\ref{eq:DCGT1}), $C_\mathrm{c}(p)$ is the capacitance of the capacitor at pressure $p$, and
$C_\mathrm{c}(0)$ that at $p = 0$~Pa and $\keff$ is the effective isothermal compressibility which
accounts for the dimensional change of the capacitor due to the gas pressure. In the low-pressure
(ideal gas) limit, the working equation can be simply derived by combining the classical ideal-gas
law and the Clausius--Mossotti equation:
\begin{equation}
  p = \frac{RT}{A_\varepsilon} \frac{\epsr - 1}{\epsr + 2},
\end{equation}
with the molar polarizability $A_\varepsilon$. For a real gas in a general
formulation including electric fields, both input equations are power
series:
\begin{equation}
  \frac{p}{\rho R T} =1 + B(T)\rho + C(T)\rho^2 + D(T)\rho^3 + \ldots,
  \label{eq:pvirial}
\end{equation}
where  $B(T)$, $C(T)$, and $D(T)$ are the second, third, and fourth density
virial coefficient, respectively, $\rho$ is the molar density, and
\begin{eqnarray}
  \frac{\epsr - 1}{\epsr + 2} &=& A_\varepsilon \rho \left( 1 + b \rho + c
  \rho^2 + d \rho^3 + \ldots \right)
  \label{eq:cm} \\
  &=&
  \rho
    \left(
    A_\varepsilon + B_\varepsilon \rho + C_\varepsilon \rho^2 +
    D_\varepsilon \rho^3 + \ldots
    \right).
    \label{eq:DCGT}  
\end{eqnarray}
In the literature, the quantities $b$, $c$, $d$ and $B_\varepsilon$, $C_\varepsilon$,
$D_\varepsilon$ are both called the second, third, and fourth dielectric virial coefficient,
respectively. The form used in Eq.~(\ref{eq:cm}) comes from the tradition of
DCGT~\cite{Gugan80,Luther96} of factoring out $A_\varepsilon$ so that $b, c$, and $d$ have the same
units as $B, C$, and $D$. Conversely, {\em ab initio} calculations naturally provide the quantities
$B_\varepsilon$, $C_\varepsilon$, and $D_\varepsilon$.

The DCGT working equation is obtained by eliminating the density using Eqs.~(\ref{eq:pvirial}) and
(\ref{eq:cm}) and substituting $\epsr$ with the relative capacitance
change corresponding to Eq.~(\ref{eq:DCGT1}).

This leads to a power expansion in terms of $\Xi =
(\Delta C_\mathrm{c} / C_\mathrm{c}) / (\Delta C_\mathrm{c} / C_\mathrm{c} + 3)$:
\begin{widetext}
\begin{equation}
  p = \left(\frac{A_\varepsilon}{R T} + \frac{\keff}{3} \right)^{-1}
  \left[
    \Xi + \Xi ^2 
    \left( \frac{A_\varepsilon}{RT} + \frac{\keff}{3} \right)^{-1}
    \left[ \frac{B-b}{RT} - \frac{\keff}{3} \left( 1 +
    \frac{B}{A_\varepsilon} \right) \right] +
    \Xi^3 
    \left( \cdots \right)    
    \right].
  \label{eq:DCGT5}
\end{equation}  
\end{widetext}

The higher-order terms contain combinations of both the dielectric and
density virial coefficients and the compressibility. Equation
(\ref{eq:DCGT5}) up to the fourth order can be found in Ref.~\onlinecite{Gaiser19}.

DCGT works as a primary thermometer if the molar polarizability $A_\varepsilon$ and virial
coefficients contained in Eq.~(\ref{eq:DCGT5}) are known from fundamental principles or independent
measurements with sufficient accuracy.  The effective compressibility $\keff$ is also required. For
classical DCGT, where isotherms are measured and the data are extrapolated to zero pressure via
least-squares fitting, only $A_\varepsilon$ and $\keff$ are mandatory. This was the way
thermodynamic temperature was determined for decades.~\cite{Gugan80,Luther96,Gaiser2015}
Consequently, in classical DCGT, {\em ab initio} data on virial coefficients serve as a consistency
check or conversely DCGT is used for determination of virial coefficients to check
theory.~\cite{Gaiser19} Since the theoretical calculations of the virial coefficients for helium
improved drastically, it is now possible to use higher-order virial coefficients from theory to
reduce the number of fitting coefficients or even to use the working equation directly without
fitting and to determine temperature at each pressure point via the rearranged working equation.
Recently, all three approaches have been tested and compared.~\cite{Gaiser21primary} Especially, the
point-by-point evaluation is a shift of paradigm and at the moment only possible for helium, where
the uncertainty of the {\em ab initio} calculations, especially of the second density virial
coefficient, is small enough. Nevertheless, for other gases not only the virial coefficients but also
the molar polarizabilities determined via DCGT have comparable or smaller uncertainties than {\em ab
  initio} calculations.~\cite{Gaiser2018} This is a field of potential improvement of theory already
started with calculations of $A_\varepsilon$ for neon~\cite{Lesiuk2020,Hellmann2022} and for
argon.~\cite{Lesiuk2023}

DCGT was operated in the temperature range from 2.5~K to about 273~K using
helium-3, helium-4, neon,~\cite{Gaiser17} and argon.~\cite{Gaiser20thermo} All noble gases have the
advantage that the molar polarizability is independent of temperature at a level of precision far
beyond that of state-of-the-art experiments.~\cite{Jentschura11}

Besides the use of dielectric measurements in primary thermometry, accurate
determinations of polarizability and virial coefficients of noble gases and
molecules using gas-filled capacitors have a much longer tradition. These
setups, very similar to DCGT, use thermodynamic temperature as one of
the input parameters. A complete overview of measurements cannot be given
here. Already a very broad overview of existing data, partly at radio
frequencies, was summarized by NBS in the 1950’s.~\cite{Maryott53} In the
following decades,~\cite{Johnston60} different
institutes with changing teams performed measurements until the early
1990’s.~\cite{Huot91} In the year 2000, NIST started measurements on gases
using capacitors resulting in the most accurate values for the measured
molecules.~\cite{Buckley2000,Schmidt03} Very recently, PTB established a setup for separate
measurement of dielectric and density virial coefficients using a
combination of Burnett expansion techniques and DCGT.~\cite{Guenz17}  The focus of
this setup is on the determination of properties of energy gases like
hydrogen-methane mixtures in the context of the transition to renewable
energies.

For primary thermometry, most significant recent improvements in DCGT
have been achieved by independent determination of $\keff$ using resonant
ultrasound spectroscopy around $0~{}^\circ$C and an optimal choice of
capacitor materials. For the Boltzmann experiment with measuring pressures
of up to 7~MPa, tungsten carbide was the ideal choice, while at low
temperatures beryllium copper was used together with an extrapolation
method. Relative uncertainties for $\keff$ in terms of temperature on the
level of 1~ppm near $0~{}^\circ$C have been achieved. Equally
important are the improvements in pressure measurement. In contrast to AGT,
where pressure is a second-order effect, in DCGT $\epsr$ is directly linked
to pressure.  Therefore, the relative uncertainty in pressure can be
transferred to a relative uncertainty in temperature. The major steps here
are discussed in section \ref{sec:p_intermediate}
regarding the mechanical pressure standard developed at PTB in the framework of the Boltzmann
constant determination.~\cite{Zandt15} These systems with relative uncertainties on the level of
1~ppm at pressures up to 7~MPa have been used to calibrate commercially available systems for
pressures up to 0.3~MPa with relative uncertainties between 3~ppm and 4~ppm. The dominant
uncertainty component in DCGT measurements is the standard deviation of the capacitance
measurement. The typical relative uncertainty in terms of temperature connected to this component is
on the order of 5~ppm for the low temperature range but was reduced to the 1~ppm level in the case
of the Boltzmann experiment at about $0~{}^\circ$C.~\cite{Gaiser17b} Finally, one problem in DCGT
using helium is the very small molar polarizability compared to all other gases and
molecules. Therefore, special care must be taken concerning impurities and here an especially severe
issue is contamination with water.

The polarizability of water at frequencies of capacitance bridges and microwave
resonators (see section \ref{sec:mol_pol}) is about a factor of 160 larger than that of 
helium. At cryogenic temperatures, water contamination in the gas phase
is naturally reduced by outfreezing but especially at room temperature the
whole measuring setup as well as the gas purifying system must be highly
developed. Furthermore, pollution with other noble gases must be treated
carefully because they cannot be extracted by getters and filters. Ideally,
a mass-spectrometer should be used for the detection of noble gases
impurities to allow for an upper estimate of the uncertainty due to gas
purity. In summary, with DCGT in the low temperature range from 4~K to 25~K
uncertainties on the level 0.2~mK for thermodynamic temperature are
achievable. At around $0~{}^\circ$C, the smallest uncertainty for DCGT was
achieved during the determination of the Boltzmann constant.~\cite{Gaiser17b} Converted
to an uncertainty for thermodynamic temperature, this would lead to about
0.5~mK.

In the intermediate range, the uncertainties are larger (at 200~K between 1~mK and
2~mK~\cite{Gaiser20thermo}). The main restriction of the present low-temperature setup is the
limited pressure range at intermediate temperatures. A measurement of high-pressure isotherms in
this range is planned. Together with improved {\em ab initio} calculations for the second virial
coefficients of argon and neon, a point-by-point evaluation could be possible. This could result in
a significant reduction in both uncertainty and measurement time.

\subsubsection{Refractive index gas thermometry}
\label{sec:rigt}

Both DCGT and RIGT are versions of polarizing gas thermometry. Both rely on
virial-like expansions of either the dielectric constant $\epsr$ or of the
refractive index $n$ in powers of the molar density $\rho$, that is Eq.~(\ref{eq:DCGT}) in the case
of DCGT, and the Lorentz--Lorenz equation 
\begin{equation}
  \frac{n^2 - 1}{n^2 + 2} = \rho
    \left(
    A_\varepsilon + A_\mu + B_\mathrm{R} \rho + C_\mathrm{R} \rho^2 + \ldots
    \right),
    \label{eq:RIGT}
\end{equation}
in the case of RIGT.  In the limit of zero frequency, $A_\mu/A_\varepsilon \approx -1.53 \times
10^{-5}$ for He, $B_\varepsilon = B_\mathrm{R}$, $C_\varepsilon = C_\mathrm{R}$, {\em
  etc.}~\cite{Garberoglio2020} Except for the small magnetic-permeability term $A_\mu$ (which is
well-known from theory for helium~\cite{Bruch02}), low-frequency measurements of $n$ and
of $\epsr$ are analyzed using the same {\em ab initio} constants. RIGT determines the thermodynamic
temperature $T$ by combining measurements of the pressure $p$ with the density virial equation of
state, Eq.~(\ref{eq:pvirial}),
and Eq.~(\ref{eq:RIGT}). The density is eliminated from both equations,
either numerically or by iteration, to obtain 
\begin{equation}
  T = \frac{p \left( A_\varepsilon + A_\mu \right) }{R}
  \frac{n^2+2}{n^2-1} + \ldots
\label{eq:T_RIGT}
\end{equation}

The constants $B$, $B_\mathrm{R}$, $C$, $C_\mathrm{R}$, {\em etc.} that appear in the higher-order
terms of Eq.~(\ref{eq:T_RIGT}) are obtained either from theory or from fitting measurements of
$n^2(p)$ on isotherms. [DCGT determines $T$ using a version of Eq.~(\ref{eq:T_RIGT}) in which
  $\epsr$ replaces $n^2$.]

Here, we focus on RIGT conducted at microwave frequencies as developed by Schmidt {\em et
  al.}~\cite{Schmidt07} and as recently reviewed by Rourke {\em et al.}~\cite{Rourke19} These
authors determined $n$ from measurements of the microwave resonance frequencies $f_\mathrm{m}$ of a
gas-filled, metal-walled, quasi-spherical cavity. Typical frequencies ranged from 2.5~GHz to 13~GHz;
for this range, the frequency dependence of $n$ in the noble gases is negligible. As discussed in
Sec.~\ref{sec:p_low}, RIGT has also been realized at optical frequencies in the context of pressure
standards.~\cite{Egan17} For helium, the corrections of $A_\varepsilon$ and $B_\mathrm{R}$ from
optical frequencies to zero frequency have been calculated {\em ab
  initio}.~\cite{Garberoglio2020,Puchalski:16}

A working equation for measuring $n$ is:
\begin{equation}
  n = \sqrt{\mu \epsr} =
  \frac{\langle f_\mathrm{m} + g \rangle_\mathrm{vacuum}}{\langle f_\mathrm{m}+g
    \rangle_\mathrm{pressure} (1 - \kappa_\mathrm{eff} p)},
\label{eq:n_working}
\end{equation}
where the brackets ``$\langle \rangle$'' indicate averaging over the frequencies of a
nearly degenerate microwave multiplet and $g$ accounts for the penetration of
the microwave fields into the cavity's walls. Usually, $g$ is determined from
measurements of the half-widths of the resonances; its contribution to
uncertainties is small. The term $\kappa_\mathrm{eff} p$ accounts for the
temperature-dependent change of the cavity's volume in response to the gas
pressure $p$. Often, the uncertainty of $\keff$ is the largest contributor to
the uncertainty of RIGT. To make this explicit, we manipulate
Eqs.~(\ref{eq:T_RIGT}) and (\ref{eq:n_working}) to obtain: 
\begin{equation}
  T = \frac{3p}{R}
  \left(
  \frac{A_\varepsilon + A_\mu}{n^2 - 1}
  \right)
  \left( 1 - 
  \frac{2 \kappa_\mathrm{eff} R T n^2}{3(A_\varepsilon + A_\mu)}
  \right)
  + \ldots,
  \label{eq:T_RIGT_2}
\end{equation}
where the term $2 \kappa_\mathrm{eff} R T
n^2/[3(A_\varepsilon + A_\mu)] \approx 0.007$ for a
copper-walled cavity immersed in helium near $T_\mathrm{TPW}$. (This
estimate assumes that the cavity's walls are homogeneous and isotropic;
therefore, $\kappa_\mathrm{eff} = \kappa_T/3$ where
$\kappa_T$ is the isothermal compressibility of copper.) Thus, a 
relative uncertainty $\ur (\kappa_\mathrm{eff}) = 0.01$ contributes the relative uncertainty
$\ur (T_\mathrm{TPW}) = 70 \times 10^{-6}$ to a RIGT determination of
$T_\mathrm{TPW}$. In the approximation $n^2 \approx 1$, this uncertainty
contribution is a function of $T \times \kappa_\mathrm{eff}(T)$, but it is 
not a function of the pressures measured on an isotherm. Because $T \times \kappa_\mathrm{eff}(T)$
decreases with $T$, RIGT is more attractive at cryogenic temperatures than
near or above $T_\mathrm{TPW}$.

Recently, two independent groups explored a two-gas method for measuring $\keff$ of assembled RIGT
resonators.~\cite{Rourke20,MadonnaRipa21} Ideally, two-gas measurements would replace measurements
of $\kappa_T$ of samples of the material comprising the resonator's wall and also models for the
cavity's deformation under pressure. Both groups relied on new, accurately measured and/or
calculated values of the density and refractivity virial coefficients of neon or
argon.~\cite{Garberoglio2020,Gaiser19} Using helium and argon, Rourke determined $\keff$ at
$T_\mathrm{TPW}$ with the remarkably low uncertainty $\ur(\keff) = 9.6 \times
10^{-4}$.~\cite{Rourke20} Madonna Ripa {\em et al.} combined helium and neon data to reduce the
uncertainty contribution from $\keff$ to their determinations of $T$ at the triple points of O${}_2$
($\approx 54$~K), Ar ($\approx 84$~K), and Xe ($\approx 161$~K).~\cite{MadonnaRipa21} They reported
``partial success'' and suggested that a revised apparatus using both gases and operating at higher pressures
($p > 500$~kPa) would obtain lower uncertainty determinations of $T$. They also noted that the
two-gas method requires twice as much RIGT data, accurate pressure measurements, and dimensional
stability between gas fillings

Rourke's review of RIGT~\cite{Rourke19} noted 5 groups implementing RIGT using microwave
technology. In contrast, we are aware of only one group (at PTB) implementing
DCGT.~\cite{Gaiser21primary} The relative popularity of RIGT results from the commercial
availability of vector analyzers that can measure microwave frequency ratios with resolutions of
$10^{-9}$. To our knowledge, using commercially available capacitance bridges, the highest
attainable capacitance ratio resolution is $70 \times 10^{-9}$.~\cite{Andeen} To attain higher
resolution for DCGT, PTB developed a unique bridge that measures capacitance ratios with a
resolution of order $10^{-8}$ in a 1~s averaging time. To achieve this specification, the PTB bridge
must operate at 1~kHz and both the standard (evacuated) capacitor and the unknown (gas-filled)
capacitor must have identical construction and be located in the same thermostat.~\cite{Fellmuth11}

\begin{figure}[h]
\center\includegraphics[width=0.9\linewidth]{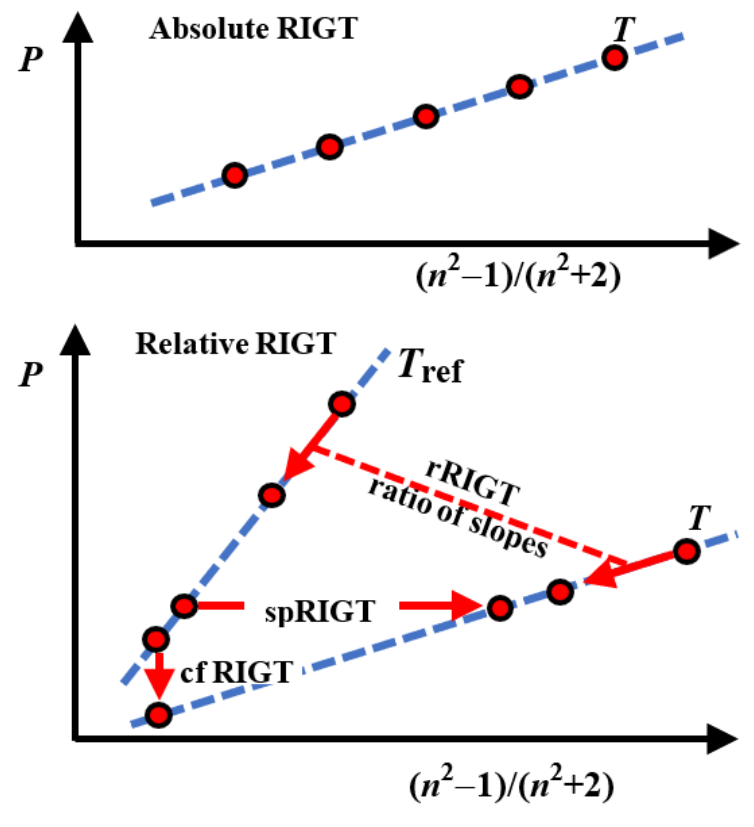} 
  \caption{Measurement trajectories for RIGT in the variables pressure ($p$)
    and refractive index ($n$). Blue dashed lines represent isotherms. Top:
    Absolute RIGT takes many data points on an isotherm at the unknown
    temperature $T$. Bottom: Relative rRIGT takes several measurements on a
    reference isotherm $\Tref$ and on an unknown isotherm $T$. Single pressure
    (spRIGT) uses data at one pressure. Constant frequency (cfRIGT) uses
    data at one value of the refractive index $n$.}
  \label{fig:RIGT}
\end{figure}

Figure~\ref{fig:RIGT} illustrates the several strategies being
explored for acquiring RIGT data. Absolute RIGT acquires many $(p, n)$ data
on an isotherm and determines $T$ via Eq.~(\ref{eq:T_RIGT_2}). This method requires
state-of-the-art, absolute pressure measurements; therefore, the pressure
gradient between the gas-filled cavity and the manometer (normally at
ambient temperature) is required.~\cite{Pan20} Uncertainty budgets for absolute
RIGT can be found in Refs.~\onlinecite{Rourke20,MadonnaRipa21}.

Relative RIGT (rRIGT) comes in several flavors, each designed to simplify
some aspect of absolute RIGT. Each flavor requires measurements
on at least two isotherms: (1) a reference isotherm $\Tref$ for which the
thermodynamic temperature is already well known, and (2) an
unknown isotherm for which $T$ will be determined. As suggested in
the lower panel of Fig.~\ref{fig:RIGT}, one flavor of rRIGT determines
$T/\Tref$ by determining the low-pressure limit of the ratio of
slopes~\cite{Schmidt07}
\begin{equation}
  \frac{T}{\Tref} =
  \lim_{p \to 0} \left(
    \frac{n^2_T - 1}{n^2_{\Tref} - 1}
    \right).
    \label{eq:TRIGT_lim}
\end{equation}
If $\Tref$ and $T$ are low temperatures, where the pressure deformation
of the cavity $\kappa_\mathrm{eff} p$ is small, this strategy circumvents
the problem of 
accurately determining $\kappa_\mathrm{eff}$.

Single-pressure RIGT (spRIGT) measures $(p, n, T)$ and $(p, n, \Tref)$ and determines $T$ from
$T/\Tref \approx (n^2_T - 1)/(n^2_{\Tref} - 1)$. This strategy entirely avoids accurate pressure
measurements; instead, the pressure in the cavity is required to be identical when $n$ is measured
at $T$ and $\Tref$ and the pressure (actually, the gas' density) must be sufficiently low that an
approximate pressure is adequate for making the virial corrections. This strategy was used by Gao
{\em et al.} for RIGT between the triple point of neon ($\Tref \approx 24.5$~K) and
5~K.~\cite{Gao20} After establishing $\Tref$ by acoustic thermometry, they claimed the uncertainties
of this implementation of RIGT were smaller than the uncertainties of ITS-90.~\cite{Pan21}

When constant-frequency RIGT (cfRIGT) is implemented, the
pressure in the cavity is changed to keep the refractive index
constant as the temperature is changed from $\Tref$ to $T$. In this case,
$T/\Tref \approx p(T,n)/p(\Tref,n)$.~\cite{Zhang23} This scheme minimizes the frequency-dependent
effects of the coaxial cables on the microwave determination of $T/\Tref$.

To economically search for measurement or modeling errors, one can obtain 3 redundant values of
$T/\Tref$ by measuring microwave frequencies at 4 judiciously chosen values of $(p, n)$. Two
measurements are made on the isotherm $\Tref$ at the values $(p_1, n_1)$ and $(p_2, n_2)$. Two other
measurements are on the isotherm $T$ at $(p_2, n_3)$ and $(p_3, n_1)$. spRIGT connects the points
$(p_2, n_2)$ and $(p_2, n_3)$. cfRIGT connects the points $(p_1, n_1)$ and $(p_3, n_1)$. All 4
points are used to approximately implement rRIGT via Eq.~(\ref{eq:TRIGT_lim}).

Compared with other forms of gas thermometry, relative RIGT has significant advantages at low
temperatures. We have already emphasized the availability of microwave network analyzers and the
possibility of avoiding state-of-the art pressure measurements.  By measuring several microwave
resonance frequencies at each state, certain imperfections of the measurements and modeling can be
detected. Comparisons of the frequencies of TE and TM microwave modes might detect the presence of
dielectric films such as oxides, oil deposits, or adsorbed water on the cavity's
walls.~\cite{May2004} Because relative RIGT relies on microwave frequency ratios, the precise shape
of the cavity is unimportant. Cavity shapes other than quasispheres may be advantageous in
particular applications.

RIGT is simpler and more rugged than relative acoustic gas thermometry (rAGT) because RIGT requires
neither delicate acoustic transducers nor acoustic ducts. However, RIGT is unlikely to replace rAGT
at ambient and higher temperatures because RIGT is more sensitive to the cavity's dimensions than
rAGT by the factor $1/(\epsr-1)$ which typically ranges from 200 to 20000. Furthermore, microwave
RIGT is especially sensitive to polar impurities. Adding 1~ppm (mole fraction) of water vapor to
dilute argon gas at 293~K will increase the gas' dielectric constant by 18~ppm and increase the
square of the gas' speed-of-sound by 0.12~ppm. If the water vapor were undetected, these changes
would reduce the argon's apparent RIGT temperature by 18~ppm and argon's apparent rAGT temperature
by $-0.12$~ppm. For helium, the corresponding temperature reductions are 145~ppm and 4~ppm.

\subsubsection{Constant volume gas thermometry}
\label{sec:cvgt}

The website of the International Bureau of Weights and Measures includes a document (``{\em Mise en
  pratique} $\ldots$'') that indicates how the SI base unit, the kelvin, may be realized in practice
using 4 different versions of gas thermometry.~\cite{MEPK} Surprisingly, this document omits
CVGT, the version of gas thermometry that was the primary basis of
ITS-90. In this section, we briefly describe the operation of a particular realization of CVGT and
the inconsistent results it generated. This may explain why CVGT was omitted from the {\em Mise en
  pratique}. We mention the post-1990 theoretical and experimental developments that suggest an
updated realization of CVGT might generate very accurate realizations of the kelvin.

CVGT at NBS/NIST began in 1928 and concluded in 1990. We denote the
most-recent realization of NBS/NIST's relative CVGT by
``CV${}_\mathrm{NIST90}$''. The heart of \CVNIST\ was a
metal-walled, cylindrical cavity (``gas bulb''; $V \sim 407$~cm${}^3$)
attached to a ``dead space'' comprised of a capillary leading from the bulb
to a constant-volume valve at ambient temperature. The valve separated the
gas bulb from a pressure-measurement system. A typical temperature
measurement using \CVNIST\ began by admitting $N_\mathrm{r} \sim
0.0023$~mol of helium into the gas bulb at a measured reference pressure
($p_\mathrm{r} \sim 13$~kPa) and a measured reference temperature
($T_\mathrm{r} \sim T_\mathrm{TPW}$).~\cite{Schooley90,Edsinger89} Then, the valve was closed to
seal the helium in the gas bulb and dead space. The bulb was moved into a
furnace that was maintained at the unknown temperature $T$ to be determined
by CVGT. After the gas bulb equilibrated, the valve was opened to measure
the pressure $p$ again. The temperature ratio $T/T_\mathrm{r}$ was determined by applying
the virial equation at each temperature:
\begin{equation}
  \frac{T}{T_\mathrm{r}} =
  \frac{p V_T}{N_T R \left(1 + (B N / V)_T
     + \ldots \right)}
  \frac{N_\mathrm{r} R (1 + (B N / V)_\mathrm{r} + \ldots)}{p_\mathrm{r} V_\mathrm{r}}
\label{eq:TCVGT}
\end{equation}
Thus, $T/T_\mathrm{r}$ is determined, in leading order, by the three ratios: $p/p_\mathrm{r}$,
$V_T/V_\mathrm{r}$, and $N_\mathrm{r}/N_T$. For \CVNIST,
$N_\mathrm{r}/N_T \neq 1$ because a tiny quantity of
helium flows from the bulb into the capillary when the bulb is
moved into the furnace. This quantity was calculated using the
measured temperature distribution along the capillary. For \CVNIST,
$V_T/V_\mathrm{r}$ was calculated using auxiliary measurements of the linear
thermal expansion of samples of the platinum-rhodium alloy
comprising the gas bulb. These samples had been cut out of the gas
bulb after completing all the CVGT measurements.

The simplicity of Eq.~(\ref{eq:TCVGT}) hides the many complications of
CVGT. We mention three examples. (1) During pressure measurements, helium
outside the gas bulb was maintained at the same pressure as the helium
inside the gas bulb. (2) Thermo-molecular and hydrostatic pressure
gradients in the capillary were taken into account. (3) At high
temperatures, creep in the gas bulb's volume was detected by time-dependent
pressure changes; the pressure was extrapolated back in time to its value
when the bulb was placed in the furnace.

We denote the second most recent realization of NBS/NIST's relative CVGT by
``\CVNBS''.~\cite{Guildner76} Both \CVNIST\ and \CVNBS\ shared apparatus and many
procedures. However, Ref. \onlinecite{Schooley90} lists 11 significant changes. Here, we mention
only one. \CVNIST's two cylindrical, gas bulbs had been fabricated
entirely from sheets of (80 wt\% Pt + 20 wt\% Rh) alloy. The sides and
bottom of \CVNBS's gas bulb were fabricated from the same alloy; however,
the top of the bulb was inadvertently fabricated from (88 wt\% Pt + 12
  wt\% Rh) alloy. Perhaps the slight differences in thermal expansions of
these alloys led to an anomalous thermal expansion of the volume of \CVNBS's
gas bulb.

Unfortunately, the results from \CVNIST\ and \CVNBS\ were inconsistent, within claimed
uncertainties, in the range of temperature overlap (505~K $\leq T \leq$ 730~K). An approximate
expression for the differences is: $T_\mathrm{NIST90} - T_\mathrm{NBS76} \approx 0.090 \times
(T/\mathrm{K}-400)$~mK. This inconsistency was not explained by the authors of \CVNIST\ nor by the
authors of \CVNBS. Furthermore, the authors did not assert the more recent \CVNIST\ results were
more accurate than the earlier \CVNBS\ results. The working group that developed ITS-90 had no other
data, from NIST or elsewhere, that were suitable for resolving the inconsistency. Therefore,
the working group required ITS-90 to be the average of $T_\mathrm{NIST90}$ and $T_\mathrm{NBS76}$ in
the overlap range.~\cite{Rusby91}

In the range 2.5~K to 308~K, ITS-90 relied, in part, on another realization of CVGT that had a
troubled history. Astrov {\em et al.}  deduced the thermal expansion of their copper gas bulb's
volume from measurements of the linear thermal expansion of copper samples taken from the block used
to manufacture their bulb.~\cite{Astrov89} However, the thermal expansion data were inconsistent
with other data for copper. Astrov's group repeated the thermal expansion measurements using another
(better) dilatometer. The more recent expansion data, published in 1995, changed the values of $T$
by more than $50 \times 10^{-6} T$ in the range 130~K~$ < T < $~180~K, where the uncertainties had
been estimated as $\leq 26\times 10^{-6} T$.~\cite{Astrov95}

Recently, a working group of the Consultative Committee for Thermometry reviewed primary thermometry
below 335~K.~\cite{Gaiser2022} Astrov’s revised CVGT values are close to the current consensus, which
is primarily based on AGT and DCGT. The working group retained three other low-temperature
realizations of CVGT. Post-1990 AGT measurements of $T-T_{90}$ near 470~K and 552~K indicate that
\CVNIST\ is indeed more accurate than \CVNBS.~\cite{Ripple07} Despite the fact that CVGT was the
primary basis for the ITS-90, the {\em Mise en pratique} does not include CVGT. We speculate that no
temperature metrology group is pursuing CVGT because: (1) CVGT is complex, (2) Astrov {\em et al.}'s
thermal expansion problem, (3) unexplained problems with NBS/NIST's CVGT, and (4) rapid advances in
other versions of gas thermometry.

We now ask: is CVGT a viable method of primary thermometry today? The gas bulb of a modern CVGT
would incorporate feedthroughs to enable measuring microwave resonance frequencies of the bulb's
cavity. The resonance frequencies would determine the bulb's volumetric thermal expansion, thereby
avoiding auxiliary measurements of linear thermal expansion and also avoiding the assumption of
isotropic expansion. If the bulb incorporated a valve and a differential-pressure-sensing diaphragm,
the dead-space corrections would vanish. (The diaphragm's motion could be detected using optical
interferometry.) Today, the {\em ab initio} values of $B(T)$ would reduce the uncertainty component
from $B(T)$ to near zero. A contemporary CVGT could operate at $\sim 5 \times$ higher helium
densities than published experiments without generating significant uncertainties from either the
virial coefficients or from pressure-ratio measurements. The higher density, together with
simultaneous pressure and microwave measurements, might enable separation of the bulb's creep from
contamination by outgassing. Most outgassing contaminants affect helium's dielectric constant,
refractivity, and speed of sound much more than they affect helium's pressure, an advantage of
CVGT. However, CVGT inherently uses fixed aliquots of gas.  Therefore, CVGT cannot benefit from
flowing gas techniques that have been used, for example, in high-temperature AGT.~\cite{Ripple07} In
summary, contemporary CVGT could be competitive with other forms of primary gas thermometry, with a
possible exception at the highest temperatures, where flowing gas might be required to maintain gas
purity.

\subsection{Pressure metrology}
\label{sec:pressure}

Traditionally, standards based on the realization of the {\em mechanical} definition of pressure,
the normal force applied per unit area onto the surface of an artifact, include pressure balances
and liquid column manometers. The combined overall pressure working range of these instruments
extends over seven orders of magnitude, roughly between 10~Pa and 100~MPa. Liquid column manometers
achieve their best performance, with relative standard uncertainty as low as 2.5~ppm, near their
upper working limit at a few hundred kPa.~\cite{Pavese16} With a few notable exceptions, the typical
relative standard uncertainty of pressure balances spans between nearly $1 \times 10^{-3}$ at 10~Pa,
the lowest end of their utilization range, down to 2 to 3~ppm in the range between 100~kPa and
3~MPa.~\cite{Pavese16,Schmidt06} One such exception is the remarkable achievement of a relative
standard uncertainty as low as 0.9~ppm for the determination of helium pressures up to
7~MPa,~\cite{Zandt15} though this achievement required the extensive dimensional characterization,
and the cross-float comparison, of the effective areas of six piston--cylinder sets manufactured to
extraordinarily tight specifications, with a research effort lasting several years. In spite of this
outstanding result, the accurate characterization of pressure balances is challenging, due to the
complexity of the dimensional characterization of the cross-sectional area of piston-cylinder
assemblies, which includes finite-element modeling of their deformation under
pressure.~\cite{Sabuga11,Sharipov16} International comparisons periodically provide realistic
estimates of the average uncertainty of realization of primary standards among NMIs. In 1999, a
comparison of primary mechanical pressure standards in the range 0.62~MPa~$ < p < $~6.8~MPa,
involving five NMIs leading in pressure metrology exchanging a selected piston-cylinder set, was
completed.~\cite{Molinar2000} The resulting differences $\Delta A_\mathrm{eff} \equiv 10^6
(A_\mathrm{eff}/\langle A_\mathrm{eff}\rangle -1)$ of the effective area $A_\mathrm{eff}$ of the
piston from the reference value $\langle A_\mathrm{eff}\rangle$ spanned beyond their combined
uncertainties with such significant spread to show that the pressure standards realized by different
NMIs were mutually inconsistent.

These inconsistencies strengthened the motivation for the development of standards realizing a {\em
  thermodynamic} definition of pressure by the experimental determination of a physical property of
a gas having a calculable thermodynamic dependence on density, combined with accurate thermometry.
This possibility was initially proposed in 1998 by Moldover,~\cite{Moldover1998} who envisaged,
already at that time, the potential of first-principles calculation to accurately predict the
thermodynamics and electromagnetic properties of helium and the maturity of experiments determining
the dielectric constant using calculable capacitors.
The metrological performance of thermodynamic pressure standards has continuously improved over the
last two decades to become increasingly competitive in terms of accuracy, providing important
alternatives which may test the exactness of the mechanical standards discussed above and eventually
replace some of them. Also, due to their reduced complexity and bulkiness, simplified versions of
thermodynamics-based standards may be more flexibly adapted to specific technological and scientific
applications of pressure metrology.  The best-performing recent realizations of gas-based pressure
standards include measurements of the dielectric constant using capacitors and of the refractive
index at microwave and optical frequencies, respectively using resonant cavities and
Fabry--P{\'e}rot refractometers. In Secs.~\ref{sec:p_low} and \ref{sec:p_intermediate}, we
separately discuss the most notable of these developments depending on the pressure range of their
application.

\subsubsection{Low pressure standards (100 Pa to 100 kPa)}
\label{sec:p_low}

In the low vacuum regime, several experimental methods are available which may provide alternative
routes for traceability to the pascal.  For the cases involving optical measurements, these methods
include: (1) refractometry (interferometry), implemented in various configurations that employ
single or multiple cavities or cells with fixed or variable path lengths; (2) line-absorption
methods. The achievements and perspectives of all these methods were recently
reviewed.~\cite{Jousten2017}

At present, Fabry--P{\'e}rot refractometry with fixed length optical cavities (FLOC) has
demonstrated the lowest uncertainty for the realization of pressure standards near atmospheric
pressure and down to 100~Pa. In principle, the uncertainty of this method is limited by several
optical and mechanical effects, most importantly by the change in the length of the cavity due to
compression by the test gas, with the same sensitivity to the imperfect estimate of the
compressibility $\kappa_T$ that affects RIGT. However, this major uncertainty contribution may be
drastically reduced, though not completely eliminated, by measuring the pressure-induced length
change of a second reference FLOC monolithically built on the same spacer, which is kept
continuously evacuated. In 2015, a dual-cavity FLOC achieved an extremely accurate determination of
the refractive index of nitrogen at $\lambda = 632.9908$~nm, $T = 302.9190$~K and 100.0000~kPa by
reference to the pressure realized by a primary standard mercury manometer, and using refractive
index measurements in helium to determine the compressibility.~\cite{Egan15} A comparison of the
pressures determined by the nitrogen refractometer with the mercury manometer below the primary
calibration point at 100~kPa down to 100~Pa showed relative differences within 10~ppm.  A direct
comparison between laser refractometry with nitrogen and a mercury manometer was realized one year
later also at NIST.~\cite{Egan16} The comparison showed relative differences between these
instruments within 10~ppm over the range between 100~Pa and 180~kPa.  The laser refractometer
outperforms the precision and repeatability of the liquid manometer and demonstrates a pressure
transfer standard below 1~kPa that is more accurate than its current primary realization.  Such
remarkably low uncertainty also favorably compares to the best dimensional characterization and
modeling of non-rotating piston-cylinder assemblies.\cite{Naris18}

In 2017, more accurate measurements in helium and nitrogen were performed between 320~kPa and
420~kPa using a triple-cell heterodyne interferometer referenced to a carefully calibrated piston
gauge, showing relative differences within 5 ppm with uncertainties on the order of
10~ppm.~\cite{Egan17} Some pressure distortion errors affecting FLOC might in principle be
eliminated by refractive index measurement with a variable length optical cavity (VLOC).  The
realization of this technique requires extremely challenging dimensional measurements, with
displacements on the order of 15~cm that must be determined with picometer
uncertainty.~\cite{Stone13}
Gas modulation techniques, with the measuring cavity frequently and repeatedly switched between a
filled and evacuated condition, have been recently developed~\cite{Silander19,Forssen22} aiming at
the reduction of the effects of dimensional instabilities and other short- and long-term
fluctuations that affect Fabry--P{\'e}rot refractometers.  A novel realization of an optical
pressure standard, based on a multi-reflection interferometry technique, has also been recently
developed, demonstrating the possible realization of the pascal with a relative standard uncertainty
of 10 ppm between 10 kPa and 120 kPa.~\cite{Mari23}
Optical refractometry for pressure measurement is also being pursued at other
NMIs.~\cite{Takei2020,Yang2021}

At microwave frequencies, the realization of a low-pressure standard requires a substantial
enhancement in frequency resolution. Recently, it was demonstrated by Gambette {\em et al.} that by
coating the internal surface of a copper cavity with a layer of niobium, and working at temperatures
below 9~K where niobium becomes superconducting, pressures in the range between 500~Pa and 20~kPa
can be realized very precisely.~\cite{Gambette21,Gambette22} The overall relative standard
uncertainty of this method is currently 0.04\%, with the largest contribution from
non-state-of-the-art thermometry, which is likely to be substantially reduced in future work.

\subsubsection{Intermediate pressure standards (0.1 MPa and 7 MPa)}
\label{sec:p_intermediate}

Differently than initially envisaged, the first realization of a thermodynamic pressure standard was
not obtained by capacitance measurements, but using a microwave resonant cavity working in the GHz
frequency range, {\em i.e.}, by a RIGT method. A main motivation for this choice was the development
of quasi-spherical microwave resonators, whose internal triaxial ellipsoidal shape slightly deviates
from that of a perfect sphere.~\cite{May2004} This particular geometry resolved the intrinsic
degeneracy of microwave modes, allowing enhanced precision in the determination of resonance
frequencies.

By 2007, Schmidt {\em et al.}~\cite{Schmidt07} demonstrated a pressure standard based on the
measurement of the refractive index of helium to achieve overall relative pressure uncertainty
$\ur(p)$ within $9\times 10^{-6}$ between 0.8~MPa and 7~ MPa. At the upper limit of the pressure
range, the uncertainty was dominated by the uncertainty of the isothermal compressibility $\kappa_T$
of maraging steel, which was determined using resonance ultrasound spectroscopy
(RUS).~\cite{Migliori97} Recently, Gaiser {\em et al.}~\cite{Gaiser2020} realized Moldover's
original proposal of a capacitance pressure standard using DCGT techniques that they had refined
during their measurements of the Boltzmann constant. They achieved the remarkably low uncertainty
$\ur(p) = 4.4 \times 10^{-6}$ near 7~MPa. Recently, the same experimental data were re-analyzed to
take advantage of the increased accuracy of the {\em ab initio} calculation of the second density
virial coefficient $B$ of He,~\cite{Czachorowski2020} reducing the overall uncertainty of the
capacitance pressure standard to $\ur(p) = 2.2 \times 10^{-6}$.~\cite{Gaiser22}

At pressures below 1~MPa, the uncertainty of the realization of a pressure standard based on DCGT or
RIGT with helium is limited by the resolution of relative capacitance or frequency
measurements. This limit would be immediately reduced by up to one order of magnitude by using,
instead of helium, a more polarizable gas like neon or argon. However, while a significant
improvement of the interaction potential, and hence of the {\em ab initio} calculated $B$, has
recently been achieved for neon~\cite{Hellmann2021} and it is well underway for
Ar,~\cite{Lesiukprivate} it is not likely that the best available calculations of the molar
polarizability $A_\varepsilon$ of neon~\cite{Hellmann2022} or argon~\cite{Lesiuk2023} can be
improved sufficiently to replace experiment in the near future. However, an experimental estimate of
$A_\varepsilon$ of both neon and argon was obtained by comparative DCGT measurements relative to
helium, with relative uncertainty of 2.4~ppm,~\cite{Gaiser2018} and may now be used for the
realization of pressure standards with other apparatus. For similar purposes, the ratio of the
refractivity of several monatomic and molecular gases, namely Ne, Ar, Xe, N${}_2$, CO${}_2$, and
N${}_2$O, to the refractivity of helium was determined at $T = 293.15$~K, $\lambda \sim 633$~nm, with
standard uncertainty within $16 \times 10^{-6}$, using interferometry.~\cite{Egan19} At pressures
higher than a few MPa, the imperfect determination of the deformation of the cavity under pressure
would impact the overall uncertainty of a pressure standard based on RIGT or DCGT. One possibility
to tackle this limit would be a two-gas scheme where the compressibility $\kappa_T$ of an apparatus
would be first precisely determined by measurements with helium along an isotherm and then used to
realize a pressure standard with a different working gas. The same strategy is also applied to
increase the upper pressure range where refractometry methods like FLOC can be applied, though use
of helium for the determination of distortion effects requires correcting for diffusion within the
glasses used for the construction of these apparatuses.~\cite{Ricker21}

\subsection{High pressures and equation of state}
\label{sec:highp}

Up to this point, we have considered interactions between temperature and pressure standards and the
rigorously calculated, low-density properties of the noble gases including the polarizability and
second and third density and dielectric virial coefficients. We now compare {\em ab initio}
calculations with measurements at pressures above 7~MPa and at correspondingly higher densities. The
literature includes temperature-dependent values of 6 density virial coefficients of
helium,~\cite{Schultz19} 7 acoustic virial coefficients of krypton,~\cite{ElHawary19} and 6 density
virial coefficients of argon.~\cite{Jaeger11} These calculations used the best {\em ab initio}
two-body and nonadditive three-body potentials that were available at the time of
publication. Many-body non-additive potentials involving four or more bodies, which are needed for
the exact calculation of virial coefficients from the fourth onwards, are not available and are
generally neglected, resulting in an uncontrolled approximation.  Here, we
compare measurements of the density of helium $\rho_\mathrm{meas}(p,T)$ with values calculated {\em
  ab initio}. This comparison avoids fitting $\rho_\mathrm{meas}(p,T)$ to the VEOS because such fits
yield highly correlated values for the separate virial coefficients, each with large
uncertainties. Later in this section, we comment on comparisons using speed-of-sound data.

Measurements of gas densities with uncertainties below 0.1\% are expensive and rare because they are
not required for chemical and mechanical engineering. The uncertainties of most process models are
dominated by imperfect models of equipment (heat exchangers, compressors, distillation columns, {\em
  etc.}) and/or imperfect knowledge of the composition of feedstocks and products. An example of a
demanding application of gas density and composition measurements is custody transfer of natural gas
as it flows through large pipelines near ambient temperature and at high pressures ({\em e.g.},
7~MPa). An international comparison among NMIs achieved a $k=2$ volumetric flow uncertainty of only 0.22\%.~\cite{Mickan13} In this context, density and composition measurements with uncertainties
of order 0.1\% are satisfactory for converting volumetric flows into mass flows and heating values.

\begin{figure}[h]
\center\includegraphics[width=0.9\linewidth]{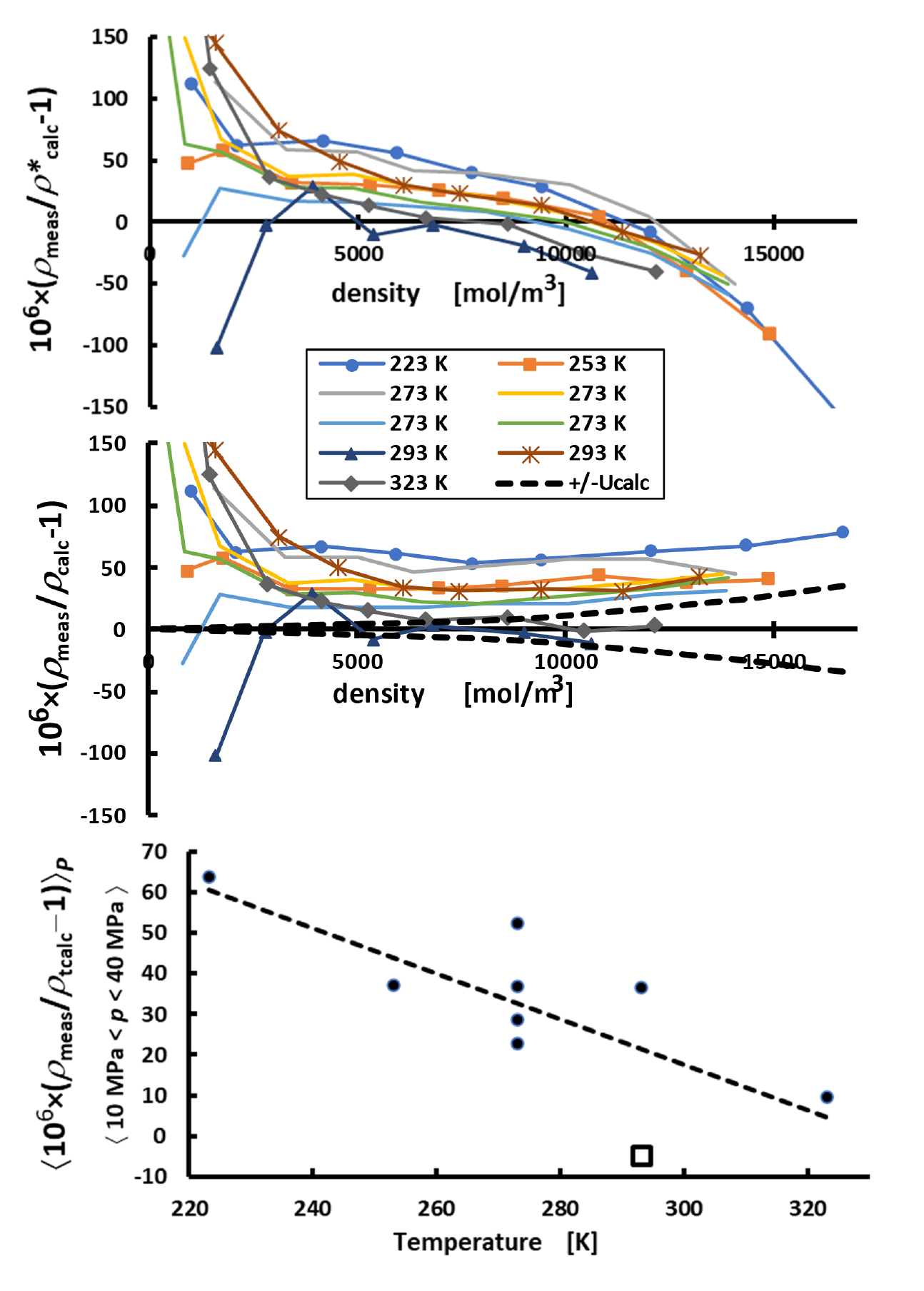} 
  \caption{Top: Differences (in parts per million) on isotherms between measured densities and the
    densities calculated using the {\em ab initio} terms $B \rho, C \rho^2, D \rho^3$, in the
    truncated VEOS. Middle: Same as top with additional theoretical terms: $E \rho^4, F \rho^5$. All
    data with $\rho_\mathrm{meas} > 4000$~ mol/m${}^3$ are within 80~ppm of
    $\rho_\mathrm{calc}$. The claimed uncertainty of $\rho_\mathrm{meas}$ is the span of the figure:
    $\pm 150$~ppm. The dashed curves (- -) bound the estimated uncertainty of $\rho_\mathrm{calc}$
    at 223~K. Bottom: Averaged deviations from $\rho_\mathrm{calc}$ for each isotherm in the
    high-pressure range 10~MPa~$< p <$~40~MPa. The dashed line fitted to the points corresponds to
    calibrating the temperature-dependence of the sinker's density using the helium VEOS. The symbol
    $\square$ identifies an anomalous isotherm that was measured with helium of lesser purity.}
  \label{fig:highp}
\end{figure}

In Fig.~\ref{fig:highp}, the remarkable data of McLinden and L{\"o}sch-Will are used to test the
{\em ab initio} VEOS of helium in the ranges 1~MPa~$< p <$~38~MPa and 223~K~$< T
<$~323~K.~\cite{McLinden07} These data were acquired using a magnetic suspension densimeter. A weigh
scale determined the buoyant forces on two ``sinkers'' immersed in the helium. The data are precise,
well-documented, and traced to SI standards with a claimed, $k=2$, density uncertainty of 0.015\% +
0.001~kg/m${}^3$ at the temperature extremes and at the highest density. These features attracted
previous comparisons with theory.~\cite{Moldover10,Schultz19,Garberoglio2021a}

For the present comparison, where recently published theoretical values of the virial coefficients
are used, we converted the measured temperatures from the ITS-90 to thermodynamic
temperatures using Ref.~\onlinecite{Gaiser2022} and we converted the measured mass densities to
molar densities using the defined value of the universal gas constant and the molar mass for
McLinden and L{\"o}sch-Will's helium sample.  At densities below $\sim 4000$~mol/m${}^3$, the
uncertainties and the values of ($\rho_\mathrm{meas}/\rho_\mathrm{calc}-1$) diverge on isotherms as
$\rho^{-1}$ and/or $p^{-1}$. (See Fig.~\ref{fig:highp}.) These low-density divergences result from
time-dependent drifts in the zeros of the densimeter and/or pressure transducer.  Because the
divergences contain more information about the apparatus than about the helium's VEOS, we do not
discuss them.

At densities above 4000~mol/m${}^3$, we compared the $\rho_\mathrm{meas}(p, T)$ data of McLinden and
L{\"o}sch-Will~\cite{McLinden07} with the values of $\rho_\mathrm{calc}^*(p, T)$ that are implicitly
defined by the truncated VEOS:
\begin{equation}
  \frac{p}{\rho_\mathrm{calc}^* R T} = 1 +
  B_\mathrm{calc} \rho_\mathrm{calc}^* +
  C_\mathrm{calc} {\rho_\mathrm{calc}^*}^2 +
  D_\mathrm{calc} {\rho_\mathrm{calc}^*}^3
  \label{eq:rhostar}
\end{equation}
The fully quantum-mechanical values of $B_\mathrm{calc}, C_\mathrm{calc}$, and $D_\mathrm{calc}$
(the latter computed neglecting four-body interactions) were taken from
Refs.~\onlinecite{Czachorowski2020,Garberoglio2011,Garberoglio2021a}, respectively. The top panel of
Fig.~\ref{fig:highp} shows that the differences trend downward as the densities increase above about
4000~mol/m${}^3$. This trend, as a function of $(p, T)$, was noted in
Ref.~\onlinecite{Garberoglio2021a}, together with the suggestion ``there may have been a small error
in the calibration for the sinkers $\ldots$'' However, the trend (Fig.~\ref{fig:highp}, top) plotted
as a function of density suggests that $\rho_\mathrm{meas}$ is sensitive to some of the truncated
virial coefficients. The truncation suggestion is confirmed by the middle panel of
Fig.~\ref{fig:highp}, which includes in $\rho_\mathrm{calc}(p,T)$ the two additional terms
$E_\mathrm{calc}(T) \rho^4$ and $F_\mathrm{calc}(T) \rho^5$ calculated semi-classically in
Ref.~\onlinecite{Schultz19}. Additional terms ({\em e.g.}, $G_\mathrm{calc}(T)\rho^6$ from
Ref.~\onlinecite{Schultz19}) are less than 1.3~ppm, too small to be visible in Fig.~\ref{fig:highp}.

The claimed $k=2$ uncertainty of $\rho_\mathrm{meas}$ is 150~ppm;~\cite{McLinden07} the span of the
upper panels of Fig.~\ref{fig:highp} is $\pm 150$~ppm. The dashed curves (- -) in the middle panel
of Fig.~\ref{fig:highp} represent upper bounds to the uncertainty of $\rho_\mathrm{calc}(T,\rho)$ at
223~K. For these upper bounds, we used the $k=2$ uncertainties of the virial coefficients $U(B),
U(C), \ldots$ provided by their authors. In Eq.~(\ref{eq:rhostar}) we replaced $B$ with $B+U(B)$; we
replaced $C$ with $C+U(C)$, {\em etc.} The uncertainties of $\rho_\mathrm{calc}(T,\rho)$ are smaller
at higher temperatures. We conclude $\rho_\mathrm{calc}$ agrees with $\rho_\mathrm{meas}$ well
within combined uncertainties.

At densities above $\sim 4000$~mol/m${}^3$, the differences
($\rho_\mathrm{meas}/\rho_\mathrm{calc}-1$) are nearly independent of the
density; however, the average densities are 34~ppm larger than their
expected values $\rho_\mathrm{calc}$. These offsets are well within the
claimed measurement uncertainties ($k=2$, $\sim 150$~ppm). However, as
shown in the lower panel of Fig.~\ref{fig:highp}, the offsets have both a
random and a systematic dependence on the temperature. The systematic
temperature-dependence can be treated as a correction to the calibration of
the sinkers' densities $\rho_\mathrm{sinker}(p,T)$.  Such a correction does
not remove the spread ($\pm 14$~ppm) among the 4 isotherms at
273~K. Possible causes of this spread are changes between runs of
temperature ($\pm 3.8$~mK) and/or of impurity content ({\em e.g.}, $\pm
2.3$~ppm of N${}_2$). In any case, the offsets are smaller than the claimed
uncertainties of $\rho_\mathrm{meas}(p, T)$.

Moldover and McLinden~\cite{Moldover10} extended McLinden and L{\"o}sch-Will's
data~\cite{McLinden07} to 500~K. The extended data are a less-stringent test of the VEOS than
Fig.~\ref{fig:highp} because they span the same pressure range ($p < 38$~MPa) at higher
temperatures; therefore, they span a smaller density range.  If McLinden's data could be extended to
lower temperatures with comparable uncertainties, they would test helium's VEOS in greater detail
and they might reach a regime where $U(\rho_\mathrm{meas}) < U(\rho_\mathrm{calc})$.  Schultz and
Kofke conducted much more detailed tests of McLinden and L{\"o}sch-Will's data.~\cite{Schultz19} We
agree with their conclusion that the data are consistent with the VEOS calculated {\em ab
  initio}.

It may be possible to significantly reduce the uncertainty of $\rho_\mathrm{meas}$ by improving
magnetic suspension densimeters, as suggested by Kayukawa {\em et al.}~\cite{Kayukawa12} They
fabricated sinkers from single crystals of silicon and germanium because these materials have
outstanding isotropy, stability, and well-known physical properties. Also, they refined the model
and the functioning of their magnetic suspension so that it was independent of the magnetic
properties of the fluid under study at the level of 1~ppm. They measured the density of a
liquid near ambient temperature and pressure with a claimed $k=1$ relative uncertainty of $5.4
\times 10^{-6}$. To date, they have not demonstrated this uncertainty far from ambient temperature
and pressure. Even if $\rho_\mathrm{meas}$ achieved such low uncertainties, tests of the VEOS would
have to solve problems arising from impure gas samples and imperfect temperature and pressure
measurements.

Alternative methods of measuring equations of state have been reviewed by
McLinden.~\cite{McLinden14} Several methods require filling a container of known volume
$V_\mathrm{cont}(p_0,T_0)$ with a known quantity of gas and then measuring the pressure as the
temperature is changed.  These methods resemble the CVGT method discussed in Sec.~\ref{sec:cvgt}.
Like CVGT, they require accurate values of $V_\mathrm{cont}(p,T)$; however, unlike CVGT, testing a
VEOS requires much higher pressures.  Determining $V_\mathrm{cont}(p,T)$ over wide ranges is complex
because: (1) containers comprised of metal alloys have anisotropic elastic and thermal expansions;
(2) containers have seals and joints or welds which have complicated mechanical properties; (3)
alloys creep and/or anneal under thermal and mechanical stresses. In summary, volumetric methods are
unlikely to replace Archimedes-type densimeters because $V_\mathrm{cont}(p,T)$ is an assembled
object subjected to complicated stresses; in contrast, the densimeter's sinkers are single objects
subjected to hydrostatic pressure.

Remarkably, the Burnett method~\cite{Burnett36} of measuring the equation of state requires neither
determining $V_\mathrm{cont}(p,T)$ nor measuring quantities of gas. This method uses two pressure
vessels with stable volumes $V_a$ and $V_b$. On each isotherm, gas is admitted into $V_a$ and the
pressure is measured. The gas is allowed to expand so that it fills both $V_a$ and $V_b$ and the
pressure is measured again. $V_b$ is evacuated and the process is repeated several times. The
measured pressures on each isotherm are fitted to the VEOS and an apparatus parameter: the volume
ratio at zero pressure $(V_{a,0}+V_{b,0})/V_{a,0}$. The pressure dependences of $V_a$ and $V_b$ must
also be known. Usually, they are estimated from elastic constants and models of the pressure
vessels; therefore, precise estimates encounter complications of estimating $V_\mathrm{cont}(p,T)$.
Perhaps this explains the large scatter in Burnett determinations of $D(T)$.~\cite{Garberoglio2021a}
A fairly recent Burnett measurement of the equations of state of nitrogen and hydrogen (353~K to
473~K; 1~MPa to 100~MPa) claimed $k=2$ uncertainties of $\rho_\mathrm{meas}$ ranging from 0.07\% to
0.24\%.~\cite{Sakoda12}

In addition to $\rho_\mathrm{meas}$, measurements of the squared speed of
sound $w^2(p,T)$ in gases have been used to critically test either the
VEOS~\cite{ElHawary19} of Eq.~(\ref{eq:pvirial}) or its acoustic analog, Eq.~(\ref{eq:agt}).
Accurate values of $w^2(p,T)$ in gases are readily available. At the low gas pressures used for
acoustic thermometry, the relative expanded uncertainties $U_\mathrm{r}(w^2(p,T))$ measured using
quasi-spherical cavity resonators are a few parts in $10^6$ and are dominated by thermometry problems
and/or impurities. However, uncertainties grow approximately linearly in pressure because of
imperfect models of the recoil of the cavity's walls in response to the resonating gas. In one study
of argon, $U_\mathrm{r}(w^2) \approx 1.2 \times 10^{-4}(p/20~\mathrm{MPa})$ except near the critical
point.~\cite{Estrada95} At pressures above $\sim 5$~MPa to $\sim 10$~MPa, pulse-echo techniques
achieve uncertainties comparable to or smaller than resonance techniques.~\cite{ElHawary19,Meier06}
Remarkably, $w^2$ from the two techniques agreed within 60~ppm to 200~ppm within a range of overlap
(argon, 250~K to 400~K, $\sim 10$~MPa to $\sim 20$~MPa~\cite{Meier06})

It is more complex to compare $w^2_\mathrm{meas}(p,T)$ to a calculated VEOS than to compare $\rho_\mathrm{meas}$ to the
same VEOS. To calculate the $n$-th acoustic virial coefficient from the $n$-th density virial
coefficient, one also needs the first and second temperature derivatives of the $n$-th virial
coefficient as well as all the lower-order density virial coefficients and their temperature
derivatives. There are several routes to conduct such a comparison. First, the temperature
derivatives of the density virial coefficients can be calculated from {\em ab initio} potentials
using, {\em e.g.}, the Mayer sampling Monte Carlo method. Second, the temperature derivatives can be
obtained from fits of the theoretically calculated temperature-dependent density virial
coefficients. Third, the virial equation of state can be transformed by thermodynamic identities
into an acoustical virial equation of state or it can be integrated to formulate a Helmholtz energy
equation, from which the speed of sound can be calculated. All of these methods are completely
equivalent. Speeds of sound calculated by either of the two resulting equations contain
contributions from terms with higher acoustic virial coefficients than those used in the density
virial equation of state, {\em i.e.}, it can be expected that the region of convergence of this
virial equation of state for the speed of sound extends to higher pressures than that of the
acoustic virial equation of state with virial coefficients derived directly from density virial
coefficients. These terms describe contributions of configurations of particles which are contained
in the low-order density virial coefficients to the higher-order acoustic virial
coefficients. Fourth, densities can be calculated from $w^2_\mathrm{meas}(p,T)$ by the method of
thermodynamic integration~\cite{Trusler91} and directly compared to the density virial equation of
state. As initial conditions for the integration, the density and heat capacity on an isobar must be
known. There are subtleties to integrating $w^2_\mathrm{meas}(p,T)$.~\cite{Trusler17} In the first
method the uncertainties of the virial coefficients and their temperature derivatives follow from
the Monte Carlo simulation and can be propagated into an uncertainty of the acoustic virial equation
of state, while in the other methods the uncertainty of the density virial coefficients or the
experimental speeds of sound can be propagated into the acoustic virial equation of state or
calculated densities, respectively.

For helium, Gokul {\em et al.}~\cite{Gokul21} calculated the acoustic virial coefficients through
the seventh order by the second method outlined above from density virial coefficients. They used
the second density virial coefficients reported by Czachorowski {\em et
  al.},~\cite{Czachorowski2020} which are based on the pair potential reported in the same work. The
higher virial coefficients were taken from the work of Schultz and Kofke.~\cite{Schultz19} They are
based on the pair potential of Przybytek {\em et al.}~\cite{Przybytek2010} and the three-body
potential of Cencek {\em et al.}~\cite{Cencek2009} Uncertainties in the density virial coefficients
were propagated into uncertainties in the acoustic virial coefficients by the Monte Carlo method
recommended in Supplement 1 to the ``Guide to the Expression of Uncertainty in
Measurement''.~\cite{GUM} Gokul {\em et al.}~\cite{Gokul21} formulated the acoustic virial equation
of state as expansion in terms of density or pressure. The uncertainty of speeds of sound calculated
with the acoustic virial equation of state was estimated from the uncertainty of the acoustic virial
coefficients.

The density expansion of Gokul {\em et al.} was compared to the experimental data of
Gammon,~\cite{Gammon76} Kortbeek {\em et al.},~\cite{Kortbeek88} and Plumb and
Cataland.~\cite{Plumb66} The data of Gammon were measured with a variable-path interferometer
operating at 0.5~MHz. They cover the temperature range between 98 K and 423 K with pressures up to
15 MPa, and according to the author have an uncertainty of 0.003\%. Gammon's data agree with the
acoustic virial equation of state within 0.01\% with a few exceptions. The data of Kortbeek {\em et
  al.} were measured with a double-path-length pulse-echo technique, cover the temperature range
from 98 K to 298 K at pressures between 100 MPa and 1 GPa, and, according to the authors, have an
uncertainty of 0.08\%, and deviate from the acoustic virial equation of state between a few tenths
of a percent at 100 MPa up to about 4\% at 298 K and 1 GPa. These rather large deviations are due to
the fact that the acoustic virial equation of state is not converged at such high pressures. The
measurements of Plumb and Cataland cover the low temperature range between 2.3 K and 20 K at
pressures up to 150 kPa. They agree with the acoustic virial equation of state of Gokul {\em et al.}
to within 0.05\% except at the lowest measured pressures of about 1.5 kPa, where the deviations
reach up to 0.18\%. 
Gokul {\em et al.} also assessed the pressure range in which the acoustic VEOS is more accurate than the available experimental data for the speed of sound. At low pressures, they observed that speeds of sound calculated with the acoustic VEOS are more accurate than the experimental data of Gammon. Gokul {\em et al.} further noticed that speeds of sound calculated with the acoustic virial equation of state are more accurate than the experimental data of Kortbeek {\em et al.} up to about 300 MPa depending on temperature. At higher pressures, they considered the experimental data of Kortbeek {\em et al.} to be more accurate than the computed virial equation of state. This conclusion appears to be too optimistic in light of the low uncertainty of 0.08\% in the experimental data and the rather large deviations of up to 2\% from the virial equation of state below 300 MPa.

Gokul {\em et al.} also examined the convergence behavior of the acoustic virial equation of state
more closely for the expansions in density or pressure. They considered a virial equation of state
converged if the value of the speed of sound calculated with it agrees with all higher orders of the
expansion within a certain tolerance.
However, the expansion in pressure fails in the supercritical region, above which
increasing the tolerance does not further extend the region of convergence. Above this pressure
limit, the expansion in pressure completely fails.

The first calculation of the third virial coefficient of argon using a first-principles three-body
potential was performed by Mas {\em et   al.}~\cite{Mas:99} using the empirical potential developed
by Aziz.~\cite{Aziz93} The results agreed almost to within combined uncertainties with the third
virial coefficient extracted from experimental data (with theoretical constraints) by Dymond and Alder.~\cite{Dymond71}
J{\"a}ger {\em et al.} calculated density virial coefficients up to seventh order for argon with
their pair and nonadditive three-body potentials.~\cite{Jaeger11} The calculated virial coefficients
were fitted by polynomials in temperature. The seventh-order VEOS was compared with the very
accurate $(p,\rho,T)$ data of Gilgen {\em et al.},~\cite{Gilgen94} which were measured with a
magnetic suspension densimeter. These data are characterized by a relative uncertainty ($k=2$) in
density of 0.02\%. Pressures calculated with the theoretical virial equation of state agree with
these data at the highest temperature of the measurements 340 K within 0.01\%.

In further work, J{\"a}ger~\cite{Jaeger13} used thermodynamic identities to calculate several
properties including the speed of sound from the virial equation of state and compared the results
with the accurate experimental data of Estrada-Alexanders and Trusler~\cite{Estrada95} and Meier and
Kabelac.~\cite{Meier06} The data of Estrada-Alexanders and Trusler~\cite{Estrada95} were measured
with a spherical resonator and cover the temperature range between 110 K and 450 K at pressures up
to 19 MPa, while the data of Meier and Kabelac were measured with a dual-path-length pulse-echo
technique and cover the temperature range between 200 K and 420 K with pressures between 9 MPa and
100 MPa. The uncertainty of these data sets was estimated to be 0.001\%-0.007\% and 0.011\%-0.036\%,
respectively. At 300 K and 400 K, the calculated speeds of sound agree with both experimental data
sets up to 100 MPa within 0.04\% and 0.08\%, respectively. At the near-critical temperature 146 K
and supercritical temperature 250 K, the deviations of the calculated values from the experimental
data of  Ref.~\onlinecite{Estrada95} increase with pressure from essentially zero in the ideal-gas limit
to about 0.3\% at 3.7 MPa and about 0.02\% at 12.2 MPa.

In another paper, J{\"a}ger {\em et al.} presented calculations of the second and third density
virial coefficient of krypton.~\cite{Jaeger15} They developed a very accurate pair potential for the
krypton dimer, and nonadditive three-body interactions were described by an {\em ab initio} extended
Axilrod--Teller--Muto potential, which was fitted to quantum chemical calculations of the
interaction energy of equilateral triangle configurations of three krypton atoms.  El Hawary {\em et
  al.}~\cite{ElHawary19} calculated density virial coefficients from the third to the eighth using
the pair potential and extended Axilrod--Teller--Muto potential of J{\"a}ger {\em et al.}  The
calculated virial coefficients were fitted to polynomials in temperature, and the virial equation of
state was integrated to formulate it as a fundamental equation of state in terms of the Helmholtz
energy. Furthermore, El Hawary {\em et al.} measured the speed of sound in liquid and supercritical
krypton between 200 K and 420 K at pressures from 6.1 MPa to 100 MPa with an uncertainty ($k=2$) of
0.005\%-0.018\%.  At 240 K, 320 K, and 420 K, the seventh-order and eighth-order virial equations of
state agree with each other within 0.02\% up to 7 MPa, 17 MPa, and 38 MPa, respectively. In the
region where the virial equation of state is sufficiently converged, the calculated speeds of sound
are systematically about 0.08\% lower than the experimental data. This small difference is probably
due to the uncertainty of the pair potential and the simplified treatment of nonadditive three-body
interactions with the extended Axilrod--Teller--Muto model.

At high density in the supercritical region where the virial equation of state does not converge and
in the liquid region, thermodynamic properties can be calculated by Monte Carlo (MC) or
molecular-dynamics (MD) simulations.~\cite{AT87} Since the generation of Markov chains in MC
simulations avoids some of the numerical errors of algorithms used to integrate the equations of
motion in MD simulations, MC simulations are the preferred method for calculating accurate values
for thermodynamic properties. In statistical mechanics, there are eight basic ensembles for
performing MC or MD simulations,~\cite{Graben93} which are characterized by a thermodynamic
potential, three independent variables, and a weight factor, which describes the distribution of
systems in the ensemble, in which Monte Carlo simulations of fluids can be performed. Str{\"o}ker
{\em et al.} [24] pointed out that the $NpT$ ensemble, in which the number of particles, the
pressure, and the temperature are the independent variables, is best suited for the calculation of
thermodynamic properties because only ensemble averages involving the enthalpy and volume, but no
derivatives of the potential energy with respect to volume, appear in the equations for
thermodynamic properties. This means that no derivatives of the potentials are needed in a
simulation.

The calculations of Mas {\em et al.}~\cite{Mas:99} described earlier were later extended by
performing $NVT$, $NpT$, and Gibbs ensemble MC simulations~\cite{Bukowski:01} along the gas-liquid
coexistence line. The parameters of the critical point agreed with experiments to within 0.8\% or
better.~\cite{Vargaftik75}

Str{\"o}ker et al.~\cite{Stroker22} carried out semiclassical Monte Carlo simulations of
thermodynamic properties of argon in the $NpT$ ensemble at the subcritical isotherm 100 K and the
supercritical isotherm 300 K at pressures up to 100 MPa. The interactions between argon atoms were
described by the pair potential of J{\"a}ger {\em et al.}~\cite{Jaeger09} and the nonadditive
three-body potential of J{\"a}ger {\em et al.}~\cite{Jaeger11} Quantum effects were accounted for by
the Feynman--Hibbs corrections to the pair potential. Calculated densities agree with the accurate
experimental data for the density measured by Gilgen {\em et al.}~\cite{Gilgen94} and Klimeck {\em
  et al.}~\cite{Klimeck98} within less than 0.01\%, while calculated speeds of sound agree with the
accurate experimental data of Estrada-Alexanders and Trusler~\cite{Estrada95} at low
pressure in the supercritical region and Meier and Kabelac~\cite{Meier06} at high pressure in the
liquid and supercritical region within less than 0.1\%.

Str{\"o}ker {\em et al.}~\cite{Stroker22Kr} also performed Monte Carlo simulations for liquid and
supercritical krypton. They employed the accurate pair potential and an extended
Axilrod--Teller--Muto potential of J{\"a}ger {\em et al.}~\cite{Jaeger15} to account for nonadditive
three-body interactions. Quantum effects were again accounted for semiclassically. Since the
potential models for krypton are not as accurate as those for argon, the deviations of the results
for the density and speed of sound from experimental data were larger than for argon, about 0.2\%
and 0.36\%, respectively.

\subsection{Transport properties and flow metrology}
\label{sec:flow}

In this section, we describe the impact of the {\em ab initio} calculations of
the zero-density limit of helium's thermal conductivity $\lamhe$ and viscosity
$\etahe$.  First, we mention the impact of $\lamhe$ and $\lamar$ on temperature metrology.
Then, we describe how accurate values of $\etahe$ have been used as standards to
reduce the uncertainty of viscosity measurements of many gases by a factor
of $10$.  We conclude by briefly considering the impact of accurate viscosity
data on metering process gases, for example, during the manufacture of
semiconductor chips. 

As discussed in Sec.~\ref{sec:AGT}, acoustic gas thermometry requires
accurate values of $\lambda$ of the working gas at low densities to account
for the effect of the thermo-acoustic boundary layer on the measured
resonance frequencies.  For example, in 2010, Gavioso {\em et al.} used a
helium-filled, quasi-spherical cavity to determine the value of the
Boltzmann constant $\kB$ prior to its definition in 2019.~\cite{Gavioso10}
They reported that a relative standard uncertainty $\ur (\lamhe) = 0.015$
generated a relative standard uncertainty of the Boltzmann constant $\ur
(\kB) = (1 ~\mathrm{to}~ 3) \times 10^{-6}$.

Today, an uncertainty of $(1 ~\mathrm{to}~ 3) \times 10^{-6}$ would be the
largest contributor to a state-of-the-art determination of the
thermodynamic temperature $T$ near $273$~K.  At low temperatures, the
uncertainty of measured values of $\lamhe$ is much larger.  Below $20$~K,
the $\lamhe$ data span a range on the order of $\pm 6\%$.~\cite{Acton77}
This large an uncertainty would lead to $\ur (T) > 10^{-5}$ for acoustic
determinations of $T$.  Fortunately, the values of $\lamhe$ calculated {\em
  ab initio} have extraordinarily small uncertainties, {\em e.g.}, $\ur (\lamhe) =
9.6 \times 10^{-6}$ at $273$~K and $\ur (\lamhe) = 7.3 \times 10^{-5}$ at
$10$~K.~\cite{Cencek2012} In essence, the calculated values of $\lamhe$
removed $\ur (\lamhe)$ from the uncertainty budgets of acoustic
thermometers based on helium-filled quasi-spherical cavities.

Cylindrical, argon-filled cavities are being developed for high-temperature acoustic
thermometry.~\cite{Zhang20,McEvoy20} These projects require low-uncertainty values of both $\lamar$
and $\etaar$. Low-uncertainty values of $\etaar$ were generated from accurate measurements of the
ratios $\etaar/\etahe$ in the range $200$~K to $653$~K and the {\em ab initio} values of
$\etahe$. Then $\lamar(T)$ was obtained by combining the ratio-deduced values of $\etaar(T)$ with
values of the Prandtl number $\prar$ calculated from model interatomic potentials. ($\Pra = C_p
\eta/\lambda$, where $C_p$ is the constant-pressure heat capacity per mass. For the noble gases,
$\Pra$ is only weakly sensitive to the potential.)~\cite{May06,Zhang13,Lin14} The measured ratios
$\etaar/\etahe$ were consistent, within a few tenths of a percent, with highly accurate measurements
made with an oscillating-disk viscometer~\cite{Vogel10} and with calculations of $\etaar$ based on
{\em ab initio} Ar--Ar potentials.~\cite{Vogel10_b} Thus, the needs of argon-based acoustic
thermometry are now met at all usable temperatures. To put this achievement in context, we note that
measuring the thermal conductivity of dilute gases is difficult, even for noble gases near ambient
temperature and pressure. Evidence for this appears in Lemmon and Jacobsen's correlation of the
``best'' measurements of $\lamar$ and $\etaar$ near ambient
temperature (270~K to 370~K) and pressure.~\cite{Lemmon04} The average absolute deviations of
selected measurements from their correlation ranged from 0.24\% to 1.0\%. Lemmon and Jacobsen
estimated the uncertainty of the correlated values of $\lamar$ was 2\% and the uncertainty of
$\etaar$ was 0.5\%. (With the benefit of {\em ab initio} calculations and ratio measurements, we now
know their correlation overestimated $\lamar$ by 0.54\% at 270~K and by 0.45\% at 370~K.)

In 2012, Berg and Moldover reviewed measurements of the viscosity of 11 dilute gases near
$25~{}^\circ$C.~\cite{Berg12} These measurements were made using 18 different instruments that used
5 different operating principles and produced 235 independent viscosity ratios during the years 1959
to 2012. Using the {\em ab initio} value of $\etahe$ at $25~{}^\circ$C as a reference, the
viscosities of the 10 other gases (Ne, Ar, Kr, Xe, H${}_2$, N${}_2$, CH${}_4$, C${}_2$H${}_6$,
C${}_3$H${}_8$, SF${}_6$) were determined with low uncertainties $\ur (\eta)$ ranging from 0.00027
to 0.00036. These ratio-derived uncertainties are less than 1/10 the uncertainties claimed for
absolute viscosity measurements, such as the measurements of $\etaar$ correlated by Lemmon and
Jacobsen.~\cite{Lemmon04} Now, any one of these gases can be used to calibrate a viscometer within
these uncertainties. Such ratio-based calibrations have reduced uncertainties of $\eta$ for many
other gases~\cite{Vogel21} and they have been extended to a very wide range of
temperatures.~\cite{Xiao20} During their study of viscosity ratios, Berg and Moldover observed that
the viscosity ratios determined using one instrument (a magnetically suspended, rotating cylinder)
were anomalous. Their observation led to an improved theory of the instrument, thereby illustrating
the power of combining a reliable standard $\etahe(T)$ with precise ratio
measurements.~\cite{Schafer15}

Accurate measurements of gas flows are required for tightly controlling
manufacturing processes ({\em e.g.}, delivery of gases to semiconductor
wafers for doping). In general, gas flow meters are calibrated using a
benign, surrogate gas over a range of flows and pressures, but only near
ambient temperature. However, calibrated meters are often used to
measure/control flows of reactive process gases [{\em e.g.},
  Ga(CH${}_3$)${}_3$, WF${}_6$] under conditions differing from the
calibration conditions. An accurate transition between gases and conditions
can be made using laminar flow meters for which there is a physical model
(similar to the model of a capillary tube). Also needed are data for the
process gas' virial coefficients and the viscosity ratio~\cite{Wright12}
$$
\frac{\eta_\mathrm{process}(p_\mathrm{process},T_\mathrm{process})}
{\eta_\mathrm{surrogate}(p_\mathrm{surrogate}, T_\mathrm{surrogate})}.
$$
Thus, there is a need for viscosity-ratio data for many
difficult-to-measure gases over a moderate range of densities. The
acquisition of such data would be facilitated by a reliable model for the
density dependence of the viscosity of surrogate gases such as SF${}_6$.

The initial density expansion of the viscosity has the form $\eta/\eta_0 =
1+\eta_1 \rho$ , where the low-density limit of the viscosity $\eta_0$
depends entirely on pair interactions and the virial-like coefficient
$\eta_1$ depends on the interactions among two and three molecules. Unfortunately,
unlike the density and dielectric virial coefficients and $\eta_0$, no
rigorous theory exists for $\eta_1(T)$. An approximate theory was developed
by Rainwater and Friend,~\cite{Friend84,Rainwater87} who presented
quantitative results based on the Lennard-Jones potential. It was later
extended with more accurate pair potentials for noble
gases.~\cite{Najafi98} While the results from the Rainwater--Friend model are
in reasonable agreement with the limited experimental data available for
the initial density dependence of the viscosity for noble
gases,~\cite{Najafi98} the error introduced by its approximations is
unknown. We note that it is a classical theory, which introduces another
source of error for light gases (such as helium) where quantum effects
might be important, even at ambient temperatures.

\section{{\em Ab initio} Electronic Structure Calculations}
\label{sec:abinitio}

\subsection{Methodology of electronic structure calculations}

In principle, solutions of equations of relativistic quantum mechanics, possibly including
quantum electrodynamics (QED) corrections, can predict all properties of matter to a precision
sufficient for thermal metrology applications.  In practice, if the goal is to match or
exceed accuracy of experiments, the range of systems reduces to few-particle ones.  The first
quantum mechanical calculations challenging experimental measurements for molecules appeared only in
the 1960s ({\em e.g.}, Ref.~\onlinecite{Kolos1965}), while the first calculations relevant to metrology
were published in the mid-1990s.~\cite{Aziz1995,Williams:96,Korona:97} Currently, the branches of
metrology discussed in this review are becoming increasingly dependent on theoretical input, as
already discussed in Sec.~\ref{sec:expt}.

Theory improvements leading to results with decreased uncertainties proceed along three main,
essentially orthogonal directions: level of physics, truncation of many-electron expansions, and
basis set size.  There exists an extended hierarchy of approaches in each direction.  For the first
direction, there exists a set of progressively more accurate physical theories that can be used in
calculations relevant for metrology, from Schr\"odinger's quantum mechanics for electrons' motion in
the field of nuclei fixed in space to relativistic quantum mechanics and to QED.  The second
direction is present for any many-electron system: one has to choose a truncation of the expansion
of the many-electron wave function in terms of virtual-excitation operators at the double, triple,
quadruple, {\em etc.} level or, equivalently in methods that use explicitly correlated bases
(depending explicitly on interelectronic distances), to take into account only correlations of two,
three, four, {\em etc.}~electrons simultaneously.  Third, for any given theory and many-electron
expansion level, there are several methods of solving quantum equations specific for this level; in
particular different types of basis sets are used to expand wave functions, resulting in different
magnitudes of uncertainties from such calculations.

The lowest theory level is Schr\"odinger's quantum mechanics for electrons moving in the field of
nuclei fixed in space, {\em i.e.}, quantum mechanics in the Born--Oppenheimer (BO) approximation.
At the next level, one usually first accounts for the relativistic effects.  Post-BO treatment of
the Schr\"odinger equation can be limited to computations of the so-called diagonal adiabatic
correction, which is the simplest method of accounting for couplings of electronic and nuclear
motions, or it can fully include nonadiabatic effects, {\em i.e.}, account for the complete
couplings of these two types of motion.  The highest level of theory applied in calculations
relevant to metrology is QED, and it can be implemented at several approximations labeled by powers of
the fine-structure constant $\alpha$.

The many-electron expansion starts at the independent-particle model, {\em i.e.}, at the Hartree--Fock
(HF) approximation, but this level is never used alone in calculations for metrology purposes.  For
systems with a few electrons (the current practical limit is about 10), one can use the FCI expansion that potentially provides exact solutions of Schr\"odinger's
equation (provided the orbital basis set is close to completeness).  In FCI, the wave function for
an $N$-electron system is represented as a linear combination of Slater determinants constructed
from ``excitations'' of the ground-state HF determinant $|\Phi_0\rangle$
\begin{eqnarray}
|\Psi \rangle &=& c_0|\Phi_0 \rangle + \sum_{r,a}c_a^r|\Phi_a^r \rangle
+ \sum_{r<s,a<b} c_{ab}^{rs}|\Phi_{ab}^{rs} \rangle 
+ \nonumber \\
& & \sum_{r<s<t,a<b<c} c_{abc}^{rst}|\Phi_{abc}^{rst} \rangle
+ \dots ,
\label{ci}
\end{eqnarray}
up to $N$-tuple excitations, where $|\Phi_a^r\rangle$ represents a singly excited Slater determinant
formed by replacing spinorbital $\phi_a$ with $\phi_r$. Similarly, $|\Phi_{ab}^{rs} \rangle$
represents a doubly excited Slater determinant formed by replacing spinorbital $\phi_a$ with
$\phi_r$ and spinorbital $\phi_b$ with $\phi_s$, and so on for higher excited determinants.  The
linear coefficients (CI amplitudes) are computed using the Rayleigh--Ritz variational principle.
While the FCI method is conceptually straightforward, the computation time  it requires scales with
the number of electrons as $N!$, where $N$ is the number of electrons, and therefore it is
computationally very costly.  One can limit the expansion in Eq.~(\ref{ci}) to a subset of
excitations (for example, retaining only single and double excitations leads to a method denoted
CISD), but truncated expansions are not size extensive.  This means that the CISD energy computed
for very large separations between two monomers (atoms or molecules) is not equal to the sum of
monomers' energies computed at the CISD level.  Only FCI is free of this problem.  Thus, truncated
CI expansions are not appropriate for calculations of interaction energies.

Another potentially exact approach is to expand the wave function in an explicitly correlated
all-electron basis set.  The set most often used in metrology-related applications is the basis set
of explicitly correlated Gaussian (ECG) functions.  If basis functions involve all electrons,
expansions in this basis approximate solutions of Schr\"odinger's equation in the BO approximation.
For He$_2$, a four-electron system, the expansion can be written as~\cite{Przybytek2017}
\begin{equation}
\label{Psi}
\Psi = {\cal A}_4 \, 
\Xi_{4}^{00}\, \hat{P} \, \left\{
c_0 \phi_0 + \sum_{k=1}^K \, c_k \, \phi_k (\br_1,\br_2,\br_3,\br_4) \right\} ,
\end{equation}
where $ {\cal A}_4$ is the four-electron antisymmetrizer,
$\Xi_{4}^{00}=(\alpha\beta-\beta\alpha)(\alpha\beta-\beta\alpha)$ is the standard four-electron
singlet spin function, $\hat{P}$ is the point-group symmetry projector,
$\hat{P}=\frac12(1+\hat{\imath})$ with $\hat{\imath}$ inverting the wave function through the
geometrical center, $c_k$ are variational parameters, and $\phi_k,\ k>0$ are ECG basis functions.
The function $\phi_0$ is the product of ECG functions for the two helium atoms.  The explicit form
of $\phi_k,\ k>0$, functions is
\begin{equation}
\label{phi}
\phi_k (\br_1,\br_2,\br_3,\br_4) = \prod_{i=1}^{4} 
\e^{ -\alpha_{ki} |{\br}_i - {\bA}_{ki}|^2 } 
\, \prod_{i>j=1}^{4} 
\e^{ -\beta_{kij} |{\br}_i - {\br}_j|^2 },
\end{equation}
where $\alpha_{ki}$, $\beta_{kij}$, and $\bA_{ki} = (X_{ki}, Y_{ki}, Z_{ki})$ are nonlinear
variational parameters.  For a given set of nonlinear parameters, the linear parameters are obtained using
the Rayleigh--Ritz variational method.  The simplest way to optimize the nonlinear ones is to use
the steepest-descent method, recalculating the linear parameters in each step of this method.  In
actual applications, significantly more advanced optimization methods are used.

Currently, the standard approach to account
for electron correlation effects is the coupled cluster (CC) method with single,
double, and noniterative triple excitations [CCSD(T)].  To reduce
uncertainties of CCSD(T), one can use the CC methods that include full triple, T, noniterative
quadruple, (Q), and full quadruple, Q, excitations.
The CC method represents the wave function in an exponential form
\begin{align}\label{cc}
|\Psi_{\rm CC}\rangle = \e^{\hat{T}} |\Phi_{0}\rangle,
\end{align}
where the operator $\hat{T}$ is the sum of excitation operators
\begin{align}\label{fcc}
\hat{T}=\hat{T}_{1} + \hat{T}_{2} + \hat{T}_{3} + \dots + \hat{T}_{N}.
\end{align}
The operators $T_i$ can be written in terms of pairs of creation ($\hat{a}^\dagger$) and
annihilation ($\hat{a}$) operators.  For the two lowest ranks, we have
\begin{eqnarray}
\hat{T}_{1} & =& \sum_{ar} t_a^r ~ \hat{r}^\dagger \hat{a}  \label{eq:t1} \\
\hat{T}_{2}& =& \frac{1}{(2!)^2}\sum_{abrs} t_{ab}^{rs} ~ 
\hat{r}^\dagger\hat{s}^\dagger \hat{b}\hat{a}. \label{eq:t2}
\end{eqnarray}
The excitation operators $\hat{r}^\dagger\hat{s}^\dagger \hat{b}\hat{a}$ acting on the ground-state
determinant produce the same excited determinants as those appearing in Eq.~(\ref{ci}).
For example,
\[ \hat{r}^\dagger\hat{s}^\dagger \hat{b}\hat{a} |\Phi_0 \rangle
             = |\Phi_{ab}^{rs} \rangle .\]
However, the amplitudes $t$ are different from the amplitudes $c$.  The
former amplitudes are obtained by using the expansion (\ref{cc}) in
the Schr\"odinger equation and projecting this equation with subsequent
determinants from Eq.~(\ref{ci}).  Since the resulting set of equations
is nonlinear, the solution is obtained in an iterative way.
If all the excitation operators are kept in Eq.~(\ref{fcc}), the method
is equivalent to the FCI method, but this expansion is almost always truncated.
The simplest CC approach is that of CC doubles (CCD),
in which $\hat{T}$ is truncated to
$$\hat{T}_{\rm CCD}= \hat{T}_2 .$$ The simplest extension of this model is obtained by including
also single excitations (CCSD), {\em i.e.},
$$\hat{T}_{\rm CCSD}= \hat{T}_1 + \hat{T}_2 .$$ The CCSD method is most often used with orbital
basis sets, but can also be used with ECGs, which are then used to expand two-electron functions
resulting from the actions of $T_2$ and are called in this context Gaussian-type geminals (GTGs).
Higher-rank approximations are
$$\hat{T}_{\rm CCSDT}= \hat{T}_1 + \hat{T}_2 + \hat{T}_3,$$
$$\hat{T}_{\rm CCSDTQ}= \hat{T}_1 + \hat{T}_2 + \hat{T}_3 + \hat{T}_4.$$ An approximation to CCSDT
is a method denoted as CCSD(T), where
the coefficients $t$ of single and double excitations in Eqs.~(\ref{eq:t1}) and (\ref{eq:t2}) are
computed iteratively while those for triple excitations are evaluated using perturbation theory.
A similar approximation, denoted CCSDT(Q), can be made for the CCSDTQ method.
In contrast to the truncated CI expansions, the truncated CC
expansions are always size extensive.  This results from the fact that the exponential ansatz of
Eq.~(\ref{cc}) can be factored for large separations between subsystems into a product of
exponential operators for subsystems.  The CC method is applied to interatomic or intermolecular
interactions in the supermolecular fashion, {\em i.e.}, subtracting monomers' total
energies from the total energy of a cluster.  Due to size extensivity, the resulting
potential-energy surface dissociates correctly. 

Another option for computing interaction energies
at theory levels similar to truncated CC is symmetry-adapted perturbation
theory (SAPT).~\cite{Jeziorski1994,Szalewicz2005ashort,Szalewicz2012,%
Szalewicz2022}
The basic assumption of SAPT is
the partitioning of the total Hamiltonian $H$ of a cluster
into the sum of the Hamiltonians of separated monomers
$$H_0 = H_A + H_B + \dots$$ and of the perturbation operator
$V$ that collects Coulomb
interactions of the electrons and nuclei of a given monomer with those
of the other monomers:
$$V = V_{AB} + V_{AC} + V_{BC} + \dots$$

The solution of the zeroth-order problem, {\em i.e.}, of the Schr\"odinger
equation with $H_0$
$$
H_0 \Phi_0 = E_0 \Phi_0,
$$
is then the product of the wave functions of free, noninteracting
monomers.  This product is not fully antisymmetric since permutations of electrons between different
monomers do not result only in a change of the sign of the wave function, {\em i.e.}, $\Phi_0$ does not
satisfy Pauli's exclusion principle.  For large intermonomer separations $R$, one can ignore this
problem and use the Rayleigh--Schr\"odinger perturbation theory (RSPT), the simplest form of
intermolecular perturbation theory.  Unfortunately, RSPT leads to unphysical behavior of the
interaction energy at short $R$ as it fails to predict the existence of the repulsive walls on the
potential-energy surfaces.  This failure is the result of the lack of correct symmetry of the
wave function under exchanges of electrons between interacting monomers.  Thus, to describe
interactions everywhere in the intermonomer configuration space, one has to perform symmetry
adaptation, {\em i.e.}, antisymmetrization, and this is the origin of the phrase ``symmetry-adapted''.
There are several ways to do it, but the simplest is to (anti)symmetrize the wave functions of the
RSPT method. This leads to the symmetrized Rayleigh--Schr\"odinger (SRS) approach, \cite{Jeziorski1978}
which is the only SAPT method used in practice.  For a dimer, the interaction energy is then
expressed as the following series in powers of $V$:
\begin{eqnarray}
E_{\rm int}^{\rm SAPT} &=& E_{\rm elst}^{(1)} 
+ E_{\rm exch}^{(1)} + E_{\rm ind}^{(2)} +
\nonumber \\
& &           E_{\rm exch-ind}^{(2)} 
              + E_{\rm disp}^{(2)} + E_{\rm exch-disp}^{(2)} + \ldots,
	          \label{eq:sapt}
\end{eqnarray}
where the superscripts denote the powers of $V$ (orders of perturbation theory) and different terms
of the same order can be identified as resulting from different physical interactions: electrostatic
(elst), exchange (exch), induction (ind), and dispersion (disp).  When SAPT is applied to
many-electron systems, monomers can be described at various levels of electronic structure theory:
from the HF level to the FCI level.  This leads to a hierarchy of SAPT levels of approximations
depending on treatment of intramonomer electron correlation.  If the monomers are approximated at an
order $n$ of many-body perturbation theory (MBPT) with the M{\o}ller--Plesset (MP) partition of the
Hamiltonians $H_A$ and $H_B$, denoted as MP$n$, we can write
\begin{align} \label{MP} 
E^{(i)} \approx \sum_{j = 0}^n E^{(ij)} ,
\end{align}
which becomes an equality when $n \rightarrow \infty$.

The third direction determining the accuracy of electronic structure calculations involves the size
of the basis sets used to expand wave functions.  In the CC and CI approaches, the standard
technique is to use products of orbital (one-electron) basis sets.  Large number of such basis sets
are available; the ones most often used in metrology-related calculations are the correlation
consistent (cc) basis sets introduced by Dunning.~\cite{Dunning1989} These basis sets are denoted by
cc-pV$X$Z: cc $p$olarized $V$\!alence ({\em i.e.}, optimized using a frozen-core approximation),
$X$-$Z$eta, where $X$ = D, T, Q, 5, $\dots$ is the so-called cardinal number, determining the
maximum angular momentum of orbitals.  Such basis sets can be augmented by an additional set of
diffused functions and are then denoted as aug-cc-pV$X$Z, or two such sets: daug-cc-pV$X$Z.  Another
option is to use explicitly correlated basis sets in the CC method or to expand the whole
many-electron wave function in such a basis set.  Explicitly correlated basis sets provide a much
faster convergence than products of orbital basis sets, but in most cases require optimizations of a
large number of nonlinear parameters.

In order to achieve some target size of uncertainties, one has to choose a proper level in each of
the three hierarchies defined earlier.  For example, it is possible to perform an FCI calculation
for a 10-electron system such as Ne. However, since FCI calculations scale factorially with the
number of orbitals, only very small basis sets can be used, resulting in a large uncertainty of the
results. Consequently, a better strategy is to use the CCSD(T) method which allows applications of
the largest orbital bases available for a system like Ne$_2$.  The computed interaction energy will
be accurate to about four significant digits relative to the CCSD(T) limit, but will have a fairly
large error, on the order of 1--2\%, with respect to the exact interaction energy at the
non-relativistic BO level.  In contrast, FCI calculations for Ne${}_2$ in the smallest sensible
basis set of augmented double-zeta size, apparently never performed, would have an error of the
order of 40\% (such calculations could still be useful in hybrid approaches discussed below).

The orbital basis sets consist of families of bases of varying size.  One usually carries
out calculations in two or more such basis sets and then performs approximate extrapolations to the
complete basis set (CBS) limit.  In addition to the standard extrapolations,
which assume the $X^{-3}$ decay of errors, extrapolations using very accurate ECG results can be
performed.~\cite{Patkowski:07a,Jeziorska:07,Jeziorska:08} CCSD(T)/CBS results may have sufficiently
small uncertainties to make calculations of relativistic and diagonal adiabatic corrections
necessary, {\em i.e.}, these corrections may be of the same order of magnitude as the uncertainties
of the CCSD(T)/CBS results.  To reduce the errors resulting from the truncation of the many-electron
expansion, one can follow CCSD(T)/CBS calculations by CCSDT(Q) or FCI ones in smaller basis sets.
These effects are then included in an incremental way, {\em i.e.}, by adding the difference between
FCI and CCSD(T) energies computed in the same (small) basis set.

Accurate solutions of quantum equations are followed by estimates of uncertainties, absolutely
necessary for metrology purposes.  The latter step is often more time consuming than the former.
One should emphasize that theoretical estimates of uncertainties are different from statistical
estimates of uncertainties of measurements and in particular one cannot assign a
rigorous
confidence level to
them, although when comparing to experiments one usually assumes 
that theoretical uncertainties are equivalent to $k=2$ expanded uncertainties
(95\% confidence level).

A theoretical estimate of uncertainty consists of several elements.  The most rigorous and reliable
estimates are those of basis set truncation errors derived from the observed patterns of convergence
in basis set.  Much more difficult are estimates of uncertainties resulting from truncations of
many-electron expansions.  Such estimates can sometimes be made by performing higher-level
calculations at a single point on a potential-energy surface, but one most often uses analogy to
similar systems for which higher-level calculations have been performed.  The same approach can be
used to estimate the neglected physical effects, for example, to estimate the uncertainty due to
relativistic effects.

Solutions of the electronic Schr\"odinger equation for a given nuclear configuration of a dimer or a
larger cluster, providing accurate quantum mechanical descriptions of such systems, are only the
first step in theoretical work of relevance to metrology, as most measured quantities discussed in
this review are either bulk properties or response properties of atoms and molecules.  In the former
case, {\em i.e.}, to predict properties of gases or liquids relevant for metrology, one needs to
know energies of such systems for a large number configurations, {\em i.e.}, for different
geometries of clusters.  This issue is approached by using the many-body expansion, where here the
bodies are atoms or molecules forming the cluster, starting from two-body (pair) interactions,
followed by three-body (pairwise nonadditive) interactions.  The approach can be continued to
higher-level many-body interactions, but so far this has not been done.  The {\em ab initio}
energies are usually fitted to analytic forms only for the two- and three-body interactions.

In addition to energies, metrology applications often require knowledge of accurate values of
various properties of atoms and molecules, mainly the static and dynamic polarizabilities and 
magnetic susceptibilities.
These quantities can be computed as analytical energy derivatives with respect to appropriate
perturbations.  Properties of a single atom or molecule change in condensed phases and the so-called
interaction-induced corrections to properties of isolated atoms or molecules are of interest to
metrology.

As already mentioned above, although Schr\"odinger's quantum mechanics at the BO-approximation level
provides the bulk of the physical values of interest to metrology, computations of various effects
beyond this level are often needed to reduce uncertainties of these properties to the magnitude
needed for metrology standards.  We will refer to these as post-BO effects.  It should be stressed 
 that we really have in mind here the post-nonrelativistic-BO level since both the relativistic and QED corrections for moleculas 
are always computed using the BO approximation.  One goes beyond this approximation when computing
adiabatic and nonadiabatic corrections.  Any reasonably detailed description of methodologies used
in post-BO calculations would be too voluminous for the present review.  Therefore, we refer the
reader to the original papers, in particular
Refs.~\onlinecite{Cencek:01,Lach:04,Cencek:05,Przybytek2010,
  Cencek2012,Przybytek:12,Piszczatowski:15,Puchalski:16,
  Przybytek2017,Puchalski:20,Czachorowski2020}.

Systems of interest to thermodynamics-based precision metrology are mainly noble-gas atoms and their
clusters, and this section will be restricted to such systems, with the majority of text devoted to
helium.  Apart from being the substance whose behavior is closest to the ideal gas, it is also the
only system where theory can currently provide results that are generally more accurate than the
measured ones.  Nevertheless, neon and argon are also of significant interest since they may be used
in secondary standards to improve instrument sensitivity or ease of use. Although for many properties
computations for neon and argon have larger uncertainties than the best measurements, such results are
still useful as independent checks of experimental work and to guide extrapolation beyond the
measured range.

\subsubsection{Importance of explicitly correlated basis sets}
\label{ECG}

The current theoretical results for helium owe their very small uncertainties mostly to the use of
explicitly correlated basis sets. The calculations involving helium atoms are probably one of the
best examples where an important science problem was solved in such a way.  To clearly show where
this field would be without the use of such basis sets, we discuss in this subsection numerical
comparisons of ECG and orbital calculations for He$_2$, performed recently in
Ref.~\onlinecite{Szalewicz:23}.  The majority of molecular electronic structure calculations are
carried out using orbital basis sets.  This means that many-electron wave functions are expanded in
products of orbitals.  The simplest example is the CI method discussed earlier, where the wave
function is a linear combination of Slater determinants built of orbitals that are usually obtained
by solutions of HF equations.  However, expansions in orbital products converge slowly due to the
difficulty of reproducing the electron cusps in wave functions.

A way around this difficulty is to use bases that depend explicitly on $r_{12} = |{\bm r}_2 - {\bm
  r}_1|$, the distance between electrons.  Bases of this type are called explicitly correlated ones.
For few-electron systems, such bases are mostly used to directly expand the $N$-electron wave
functions of the nonrelativistic BO approximation.  The explicitly correlated bases are also often
used for many-electron systems within a perturbative or CC approach.~\cite{Szalewicz2010,Kong:12}
For two-electron systems, one mostly uses variants of Hylleraas--Slater bases with linear and/or
exponential dependence on powers of $r_{12}$, including the first power.  For more than two
electrons, integrals in such bases become very expensive and bases dependent on $\e^{-r_{12}^2}$,
{\em i.e.}, ECG bases, are mostly used.  For a review of the ECG approach, see
Refs.~\onlinecite{Szalewicz2010,Mitroy2013}.

Since expansions in explicitly correlated bases of the type described above approach solutions of
the Schr\"odinger equation, the equivalent orbital calculations should be performed at the FCI level.
As already mentioned, FCI calculations scale as $N!$ with the number of electrons and therefore are
the most expensive of all orbital calculations.  Even for He$_2$, FCI calculations cannot be
performed using the largest available orbital basis sets.  Therefore, the optimal orbital-based
strategy is a hybrid one consisting of performing calculations in the largest basis sets at a lower
level of theory, for example, at the CCSD(T) level, and adding to these results FCI corrections
computed in smaller basis sets.

The BO energies computed in ECG basis sets in Ref.~\onlinecite{Przybytek2017} established a new
accuracy benchmark for the helium dimer; see the description of these calculations in
Sec.~\ref{sec:BO}.
These ECG interaction energies were compared in Ref.~\onlinecite{Szalewicz:23} to those computed in
orbital bases at the hybrid CCSD(T) plus FCI level.  The largest available basis sets were applied.
For most points, the CCSD(T)+$\Delta$FCI approach gives errors nearly two orders of magnitude larger
than the ECG estimated uncertainties.  For a couple of points, the CCSD(T)+$\Delta$FCI results are
fairly close in accuracy to the ECG results, but this is mainly due to the former method
overestimating the magnitude of the interaction energy at small $R$ and underestimating at large
$R$.  Since these points are near the van der Waals minimum, some previous evaluations of the
performance of orbital methods restricted to this region might have been overoptimistic.  When the
whole range of $R$ is considered, CCSD(T)+$\Delta$FCI is no match for the ECG approach.  One should
also realize that any improvements of accuracy of the CCSD(T)+$\Delta$FCI approach would require a
huge effort, in particular one would have to develop quadruple-precision versions of all needed
orbital electronic structure codes.

\subsection{Helium atom polarizability}
\label{helpol}

One of the properties of helium required by precision measurement
standards~\cite{Moldover:16,Gavioso:16,SI:18} is the helium atom polarizability, both static and
dynamic (frequency-dependent).  Non-relativistic calculations of the static polarizability date back
to the 1930s and reached an accuracy of 0.1 ppb in 1996 calculations using Hylleraas basis
sets.~\cite{Yan:96} However, the relativistic correction, which is proportional to $\alpha^2$, could
be expected to contribute at the 80 ppm level.  Unfortunately, the values of these corrections
published before 2001 differed significantly from one another.  These discrepancies were resolved by
accurate calculations of Refs.~\onlinecite{Cencek:01} (using GTGs) and~\onlinecite{Pachucki:01}
(using Slater geminals) with uncertainties of 20 ppb relative to the total polarizability.  This
work used the Breit--Pauli operator,~\cite{Bethe:57} whose expectation values were computed with
appropriate components of the nonrelativistic wave function for the Hamiltonian depending on static
electric field.

The authors of Ref.~\onlinecite{Pachucki:01} also computed the QED corrections of order $\alpha^3$,
which turned out to be significant, amounting to 30 ppm relative to the total polarizability.
However, an important part of the QED correction, resulting from the so-called Bethe logarithm term
in the Hamiltonian, was only estimated.  The first calculation of this term was performed in
Ref.~\onlinecite{Lach:04}. This work used the basis set of explicitly correlated Slater geminals.
Bethe's logarithm is a particularly difficult term in the QED operator since it is the only term
that depends on the total nonrelativistic Hamiltonian, {\em i.e.}, it depends therefore on the
electric field.  Since the polarizability is proportional to the second derivative of energy with
respect to the field, one has to compute such second derivative of the Bethe logarithm.  This had
never been done before the work of Ref.~\onlinecite{Lach:04}, so the algorithms and their numerical
implementations had to be developed from scratch.  The final result was a new value of the QED
correction to the helium atom polarizability with uncertainty of 3 ppb relative to the total
polarizability. The contribution from the derivative of Bethe's logarithm was as large as 0.2 ppm
and its uncertainty was the major part of the overall uncertainty.  With such an accurate QED
contribution, the main source of uncertainty became the $\alpha^4$ terms which were estimated to
amount to 0.5 ppm, with uncertainty of 0.2 ppm.

In 2015, calculations of Refs.~\onlinecite{Cencek:01}
and~\onlinecite{Lach:04} were extended to frequency-dependent
polarizabilities in Refs.~\onlinecite{Piszczatowski:15,Puchalski:16}.
This polarizability was expanded in inverse powers of the wavelength
$\lambda$ up to $\lambda^{-8}$.  Different levels of theory were used for
each power of $\lambda$: up to $\alpha^4$ for the static term,
$\alpha^2$ for inverse powers 2 through 6 (only even powers contribute),
and nonrelativistic for 8.  The dynamic polarizability at the He-Ne
laser wavelength of 632.9908 nm had an uncertainty of 0.1 ppm.  This
uncertainty results entirely from the uncertainty of the static
polarizability.  The latter was reduced compared to
Ref.~\onlinecite{Lach:04} mainly because work of
Ref.~\onlinecite{Pachucki:06b} has shown that the error of the so-called
one-loop approximation that was used to evaluate the $\alpha^4$ terms is
only about 5\% when applied to the excitation energies of helium.
Another small change in the static polarizability was due to a slightly improved
value of the Bethe-logarithm contribution.

Further improved accuracy of helium's static polarizability was achieved in
Ref.~\onlinecite{Puchalski:20}.  This work concentrated on the second derivative of the Bethe
logarithm with respect to the electric field.  This quantity can be obtained in a couple of ways,
with completely different algorithms.  The goal was to achieve agreement between two such approaches
and also with Ref.~\onlinecite{Lach:04}.  This goal was met, providing a reliable cross-validation
for both approaches.  The agreement with Ref.~\onlinecite{Lach:04} was to within 5\% and the reasons
for this discrepancy were found.  The final results are given in Table~\ref{tab:alpha}.

\begin{table}[!htbp]
\caption{\label{tab:alpha} Static polarizability of ${}^4$He (in $a_0^3$, where $a_0$ is the Bohr
  radius) including relativistic and QED corrections.  When no uncertainty is given, the last digit is
  certain.  $m$ is the mass of the helium nucleus. $\partial_\epsilon^2 \ln k_0$ denotes second
  derivative of the Bethe logarithm with respect to the electric field} \begingroup
\setlength{\tabcolsep}{0.3em} \renewcommand{\arraystretch}{0.7}
\begin{tabular}{ld}
\hline\hline
Contribution & \multicolumn{1}{c}{\rm Value} \\
\hline
\hline
Nonrelativistic                  &  1.383\,809\,986\,4 \\
$\alpha^2$ relativistic          & -0.000\,080\,359\,9 \\
$\alpha^2/m$ relativistic recoil & -0.000\,000\,093\,5(1) \\
$\alpha^3$ QED $-\ \partial_\epsilon^2 \ln k_0$ term  
                                 & 0.000\,030\,473\,8 \\
$\partial_\epsilon^2 \ln k_0$ term          
                                 & 0.000\,000\,182\,2 \\
$\alpha^3/m$ QED recoil          & 0.000\,000\,011\,12(1) \\
$\alpha^4$ QED                   & 0.000\,000\,56(14) \\
Finite nuclear size              & 0.000\,000\,021\,7(1) \\
Total                            & 1.383\,760\,78(14) \\
\hline
\end{tabular}
\endgroup
\end{table}

The values of polarizabilities computed in
Refs.~\onlinecite{Cencek:01,Lach:04,Piszczatowski:15,Puchalski:16} had uncertainties orders of
magnitude smaller than the best experimental results.  However, recently a new, very accurate
measurement of this quantity was published.~\cite{Gaiser2018} The measured value of the molar
polarizability, 0.517\,254\,4(10) cm$^3$/mol, is consistent with the theoretical molar
polarizability computed from the atomic one listed in Table~\ref{tab:alpha} and equal to
0.517\,254\,08(5) cm$^3$/mol: the combined uncertainty is more than three times the difference and
the experimental uncertainty is 20 times larger than the theoretical one.

When a helium atom is in a gas or condensed phase, its polarizability changes due to interactions
with other atoms. In other words, the polarizability of a helium cluster is not equal to the sum of
polarizabilities of helium atoms.  This change is often referred to as collision-induced
polarizability and for atoms is a function of relative distance $R$.  This quantity was computed in
Ref.~\onlinecite{Cencek2011}.  Reconciling previously published inconsistent calculations, the
results of Ref.~\onlinecite{Cencek2011} were used to compute the second~\cite{Garberoglio2020} and
third~\cite{Garberoglio2021} dielectric virial coefficients of helium.  Very recently, the
collision-induced three-body polarizability of helium was computed.~\cite{a3_2023}

A system consisting of one or two helium atoms cannot have any dipole moment in the BO approximation
due to rotational symmetry. However, configurations of three or more atoms might have a non-zero
such moment, which in turn influences the value of the third dielectric virial
coefficient.~\cite{Garberoglio2021} Presently, the only {\em ab initio} description of the
three-body dipole moment of noble gases is the one developed by Li and Hunt.~\cite{Li97} However,
these results apply only at large separations, and do not have associated uncertainties. A
dipole-moment surface for the helium trimer with rigorously defined uncertainty is currently being
developed.~\cite{m3_2023}

\subsection{Helium dimer potential}
\label{sec:He2}

\subsubsection{Born--Oppenheimer level}
\label{sec:BO}

\begin{figure}[h]
\includegraphics[width=0.8\linewidth]{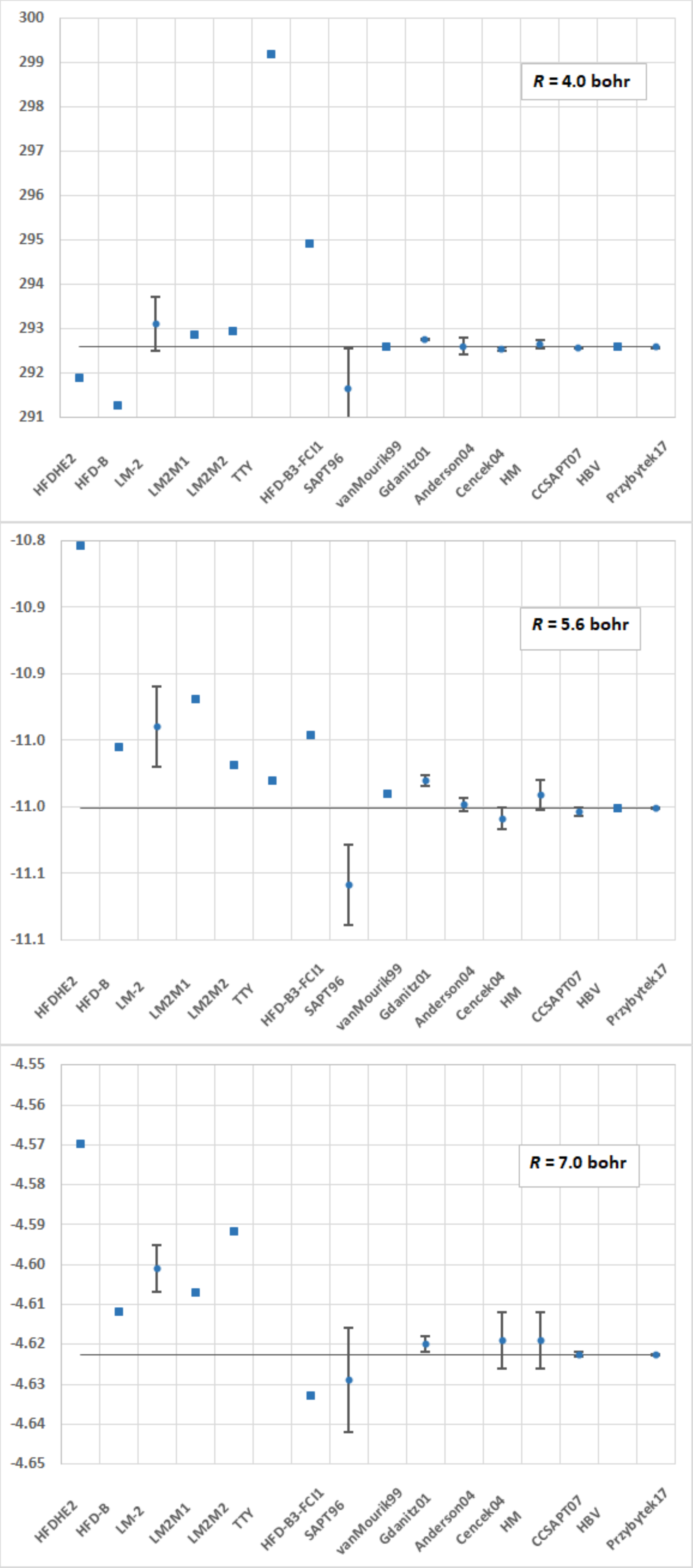}
\caption{%
Comparison of ECG BO interaction energies $E_{\rm int}$ (in kelvin) at $R$ = 4.0, 5.6, and 7.0 bohr
with those from selected earlier potentials.  For empirical potentials (HFDHE2 and HFD-B), the sum
of post-BO corrections was subtracted in each case.  The energies are plotted as error bars from
$E_{\rm int}-\sigma$ to $E_{\rm int}+\sigma$ (with dots at $E$) whenever uncertainty estimates
$\sigma$ are available and as squares otherwise (for three cases at $R = 7.0$~bohr, the energies are
not available).  The horizontal lines denote the positions of the BO energies calculated in
Ref.~\onlinecite{Przybytek2017}.  For acronyms, see the text.  }
\label{fig-history}
\end{figure}

The interest in the helium dimer potential is nearly as old as quantum mechanics.  In 1928,
Slater~\cite{Slater:28} developed the first potential for this system, which gave the interaction
energy of $-8.8$ K at the internuclear distance $R$ = 5.6 bohr (1 bohr $\approx 52.91772109$~pm).
There is a wide range of helium dimer potentials available in the literature (see
Ref.~\onlinecite{Spirko:13} for a comparison of bound-state calculations using a large number of
potentials).  Figure~\ref{fig-history} illustrates the remarkable
progress in accuracy of predictions achieved since 1979.
Empirical potentials dominated the field until the end of the 1980s; the two most widely used ones,
HFDHE2~\cite{Aziz:79} and HFD-B,~\cite{Aziz:87} developed by Aziz {\em et al.\/} The first really
successful {\em ab initio\/} potential was LM-2 by Liu and McLean.~\cite{Liu:89}  Those authors
performed CI calculations and, by analyzing the configuration space and basis set convergence,
obtained extrapolated interaction energies with estimated uncertainties (several versions of the
LM-2 potential were proposed).

Aziz and Slaman~\cite{Aziz:91} used the HFD-B functional form with refitted parameters to ``mimic''
the behavior of LM-2, of the unpublished {\em ab initio\/} data computed by Vos {\em et
  al.\/},~\cite{Vos} and of the small-$R$ Green-function Monte Carlo (GFMC) data~\cite{Ceperley:86}
to obtain potentials denoted as LM2M1 and LM2M2, differing by assuming the smallest and largest LM-2
potentials well depths, respectively.  The parameters of these potentials were not fitted directly
to {\em ab initio} data, but chosen by trial and error to reproduce both theoretical data and
measured quantities to within their error bars.  The LM2M2 potential was considered to be the best
helium potential until the mid-1990s, when purely {\em ab initio\/} calculations took the lead.
Among the latter ones, TTY is a remarkably simple potential based on perturbation theory developed
by Tang {\em et al.}~\cite{Tang:95} The HFD-B3-FCI1 potential was obtained by Aziz {\em et
  al.\/},~\cite{Aziz1995} who used the HFD-B functional form with its original parameters adjusted
so that the new potential runs nearly through the {\em ab initio} data points.  These points were
GFMC results of Ref.~\onlinecite{Ceperley:86} and the FCI results of van~Mourik and
van~Lenthe.~\cite{Mourik:95} No uncertainties were assigned to HFD-B3-FCI1 and
Fig.~\ref{fig-history} shows that it was about as accurate as LM1LM2.

The SAPT96 potential~\cite{Williams:96,Korona:97} opens an era of helium potentials based mostly on
calculations with explicitly correlated functions.  It was the first fully first-principles He$_2$
potential with reliable estimates of uncertainties.  The potential was obtained using a two-level
incremental strategy.  The leading SAPT corrections (the complete first-order and the bulk of the
second-order interaction energies) were computed using GTG basis
sets.~\cite{Szalewicz1979,Szalewicz2010,Mitroy2013} The GTG-based variant of SAPT was developed in
Refs.~\onlinecite{Chalasinski1977b, Szalewicz1979,Rybak1987,Rybak1989,Jankowski1990}.  Higher-order
SAPT corrections were computed using the general SAPT program based on orbital
expansions.~\cite{Rybak1991,Moszynski1993a,Moszynski1994c,Moszynski1994a, Williams:95a} Large
orbital basis sets including up to $g$-symmetry functions and midbond functions (placed between the
nuclei)~\cite{Williams:95} were used.  The remaining many-electron effects were computed using both
SAPT based on FCI monomers, with summations to a very high order of perturbation theory (using
He$_2$-specific codes), and supermolecular FCI calculations in small orbital basis sets.  It is
interesting to note that the errors of SAPT96 turned out to be completely dominated by the residual
orbital (rather than GTG) contributions.  For instance, at $R = 5.6$~bohr, the orbital part
constitutes only $-$1.81 K out of $-$11.00 K, but its error was $-$0.05 K out of the total SAPT96
error of $-$0.06 K. The underestimation of the uncertainties seen in Fig.~\ref{fig-history} for $R =
5.6$ bohr was entirely due to this issue.  With an added retardation correction, SAPT96 was used
(under the name of SAPT2) by Janzen and Aziz~\cite{Janzen:97} to calculate properties of helium and
found to be the most accurate helium potential at that time.

Van Mourik and Dunning~\cite{Mourik:99a} calculated CCSD(T) energies in
basis sets up to daug-cc-pV6Z, CCSDT $-$ CCSD(T)
differences in the daug-cc-pVQZ basis set, and 
FCI~$-$~CCSDT differences in the daug-cc-pVTZ basis set.  The
CCSD(T) energies were CBS-extrapolated and then refined by adding a
correction equal to the $R$-interpolated differences between highly
accurate CCSD(T)-R12 results (available at a few distances in
Ref.~\onlinecite{Noga:97}) and the obtained CBS limits.
The CC-R12 methods are analogous to CC-GTG methods, but the explicit
correlation factor enters linearly.~\cite{Kong:12}

Supermolecular ECG-based calculations for He$_2$ started to appear in the late
1990's,~\cite{Komasa:97,Komasa:99} and were initially aimed at providing
upper bounds to the interaction energies (by subtracting essentially
exact monomer
energies), as the authors did not attempt to extrapolate their results
to the CBS limits.  Another application of explicitly correlated
functions to the helium interaction was a series of papers by
Gdanitz,~\cite{Gdanitz:99,Gdanitz:00,Gdanitz:01} who used the
multireference averaged coupled-pair functional method with linear
$R_{12}$ factors, $R_{12}$-MR-ACPF.  The extrapolated results from the
last paper of the series, Ref.~\onlinecite{Gdanitz:01} (denoted ``Gdanitz01''
in Fig.~\ref{fig-history}), were among the most accurate results
available at that time.  However, the reported uncertainties were
strongly underestimated at shorter distances (as much as 5 times at 5.6
bohr and 17 times at 4.0 bohr).  

Another important series of papers was published by Anderson {\em et
al.\/},~\cite{Anderson:93,Anderson:01,Anderson:04} who reported quantum
Monte Carlo energies with progressively reduced statistical
uncertainties.  Although these results were obtained only for a few
internuclear distances, they represented very valuable benchmarks for
mainstream electronic structure methods.  In fact, until the publication of
the CCSAPT07 potential, the result from Ref.~\onlinecite{Anderson:04},
$-$10.998(5) K (see ``Anderson04'' in Fig.~\ref{fig-history}), was the
most accurate value available at 5.6 bohr.  

In Refs.~\onlinecite{Jeziorska:03,Cencek:04}, a hybrid supermolecular
ECG/orbital method was applied to the helium dimer.  The bulk of the
correlation effect on the interaction energy, at the CCSD level, was
evaluated using GTG functions and the method developed in
Refs.~\onlinecite{Szalewicz:82,Szalewicz:83a,Szalewicz:83b,Jeziorski:84,%
Szalewicz:84,Wenzel:86,Bukowski:94,Bukowski:95,Bukowski:96,%
Jeziorski:97,Bukowski:99,Przybytek:09}.
The nonlinear parameters were optimized at the
MP2 level.  The effects of
noniterative triple excitations (the ``(T)'' contribution),
{\em i.e.\/}, the differences between
CCSD(T) and CCSD energies, were calculated using large orbital basis
sets (up to aug-cc-pV6Z with bond functions and
daug-cc-pV6Z) and extrapolated to the CBS limits.
Finally, the FCI corrections (differences between FCI and CCSD(T)
energies) were obtained in basis sets up to aug-cc-pV5Z
with bond functions and daug-cc-pV5Z, and also
extrapolated.  Results for three distances were reported in
Ref.~\onlinecite{Cencek:04} (see ``Cencek04'' in Fig.~\ref{fig-history}).

Hurly and Mehl (HM) analyzed the best existing {\em ab initio\/} data
for the helium dimer and created a new potential~\cite{Hurly2007}
representing a compromise based on uncertainties of existing data and
their mutual agreement (for instance, as can be seen in
Fig.~\ref{fig-history}, the result from Ref.~\onlinecite{Cencek:04} was
used  at $R$ = 7.0 bohr).  The diagonal adiabatic corrections from
Ref.~\onlinecite{Komasa:99a} were added to the final potential, which
was then used to calculate the second virial coefficient, viscosity, and
thermal conductivity of helium.

The CCSAPT07 potential~\cite{Jeziorska:07} based on the hybrid GTG/orbital method, published in
2007, was a significant improvement over the previous complete potential of this type, {\em i.e.},
the SAPT96 potential.~\cite{Williams:96,Korona:97} CCSAPT07 combined three different computational
techniques, according to the criterion of the lowest uncertainty available for a given internuclear
distance.  Variational four-electron ECG calculations were used for $R \le 3.0$ bohr and SAPT+FCI
was employed for $R>6.5$ bohr.  At intermediate distances, the hybrid supermolecular method
developed in Refs.~\onlinecite{Jeziorska:03,Cencek:04} and described above provided the highest
accuracy.  Compared to Refs.~\onlinecite{Jeziorska:03,Cencek:04}, several computational
improvements were introduced,~\cite{Patkowski:07a} resulting in significantly reduced uncertainties.
The SAPT calculations~\cite{Jeziorska:07} of CCSAPT07 followed the SAPT96 recipe, but also with
larger basis sets and some computational improvements.  The uncertainties of this potential were
smaller than some effects that are neglected at the nonrelativistic BO level.  Calculations of these
effects will be discussed in Sec.~\ref{sec:postBO}

Another highly accurate potential, by Hellmann, Bich, and Vogel (HBV),~\cite{Hellmann:07} appeared
almost at the same time as CCSAPT07.  Those authors used very large basis sets (up to daug-cc-pV8Z
with added bond functions at the CCSD level, progressively smaller for higher levels of theory up to
FCI) and CBS extrapolations.  After augmenting the HBV potential with adiabatic, approximate
relativistic, and retardation corrections, the authors used it to calculate thermophysical
properties of helium.~\cite{Bich:07} However, the uncertainties of the HBV potential were not
estimated, which restricts its usefulness.  A direct accuracy comparison between the pure BO
component of HBV and CCSAPT07 is now possible because of the much higher accuracy of the present-day
benchmark energies,~\cite{Przybytek2017} and we performed such analysis using the values reported in
the last column of Table 3 in Ref.~\onlinecite{Hellmann:07}.  Out of 11 distances for which all
three energies are available, the largest relative error (with respect to the results of
Ref.~\onlinecite{Przybytek2017}), equal to 0.90\%, occurs for the CCSAPT07 energy at 5.0 bohr, while
the error of the HBV energy at this distance is only 0.48\%.  If one excludes this distance, which
is close to where the helium potential crosses zero, and calculates the average relative error at
the remaining distances, one obtains 0.007\% for CCSAPT07 and 0.011\% for HBV.  Therefore, both
potentials exhibit a similar level of accuracy.

The current most accurate nonrelativistic BO potential for the helium
dimer was published in Ref.~\onlinecite{Przybytek2017}; see
Ref.~\onlinecite{Szalewicz:23} for details of these calculations.
The significant improvement over all previous potentials
was achieved by a combination of
three factors.  First, a pure ECG approach was used, {\em i.e.\/}, with
all four electrons explicitly correlated and no contributions calculated with
orbital methods.  Indeed, the residual errors of the older hybrid
ECG-orbital potentials, SAPT96~\cite{Williams:96,Korona:97}
and CCSAPT07,~\cite{Jeziorska:07} were dominated by
insufficient basis set saturation of the relatively small orbital
contributions.
Second, the use of the monomer contraction method,~\cite{Cencek:05,Cencek08a} {\em i.e.}, the use of
the product of helium atoms wave functions as one of the functions in the basis set, dramatically
improved the energy convergence with respect to the ECG expansion size. Furthermore, an optimized
contraction~\cite{Patkowski2008,Cencek2011} replacement of the simple product of monomer wave functions by a more compact sum of
four-electron functions optimized for two noninteracting helium atoms, reduced the computational
cost at the nonlinear parameters optimization stage.  Third, a near-complete optimization of
nonlinear parameters in large basis set expansions was possible due to this reduced cost and due to
other improvements of the optimization algorithm.

\subsubsection{Physical effects beyond the nonrelativistic Born--Oppenheimer level}
\label{sec:postBO}

With the small uncertainties of the CCSAPT07 BO potential,~\cite{Jeziorska:07} it became clear that
a further reduction of uncertainties required inclusion of post-BO effects.  The first calculation
of all relevant such effects for the whole potential-energy curve was presented in
Ref.~\onlinecite{Przybytek2010} and later improved in
Refs.~\onlinecite{Cencek2012,Przybytek2017,Czachorowski2020}.  Some post-BO effects for the whole
curve were included even earlier in Refs.~\onlinecite{Hellmann:07,Bich:07}, but this work omitted
non-negligible two-electron terms in the $\alpha^2$ relativistic and $\alpha^3$ QED corrections.
The current helium dimer potentials at the post-BO level include the diagonal adiabatic correction,
relativistic corrections (earlier computed in Ref.~\onlinecite{Cencek:05}, but for the minimum
separation only), the QED correction, and the retardation effect (a long-range QED correction)
\[ 
   V(R) = V_{\rm BO} + V_{\rm ad} + V_{\rm rel} + V_{\rm QED} + V_{\rm ret} .
\]

In Refs.~\onlinecite{Przybytek2010,Cencek2012}, all the post-BO corrections were
computed in the supermolecular way as the differences of expectation
values of appropriate operators with the dimer and monomer wave
functions, except at the CCSD(T) level, see below.
The kinetic energy of the nuclei operator was used for the adiabatic
correction, the $\alpha^2$ Breit--Pauli operator~\cite{Bethe:57}
for the relativistic
correction, and the $\alpha^3$ QED operator~\cite{Pachucki:06}
for the QED correction.
In the latter case, one approximation was made in the operator. In the term
\[
      -\frac{8\alpha}{3\pi} \hat{D}_1 \ln k_0, \\
\]
with
\[
       \hat{D}_1  =  \frac{\pi}2 \,\alpha^{2}
       \sum_I \sum_{i=1}^4 Z_I \delta({\bm r}_{i}-{\bm r}_I),
\]
where the sum over $I$ is over the nuclei, the value of $\ln k_0$ should be computed for each $R$,
but instead a constant value was taken, equal to the value of $\ln k_0$ for the helium atom.  Thus,
the $R$ dependence originated entirely from that of the $\hat{D}_1$ operator.

All corrections were computed using both four-electron ECG basis sets
and orbital basis sets (except for the Araki--Sucher part of the QED
operator where only ECG functions were used).
The calculations with smaller uncertainties
were selected for the final potential.  Orbital calculations were performed
using a combination of CCSD(T) and FCI approaches or FCI alone.
For the adiabatic correction, only FCI was used.  The calculations of
the average values of the operators listed above with ECG and FCI wave
functions are straightforward (although regularization techniques have
to be used for singular operators).  However, the CCSD(T) wave function
needed to compute expectation values is not available (not defined) and
instead the CCSD(T) linear response method was used,~\cite{Coriani04} {\em i.e.}, first-order
analytic derivatives of CCSD(T) energy (with the perturbation operator
included) were computed.  The retardation effects of long-range
electromagnetic interactions were computed from the Casimir--Polder
expression~\cite{Casimir:48} dependent on the dynamic polarizability
of the helium atom.

Calculations of Ref.~\onlinecite{Przybytek2017} dramatically improved the accuracy of the helium
dimer potential with uncertainties reduced by an order of magnitude compared to those of
Refs.~\onlinecite{Jeziorska:07,Przybytek2010}.  As already discussed, the main improvement was due
to the use of larger and better optimized ECG wave functions at the nonrelativistic BO level of
theory for all $R \leq 9$~bohr.  Accuracy of the adiabatic and relativistic corrections was also
improved by using larger basis sets than in Refs.~\onlinecite{Przybytek2010,Cencek2012}.  A major
theoretical advance was the calculation of the properties of the very weakly bound state of
He$_2$ (the so-called halo state) with full inclusion of nonadiabatic effects.

The accuracy of relativistic and QED contributions was further improved
in Ref.~\onlinecite{Czachorowski2020}.  The contributions to the interaction
energy at the van der Waals minimum are presented in
Table~\ref{tab:V5.6}.  Clearly, with the uncertainty of the BO
contribution of $0.00020$~K, all the included post-BO contributions are
relevant, except for the retardation contribution, but this
contribution does become important at very large
separations.~\cite{Przybytek:12} One can also see that uncertainties
of the adiabatic,
relativistic, and QED terms are almost negligible compared to the
uncertainty of the BO term.  The potential of
Ref.~\onlinecite{Czachorowski2020} was used
to compute the second virial
coefficient and the second acoustic virial coefficient of helium.

\begin{table}[!htbp]
\caption{\label{tab:V5.6}
Contributions to the interaction energy of helium dimer (in K) at the van der
Waals minimum at $R = 5.6$ bohr. Results from
Refs.~\onlinecite{Przybytek2017,Czachorowski2020}.
}
\begingroup
\setlength{\tabcolsep}{0.3em}
\renewcommand{\arraystretch}{0.7}
\begin{tabular}{ldl}
\hline\hline
Contribution & \multicolumn{1}{c}{\rm Value} & Uncertainty \\
\hline
\hline
$V_{\rm BO}$     & -11.00071      &  0.00020 \\
$V_{\rm ad}$     &  -0.0089048    &  0.0000097 \\
$V_{\rm rel}$    &   0.0153911    &  0.0000154\\
$V_{\rm QED}$    &  -0.0013327    &  0.0000018 \\
$V_{\rm ret}$    &   0.000012 \\
\hline
\end{tabular}
\endgroup
\end{table}

\subsection{Nonadditive helium potentials}
\label{sec:he_u3}

In any fluid, the total interaction energy includes terms beyond
pairwise-additive interactions between monomers.  These so-called
nonadditive contributions begin with three-body
nonadditive terms defined as the part of the trimer interaction energy
that cannot be recovered by the sum of two-body interaction.
The additive and nonadditive interactions form a series called the many-body
expansion of interaction energy.  Fortunately, for all fluids consisting
of monomers interacting via noncovalent forces, this expansion converges
very fast and usually it is sufficient to limit calculation to two- and
three-body terms.  For a review of the many-body expansion, see
Ref.~\onlinecite{Szalewicz:05}.  For metrology purposes, the three-body
potential is needed to calculate the third virial coefficient.

A pairwise-nonadditive potential for helium was developed in Ref.~\onlinecite{Cencek:07} and
improved in Ref.~\onlinecite{Cencek2009}.  In the earlier work, two independent potentials were
obtained.  One was based on three-body SAPT~\cite{Lotrich:97a,Lotrich:97b,Lotrich:97c,Lotrich:98,
  Lotrich:00} and the other on the supermolecular CCSD(T) approach.  Orbital basis sets up to
aug-cc-pV5Z were used.  The two potentials were in very good agreement.  In
Ref.~\onlinecite{Cencek2009}, the CCSD(T) potential was improved by calculating the FCI correction
in an incremental approach and increasing the number of grid points, with CCSD(T) values taken from
Ref.~\onlinecite{Cencek:07} except for the new grid points.  Near the minimum of the total
potential, the three-body contribution is only $-0.0885$~K, which should be compared to the total
interaction energy of about $33$~K, but the three-body contribution is much larger than the
uncertainty resulting from the two-body term, which is $0.0006$~K.  The uncertainty of the
three-body term at the minimum of the total potential was estimated to be $0.002$~K.

Recently, the three-body potential for helium was further improved~\cite{u3_2023} by adding
relativistic and nuclear-motion corrections, using a new set of correlation-consistent basis sets
specifically developed for helium atoms,~\cite{Przybytek2017} and developing an improved analytic
form of the potential at large distances.
In particular, new terms were developed for the case when two atoms remain close while the third is
progressively more distant.  These refinements resulted in a reduction of the uncertainty by a
factor of about $5$ overall. In particular, the uncertainty at the minimum was reduced to $0.5$~mK,
a factor of 4 smaller than that of the previous work.~\cite{u3_2023}

\subsection{Heavier noble-gas atoms}
\label{sec:heavy}

While theory is superior to experiment for the helium atom and helium clusters, this is not the case
for neon, and even more so for argon.  The simple reason is the number of electrons: 2, 10, and 18,
respectively.  While for the helium atom and small helium clusters the $N$-electrons explicitly
correlated bases can reach ppm or smaller uncertainties, and FCI calculations can be performed in
fairly large bases, for the neon atom neither type of calculation can be performed in large enough
bases to get meaningful results.  To quantify this statement, let us examine the most accurate
calculations for the neon dimer,~\cite{Hellmann2021} see Table~\ref{tab:Ne2}.  The calculations
at the CCSD(T) level of theory were performed in the largest available basis sets: modified
daug-cc-pV8Z with bond functions.  The uncertainty of this result is about 200 ppm, which is only
10 times larger than the 20 ppm uncertainty of the He$_2$ BO interaction energy.  The main problem
comes from higher excitations.  As Table~\ref{tab:Ne2} shows, the uncertainties of the consecutive
terms increase, and the final $\Delta$(P) contribution increases the uncertainty of the total value
to about 1000 ppm.  The increase of uncertainties is due to the use of smaller and smaller basis
sets as the number of excitations increases: at the CCSDTQ(P) level of theory only the daug-cc-pVDZ
basis set could be used.  Even worse, the values of the consecutive contributions do not seem to
converge well, so it is not really possible to estimate the errors due the truncation of theory at the
CCSDTQ(P) level.  FCI calculations cannot currently be performed even in bases of the size used for
CCSDTQ(P), so FCI is not a solution either.

\begin{table}[!htbp]
\caption{\label{tab:Ne2} Contributions to the interaction energy of neon dimer (in K) at the van der
  Waals minimum, $R = 3.1$ \AA, from Ref.~\onlinecite{Hellmann2021}. (P) denotes noniterated
  quintuple excitations.  } \begingroup \setlength{\tabcolsep}{0.3em}
\renewcommand{\arraystretch}{0.7}
\begin{tabular}{ldl}
\hline\hline
Contribution & {\rm Value} & Uncertainty \\
\hline
\hline
CCSD(T)          & -41.3301      &  0.0100 \\
$\Delta$T        &  0.5730       &  0.0115 \\
$\Delta$Q        & -0.1645       &  0.0121 \\
$\Delta$(P)      &  0.1179       &  0.0589 \\
\hline
\end{tabular}
\endgroup
\end{table}

Similar calculations at the limits of the available technology were
performed for Ar$_2$ in Ref.~\onlinecite{Patkowski:10} (see also earlier
calculations~\cite{Patkowski:05} with accurate treatment of very small
$R$).  The final value of the interaction energy at the minimum is
($-142.86 \pm 0.46$)~K.  Thus, the estimated uncertainty is 3000 ppm (or
0.3\%).  The uncertainty does not include an estimate of effects beyond
CCSDTQ since in the case of Ar$_2$ the highest-excitation contribution was
only 0.12 K.  The results of Ref.~\onlinecite{Hellmann2021} indicate, however,
that the post-CCSDTQ contribution may be not negligible.

The first first-principles three-body potential for argon was developed in
Ref.~\onlinecite{Lotrich:97b} using three-body SAPT.  It was then used to compute the third virial
coefficient of argon~\cite{Mas:99} and to simulate vapor-liquid equilibria.~\cite{Bukowski:01} An
improved three-body potential for argon was developed in Ref.~\onlinecite{Cencek2013} using the CCSDT(Q)
level of theory and including correlation effects.  Uncertainties of the potential were estimated.
This work also computed the third virial coefficient, getting good overall agreement with
experiments.  In particular, in some regions of temperature, theoretical values have smaller
uncertainties than experiment and comparisons with theory allow evaluation of different
experiments.  When the experimental data were refitted by a new model that included an approximate
fourth virial coefficient, the agreement with theory improved, which can be considered to be a
validation of the new model.  The work of Ref.~\onlinecite{Cencek2013} shows that despite
limitations of accuracy, for some properties theory may provide information relevant for metrology
experiments and its accuracy may be competitive to experimental accuracy.

\subsection{Magnetic susceptibility}
Magnetic susceptibilities of atoms are relevant for RIGT (see
Eq.~(\ref{eq:RIGT})). In general, the magnetic susceptibility is several orders of magnitude smaller
than the electric susceptibility. Therefore, it is usually sufficient to compute it at the BO
level of theory. The contribution of relativistic effects can be estimated using perturbation
theory. These corrections can also be considered as an additional contribution to the
uncertainty of the calculation.

For the helium atom, the magnetic susceptibility was computed quite some time ago.~\cite{Bruch02}
Accurate calculations for neon~\cite{Lesiuk2020,Hellmann2022} and argon~\cite{Lesiuk2023} were
performed only recently.

\section{From Electronic Structure to Thermophysical Properties}
\label{sec:thermo}

Virial expansions are exact results from quantum statistical mechanics which enable a systematically
improvable evaluation of various thermophysical properties as a power series in density starting
from the ideal-gas reference system. The coefficients appearing in the $N$-th term of the series can
be computed from the knowledge of the interaction of clusters of $N$ particles.

In the case of the equation of state -- {\em i.e.}, the expansion of the pressure $p$ as a function
of density $\rho$ -- one obtains Eq.~(\ref{eq:pvirial})~\cite{Hirschfelder1954,Hill1987} together
with rigorous expressions for the virial coefficients $B(T)$, $C(T)$, $D(T)$, {\em etc.}, which turn
out to be functions of temperature only and are given by
\begin{eqnarray}
  \frac{B(T)}{\NA} &=& -\frac{1}{2V} \left(Z_2 - Z_1^2\right)
  \label{eq:B} \\
  \frac{C(T)}{\NA^2} &=& \frac{\left(Z_2 - Z_1^2\right)^2}{V^2} - \frac{1}{3V} \left(
  Z_3 - 3 Z_2 Z_1 + 2 Z_1^3 \right)
  \label{eq:C} \\
  \frac{D(T)}{\NA^3} &=& -\frac{Z_4 - 4 Z_3 Z_1 - 3 Z_2^2 + 12 Z_2 Z_1^2 - 6 Z_1^4}{8V} +
  \nonumber \\
  & & \frac{3(Z_2-Z_1^2) (Z_3 - 3 Z_2 Z_1 + 2 Z_1^3)}{2 V^2} - \nonumber \\
  & & \frac{5 (Z_2-Z_1^2)^3}{2 V^3},
  \label{eq:D}
\end{eqnarray}
with
\begin{equation}
\frac{Z_N}{N!} = \frac{Q_N(V,T) ~ V^N}{Q_1(V,T)^N},
\label{eq:ZN}  
\end{equation}
where $Q_N(V,T)$ is the partition function of a system of $N$ particles
evaluated in the canonical ensemble and $\NA$ is the Avogadro constant. These partition functions can be
calculated once the interaction potential $U_N(\mbx_1, \ldots, \mbx_N)$
among $N$ particles is known; the potential is generally expressed as
\begin{equation}
  U_N = \sum_{i<j}^N u_2(\mbx_i, \mbx_j) + \sum_{i<j<k}^N u_3(\mbx_i, \mbx_j,
  \mbx_k) + \ldots,
  \label{eq:UN}
\end{equation}
where $u_2$ is the pair potential, $u_3$ is the non-additive contribution to the three-body
potential, and so on.  In Eq.~(\ref{eq:UN}) we specialized to the case of atomic systems, which will
be principal topic of this review; in this case $\mbx_i$ represent the position of the $i$-th
atom. In the case of molecules, which we will discuss in Sec.~\ref{sec:molecules}, the various potentials
appearing in Eq.~(\ref{eq:UN}) depend also on coordinates $\mby_i$ that describe the intramolecular
configuration of molecule $i$. In particular, a single-body potential $u_1(\mby_1)$ will also appear
in Eq.~(\ref{eq:UN}).  The potentials and their uncertainties can be computed from first principles
using the methods described in Sec.~\ref{sec:abinitio}.  The most general expression for $Q_N(V,T)$
in quantum statistical mechanics is given by
\begin{eqnarray}
  Q_N(V,T) &=& {\sum_i}' \langle i | \e^{-\beta H_N} | i
  \rangle \label{eq:Q} \\
  &=& \frac{1}{N!} \sum_{j=1}^{N!} \sum_i \langle i | \e^{-\beta H_N} {\cal
    P}_j | i \rangle, \label{eq:Q_MC}
\end{eqnarray}
where $\beta = (\kB T)^{-1}$ and the primed sum in Eq.~(\ref{eq:Q}) is on a complete set of states
$|i\rangle$ of the $N$-body Hamiltonian $H_N$ with the proper symmetry upon particle exchange due to
the bosonic or fermionic nature of the particles involved. Equation~(\ref{eq:Q_MC}) is an equivalent
expression where the sum over the states has no restriction on the symmetry and the operators ${\cal
  P}_j$ represent the $j$-th permutation of particles in the Hilbert space, including the sign of
the permutation in the case of fermions. The latter expression will be the most convenient when
discussing the path-integral Monte Carlo approach for the calculation of virial
coefficients.~\cite{Feynman1965,Ceperley1995} The non-relativistic $N$-body Hamiltonian is
conveniently written as
\begin{eqnarray}
  H_N = \sum_{i=1}^N \frac{\mbpi_i^2}{2m_i} + U_N
  \equiv T_N + U_N,
  \label{eq:HN}
\end{eqnarray}
where we have introduced the momentum operator $\mbpi_i$ and mass $m_i$ for
the $i$-th particle and the second equality defines the $N$-body kinetic
energy $T_N$.

Virial expansions of the form of Eq.~(\ref{eq:pvirial}) have
been derived for several other quantities measured by the gas-based devices
described in Sec.~\ref{sec:expt}: the speed of sound in Eq.~(\ref{eq:agt}), the dielectric constant
in Eq.~(\ref{eq:DCGT}), and the index of refraction in Eq.~(\ref{eq:RIGT}).
The coefficients appearing in Eq.~(\ref{eq:agt}) are given by~\cite{Gillis1996}
\begin{eqnarray}
  \betaa(T) &=& 2 B + 2(\gamma_0-1) T \frac{\D B}{\D T} +
  \frac{(\gamma_0-1)^2}{\gamma_0} T^2 \frac{\D^2B}{\D T^2} \label{eq:beta_a}\\
  \RTg(T) &=& \left(\frac{\gamma_0 -1}{\gamma_0} Q^2 - \betaa(T) B(T)\right) + 
  \frac{2 \gamma_0 + 1}{\gamma_0} C + \nonumber \\
  & & \frac{\gamma_0^2-1}{\gamma_0} T \frac{\D C}{\D T} +
  \frac{(\gamma_0 -1)^2}{2\gamma_0} T^2 \frac{\D^2C}{\D T^2},
  \label{eq:gamma_a}
\end{eqnarray}
where the quantity $Q$ is
\begin{equation}
  Q = B + (2\gamma_0-1) T \frac{\D B}{\D T} + (\gamma_0-1) T^2 \frac{\D^2B}{\D T^2}.
 \label{eq:Qa}
\end{equation}

The density expansion of the dielectric constant $\epsr$ is generally given as a
generalization of the Clausius--Mossotti equation in one of the two equivalent forms given by
Eqs.~(\ref{eq:cm}) and (\ref{eq:DCGT}).
Until recently, derivations for the coefficients appearing in these equations
would agree on the expression for the second dielectric virial coefficient, $B_\varepsilon$, but
differ in the case of the higher-order coefficients.~\cite{Hill1958,Moszynski1995,Gray2011} A
systematic review of the dielectric expansion showed that the correct expressions
are~\cite{Gray2011,Garberoglio2021}
\begin{eqnarray}
  A_\varepsilon &=& \frac{4 \pi \alpha_1}{3} \NA \label{eq:Aeps} \\
  B_\varepsilon(T) &=& \frac{2 \pi \kB T}{3 V} \NA^2
  \frac{\partial^2 Z_2(V,T,E_0)}{\partial E_0^2} \label{eq:Beps} \\
  C_\varepsilon(T) &=& -\frac{2 \pi \kB T \NA^3}{3} \left[
    \frac{2}{V^2} \left( \frac{\partial Z_2}{\partial E_0} \right)^2 +
    \frac{2(Z_2-V^2)}{V^2} \frac{\partial^2 Z_2}{\partial E_0^2}
    \right. \nonumber \\
  & & \left. -\frac{1}{3V} \left(
  \frac{\partial^2 Z_3}{\partial E_0^2} - 3 V \frac{\partial^2
    Z_2}{\partial E_0^2}
  \right) \right],
  \label{eq:Ceps}  
\end{eqnarray}
where $\alpha_1$ is the atomic polarizability and the functions $Z_N$ are
given by expressions similar to Eq.~(\ref{eq:ZN}), where the interaction
Hamiltonian among the constituent particles of Eq.~(\ref{eq:HN}) is
extended with two terms in order to include the effect of the interactions
of the dipole moment and
the electronic polarizability of the system with an external electric field
of magnitude $E_0$. In Eqs.~(\ref{eq:Aeps})--(\ref{eq:Ceps}), the
derivatives are to be evaluated at $E_0 = 0$.
The two additional terms in the Hamiltonian are
\begin{eqnarray}
  H^\mathrm{dip}_N &=& - \left(
  \sum_{i=1}^N \mbm_1(i) + \sum_{i<j} \mbm_2(i,j) + \right . \nonumber \\
  & & \left.
  \sum_{i<j<k} \mbm_3(i,j,k) + \ldots  \right) \cdot \mbE_0 \label{eq:Hdip} \\
  H^\mathrm{pol}_N &=& -\frac{1}{2} \mbE_0 \cdot \left(
  \sum_{i=1}^N \mbalpha_1(i) + \sum_{i<j} \mbalpha_2(i,j) + \right.
  \nonumber \\
  & & \left. 
  \sum_{i<j<k} \mbalpha_3(i,j,k) + \ldots \right) \cdot \mbE_0,
  \label{eq:Hpol}  
\end{eqnarray}
where $\mbm_n$ and $\mbalpha_n$ are the (non-additive) dipole moments and the
(non-additive) electronic polarizabilities of a system of $n$ particles. In the case of
atoms, $\mbm_1$ and $\mbm_2$ are both zero, but a system of three particles
has, in general, $\mbm_3 \neq 0$.~\cite{Martin74,Bruch78}

An expression analogous to the Clausius--Mossotti equation (\ref{eq:DCGT}) was derived by
Lorentz and Lorenz for the refractive index $n$ and is reported in Eq.~(\ref{eq:RIGT}).
The Lorentz--Lorenz equation (\ref{eq:RIGT}) is relevant
to those experiments where the refractive index is measured by optical methods. In this case,
the refractive virial coefficients are a function of the angular frequency $\omega$ of the
electromagnetic radiation as well as the temperature.
Usually, the frequency dependence is approximated as a power-law expansion of the form
\begin{equation}
  B_\mathrm{R}(T) = B_\varepsilon + \omega^2 B_\mathrm{R}^{(2)},
\end{equation}
where $B_\mathrm{R}^{(2)}$ depends on the interaction-induced Cauchy
moment $\Delta S(-4)$.~\cite{Koch99}

\subsection{Classical limit}
\label{sec:classical}

Although the focus of this review is on calculations with no uncontrolled
approximation, let us briefly discuss the classical limit of the approach
we have outlined. Classical expressions can be computed
relatively easily, and provide a useful high-temperature check for the more
involved calculations described below.

Since quantum exchange effects are absent in classical mechanics, the only term that
remains in Eq.~(\ref{eq:Q_MC}) is the one corresponding to the identity
permutation, giving rise to the ``correct Boltzmann counting'' factor of
$1/N!$ in the partition functions.~\cite{Hill1987}
In the same limit, the kinetic term in the Hamiltonian~(\ref{eq:HN})
commutes with the potential energy $U_N$ as well as with $H^\mathrm{dip}_N$
and $H^\mathrm{pol}_N$. Its contribution can be integrated exactly,
resulting in a factor of the form $V^N / \Lambda_m^{3N}$ where $\Lambda_m = h /
\sqrt{2\pi m \kB T}$ is the thermal de~Broglie wavelength of the atoms
under consideration. Putting all of this together, one obtains
\begin{equation}
  Z_N^{[\mathrm{class}]}(V,T,E_0) = \int \e^{-\beta\left(U_N +
    H^\mathrm{dip}_N + {H^\mathrm{pol}}'_N\right)} ~ \D\mbX_N,
\label{eq:ZN_class}
\end{equation}
where the ${H^\mathrm{pol}}'_N$ is the same as Eq.~(\ref{eq:Hpol}), but
without the terms corresponding to $\mbalpha_1$. Additionally, we have denoted
by $\D \mbX_N$ the integration element in the space of all the coordinates
needed to describe a system of $N$ atoms, {\em e.g.}, the Cartesian coordinates
$\mbx_1, \ldots, \mbx_N$.
Since the system is translationally invariant, the integration produces a
factor of $V$ with the understanding that one particle, usually labelled as
$1$, is fixed at the origin of the coordinate system.
Using rotational invariance, one can further write for the integration
elements
\begin{eqnarray}
  \D \mbX_2 &=& V ~ 4 \pi r_{12}^2 ~ \D r_{12} \\
  \D \mbX_3 &=& V ~ 8 \pi^2 (r_{12} r_{13})^2 ~ \D r_{12} \D r_{13} \D \cos\theta_{23} \\
  \D \mbX_4 &=& V ~ 8 \pi^2 (r_{12} r_{13} r_{14})^2 \times \nonumber \\
  & & \D r_{12} \D r_{13} \D r_{14} \D \cos\theta_{23} \D \cos\theta_{14} \D \phi,
  \label{eq:DX4}
\end{eqnarray}
where we have denoted by $r_{ij} = |\mbr_{ij}| = |\mbx_i - \mbx_j|$ and
$\theta_{ij}$ the angle between the vectors $\mbr_{i1}$ and $\mbr_{j1}$. In
Eq.~(\ref{eq:DX4}) the angle $\phi$ is the polar angle corresponding to the
vector $\mbr_{14}$ in spherical coordinates.

Using Eqs.~(\ref{eq:B}), (\ref{eq:beta_a}), and (\ref{eq:Beps}), one obtains
the classical expressions
\begin{eqnarray}
  B_\mathrm{cl} &=& -2\pi \NA \int \left( \e^{-\beta U_2(r_{12})} - 1 \right)~
  r_{12}^2 \D r_{12} \\ 
  \beta_\mathrm{a,cl} &=& -2 \pi \NA \int r_{12}^2 \left[
    2 \left( \e^{-\beta U_2} - 1 \right) + \right. \nonumber \\
    & & 2(\gamma_0-1) \beta U_2 \e^{-\beta U_2} + \nonumber \\
    & & \left. \frac{(\gamma_0-1)^2}{\gamma_0}
    \beta U_2 (\beta U_2-2) \e^{-\beta U_2} \right] \D r_{12} \\
  B_{\varepsilon,\mathrm{cl}} &=& \frac{8\pi^2}{3} \NA^2 \int
  \Delta\alpha_2(r_{12}) \e^{-\beta U_2(r_{12})}   \D r_{12},
  \label{eq:Beps_cl}
\end{eqnarray}
for the second density, acoustic, and dielectric second virial coefficient,
respectively. In Eq.~(\ref{eq:Beps_cl}) we have defined $\Delta \alpha_2 =
\frac{1}{3}\mathrm{Tr}(\mbalpha_2)$, which is the average of
the interaction-induced pair polarizability.
The classical expression for $B_\mathrm{R,cl}^{(2)}$ is analogous to Eq.~(\ref{eq:Beps_cl}), where
$\Delta \alpha_2$ is substituted by the Cauchy moment $\Delta S(-4)$.

In the same way, one can derive expressions for the classical limit of the third density, acoustic, and
dielectric virial coefficients using Eqs.~(\ref{eq:C}), (\ref{eq:gamma_a}), and
(\ref{eq:Ceps}). After some lengthy, but straightforward, evaluation, they turn out to be
\begin{widetext}
  \begin{eqnarray}
    C_\mathrm{cl} &=& -\frac{8 \pi^2}{3} \NA^2 \int \left[
      \e^{-\beta U_3} - \sum_{i<j} \e^{-\beta U_2(r_{ij})}
       + 2 - 3(\e^{-\beta U_2(r_{12})} - 1) (\e^{-\beta U_2(r_{13})} - 1) 
      \right] ~ \D \Omega_3 \\
  C_{\varepsilon,\mathrm{cl}} &=& \frac{16\pi^2}{9} \NA^3 \int \left[
    \left( \frac{\beta |\mbm_3|^2}{3} + \Delta\alpha_3 \right)
    \e^{-\beta U_3} - \sum_{i<j} \Delta\alpha_2(r_{ij})
    \e^{-\beta U_2(r_{ij})} - \right. \nonumber \\
    & & \left.
    6 \left(\e^{-\beta U_2(r_{12})}-1\right)
    \Delta\alpha_2(r_{13}) \e^{-\beta U_2(r_{13})} 
    \right] ~ \D \Omega_3,
  \end{eqnarray}
\end{widetext}
where $\D \Omega_3 = (r_{12} r_{13})^2 ~ \D r_{12} \D r_{13} \D \cos\theta_{23}$,
$r_{23} = \sqrt{r_{12}^2 + r_{13}^2 - 2 r_{12} r_{13} \cos\theta_{23}}$,
and $\Delta \alpha_3 = \frac{1}{3} \mathrm{Tr}(\mbalpha_3) +
\displaystyle\sum_{i<j} \Delta \alpha_2(r_{ij})$. The classical expression
for the third acoustic virial coefficient $\gammaa$ is more involved and is given in
Appendix~\ref{app:RTg}.

\subsection{Quantum calculation of virial coefficients}
\label{sec:quantum_virials}

The classical approach can be expected to be valid when $\Lambda / \sigma \ll 1$, where $\sigma$ can
be taken as the size of the hard-core repulsive region of atoms (which is around $0.3$~nm for the
noble gases): this implies that the classical formulae will be asymptotically valid for high
temperatures and heavy atoms. However, in the case of helium this approximation is too drastic even
at room temperature.

The inclusion of quantum effects in the calculation of virial coefficients (density, acoustic, or
dielectric) requires evaluating the $N$-body partition functions $Q_N$ of Eq.~(\ref{eq:Q}) in a
quantum framework. A straightforward approach would be to consider in Eq.~(\ref{eq:Q}) the
eigenstates $|i\rangle$ of the $N$-body Hamiltonian, $H_N|i\rangle = E_i |i\rangle$, so that
Eq.~(\ref{eq:Q}) becomes a simple sum. To the best of our knowledge, this method has been
demonstrated to date only in the case of the second dielectric virial
coefficient.~\cite{Garberoglio2020}

In the case of $Q_2$ (which enables the calculation of virial coefficients of order 2), a very
fruitful approach dating back to the late 1930s~\cite{DeBoer1939,Hirschfelder1954} is to rewrite it
as the sum of three terms: one depending on the bound-state energies, one depending on the phase
shifts of the scattering states, and one depending on the bosonic or fermionic nature of the atoms
involved. The expression of $B(T)$ becomes then
\begin{widetext}
\begin{eqnarray}
  B(T) &=& B_\mathrm{th}(T) + B_\mathrm{bound}(T) + B_\mathrm{xc}(T) \label{eq:Bps}\\
  B_\mathrm{th}(T) &=& -\frac{\NA \Lambda_\mu^3}{\pi} \int \e^{-\beta E} S(E) ~\beta \D E\\
  S(E) &=& \sum_l (2l+1) f(I,l) \delta_l(E) \label{eq:SE} \\
  B_\mathrm{bound}(T) &=& -\frac{\NA \Lambda_\mu^3}{2} \sum_{l,n} (2l+1)
  f(I,l)
  \left( \e^{-\beta E_{nl}^\mathrm{bound}} -1 \right)\\
  B_\mathrm{xc}(T) &=& -\NA \frac{(-1)^{2I}}{2I+1} \frac{\Lambda_\mu^3}{16}, \label{eq:Bxc}
\end{eqnarray}  
\end{widetext}
where $\mu$ is the reduced mass of the pair of atoms considered, $E_{nl}^\mathrm{bound}$ is the
energy of the $n$-th bound state with relative angular momentum $l$, and $f(I,l) =
1+(-1)^{2I+l}/(2I+1)$ with $I$ the nuclear spin in the case of identical atoms (the case of
different atoms can be recovered by letting $I \to \infty$). The quantity $\delta_l(E)$ in
Eq.~(\ref{eq:SE}) is the {\em absolute} scattering phase shift for two particles with relative
energy $E$ and angular momentum $l$. Absolute phase shifts are continuous functions of $E$ that tend,
in the limit of $E \to 0$, to $\pi$ times the number of bound states at angular momentum $l$.  With
the advent of electronic computers, the use of Eqs.~(\ref{eq:Bps}) to (\ref{eq:Bxc}) enabled the
calculation of accurate numerical values~\cite{Kilpatrick1954,Boyd1969} and it is still the most
efficient way to compute the second virial coefficient of atomic
species.~\cite{Cencek2012,Czachorowski2020} 
One important benefit of this method is that once the energies of all the bound states have been
computed and phase shifts are known for a sufficiently high number of total angular momenta and
scattering energies, the values of $B(T)$ and its derivatives, and hence $\betaa(T)$, can be easily
computed at all temperatures; knowledge of the collision-induced pair polarizability also enables
the calculation of $B_\varepsilon$.~\cite{Garberoglio2020} Additionally, transport properties such
as the viscosity and the thermal conductivity -- see Sec.~\ref{sec:transport} below -- can be
computed in a straightforward manner.

Unfortunately, this approach cannot be easily extended to higher-order
coefficients. Some attempts in this direction were made in the
1960s,~\cite{Pais59,Larsen63} but all of them required the introduction of
some uncontrolled approximations and were not developed to take into account the
non-additive parts of the many-body potential.

\subsubsection{Path integral approach}

At the same time, the path-integral approach to quantum statistical
mechanics~\cite{Feynman1965} was shown by Fosdick and Jordan to provide a systematic way to
compute virial coefficients of any order without any uncontrolled
approximation.~\cite{Fosdick1966,Jordan1968}
The path-integral formulation is based on a controlled approximation
of the exponential of the $N$-body Hamiltonian, that is
\begin{eqnarray}
  \e^{-\beta H_N}
  &=&
  \left( \e^{-\beta (T_N + U_N) / P}\right)^P \\
  &\simeq&
  \left(
  \e^{-\beta T_N  / P}
  \e^{-\beta U_N  / P}
  \e^{-\beta O / P}  
  \right)^P,
  \label{eq:TI}
\end{eqnarray}
where
\begin{equation}
  O = \frac{\beta^2 \hbar^2}{24 P^2 m} \sum_{i=1}^N |\nabla_i U_N|^2.
  \label{eq:O}
\end{equation}

Equation~(\ref{eq:TI}) is the Li--Broughton expansion of the exponential of the sum,~\cite{LB1987}
which was independently discovered by Kono {\em et al.}~\cite{KTL88} based on an initial idea by
Takahashi and Imada.~\cite{TI84} It can be shown that Eq.~(\ref{eq:TI}) becomes an exact equality in
the case $P \to \infty$, although in practice satisfactory convergence is reached for a finite value
of the parameter $P$.  Actually, Eq.~(\ref{eq:TI}) becomes an equality in the $P \to \infty$ limit
also when $O$ is omitted in Eq.~(\ref{eq:TI}) (this is the original Trotter--Suzuki approach),
although in this case convergence is attained at higher values of $P$; this approach is called the
{\em primitive approximation},~\cite{Ceperley1995} and, for the sake of simplicity, will be used
throughout this review.

The path-integral approach is obtained by using Eq.~(\ref{eq:TI}) in Eq.~(\ref{eq:Q_MC})
and inserting $P-1$ additional completeness relations between the $P$
factors in Eq.~(\ref{eq:TI}). Additionally, one uses as a complete set the
(generalized) position eigenstates $|i\rangle = |\mbX_N^{(1)}\rangle$,
where
we have included a superscript $(1)$ for later convenience.
In this case, the sum over $i$ in Eq.~(\ref{eq:Q_MC}) becomes an integral
over the $3N$ coordinates $\mbX_N^{(1)}$ and the $P-1$ completeness relations can be written as
\begin{equation}
  \mathbf{1} = \int | \mbX_N^{(k)}\rangle \langle \mbX_N^{(k)} | ~ \D \mbX_N^{(k)},
  \label{eq:completeness}
\end{equation}
with $k=2, \ldots, P-1$. Notice that in this case the effect of the
permutation operators ${\cal P}_j$ is to exchange atomic coordinates in the
rightmost ket. For example, if ${\cal P}(12)$ denotes the permutation of particles $1$ and $2$
(assumed to be bosons), one has
\begin{equation}
  {\cal P}(12)|\mbx_1^{(1)}, \mbx_2^{(1)}, \ldots, \mbx_N^{(1)}\rangle =
  |\mbx_2^{(1)}, \mbx_1^{(1)}, \ldots, \mbx_N^{(1)}\rangle. 
\label{eq:P12} 
\end{equation}

Let us first proceed assuming that ${\cal P}$ is the identity permutation
(that is, we are considering Boltzmann statistics; this approximation is essentially exact for $T
\gtrsim 10$~K even in the case of helium) and the case of
density virials of pure species (so that our Hamiltonian is given by
Eq.~(\ref{eq:HN}) with $m_i = m$). The operators $U_N$
(and, if needed, $O$) of Eq.~(\ref{eq:TI}) are diagonal in the position
basis. The matrix elements of the exponential of the kinetic energy
operators can be calculated exactly~\cite{Tuckerman10} and are given by
\begin{equation}
  \langle \mbx^{(k+1)}_i |
  \e^{-\frac{\beta \mbpi_i^2}{2mP}}
  |\mbx^{(k)}_i \rangle =
  \frac{P^{3/2}}{\Lambda^3} \exp\left(-\frac{\pi P}{\Lambda^2}
  \left| \mbx^{(k+1)}_i - \mbx^{(k)}_i \right|^2 \right),
\label{eq:KE_PIMC}
\end{equation}
so that $Z_N$ can be written as
\begin{eqnarray}
  Z_N = \int \e^{-\beta \overline{U_N}}
  \prod_{i=1}^N F_i \prod_{k=1}^P \D \mbX_i^{(k)},
  \label{eq:ZN_PI}
\end{eqnarray}
where
\begin{eqnarray}
  \overline{U_N} &=& \frac{1}{P} \sum_{k=1}^P U_N(\mbX_N^{(k)})
  \label{eq:UNbar}\\
  F_i &=& \Lambda^3 \left( \frac{P^{3/2}}{\Lambda^3} \right)^P
  \exp\left(-\frac{\pi P}{\Lambda^2} \sum_{k=1}^P \Delta {\mbr_i^{(k)}}^2 \right)
  \label{eq:F} \\
  \Delta {\mbr_i^{(k)}}^2 &\equiv&
  \left|\mbx_i^{(k+1)} - \mbx_i^{(k)} \right|^2,
\end{eqnarray}
with the understanding that $\mbx_i^{(P+1)} = \mbx_i^{(1)}$.  Equations
(\ref{eq:ZN_PI})--(\ref{eq:F}), which correspond exactly (in the $P \to
\infty$ limit) to the original quantum statistical formulation, can be
interpreted as the partition function of a {\em classical}
system.~\cite{Tuckerman10} For each of the original $N$ particles of
coordinates $\mbx_i^{(1)}$, one has introduced $P-1$ copies of coordinates
$\mbx_i^{(k)}$, which, as one can see from Eq.~(\ref{eq:F}), are connected
via harmonic potentials. The equivalent classical system is then made by
$N$ ring polymers of $P$ monomers each.  As shown by Eq.~(\ref{eq:UNbar}), these
polymers interact with the original potential averaged over all the
monomers. It can be shown that the functions $F_i$ of Eq.~(\ref{eq:F})
represent probability distributions.~\cite{Garberoglio2008} Although they
are not Gaussian probabilities, because of the ring-polymer condition
$\mbx_i^{(P+1)} = \mbx_i^{(1)}$, they can be sampled exactly using an
interpolation formula due to Levy~\cite{Fosdick1966,Jordan1968} (also known
as ``the Brownian bridge'').
The harmonic intra-polymer interaction, which ultimately
comes from the kinetic energy term $T_N$ of the quantum Hamiltonian
(\ref{eq:HN}), has the effect that the average ``size'' of the ring-polymer
corresponding to each particle is of the order of the de~Broglie thermal
wavelength $\Lambda$, thus taking into account quantum diffraction (that
is, the Heisenberg uncertainty principle).

In order to compute the functions $Z_N$ (and, hence, the virial
coefficients), it is convenient to separate the $NP$ vector coordinates
$\mbx_i^{(k)}$ as follows: first of all, we notice that the energy of the
equivalent classical system is invariant upon an overall rigid rotation or
rigid translation. We can use the latter property to extract a factor of
$V$ and at the same time pin one of the coordinates -- conventionally the
first monomer of particle 1, that is $\mbx_1^{(1)}$ -- at the origin of the
coordinate system. The rotational invariance can be taken into account by
assuming that the first monomer of one particle (particle $2$, say) lies
along the $x$ axis of the coordinate system and that the first monomer of
another particle (particle $3$) lies in the $xy$ plane. This convention
brings about a factor of $4 \pi$ when $N=2$ (corresponding to the
integration over the two polar angles describing $\mbx_2^{(1)}$) and a
factor of $8 \pi^2$ (that is the integration over the two polar angles describing
$\mbx_2^{(1)}$ and the azimuthal angle of $\mbx_3^{(1)}$) when $N \geq
3$. The remaining $3NP-6$ coordinates (or $3NP-5$ in the case of $N=2$) can
be conveniently divided into
\begin{enumerate}
\item The coordinates of the first bead of all the particles, that is
  $r_{12} = |\mbx_2^{(1)} - \mbx_1^{(1)}| $ and, for $N\geq3$, $r_{13} =
  |\mbx_3^{(1)} - \mbx_1^{(1)}|, \cos\theta_{23}$ and $\mbx_i^{(1)}$ (the
  latter only for $N \geq 4$), where $\theta_{23}$ is the angle between the
  position of particles $2$ and $3$ in the $xy$ plane.
\item The relative coordinates $\Delta \mbr_i^{(k)}$ ($k=1, \ldots, P-1$).
\end{enumerate}

Since the functions $F_i$ depend only on $\Delta \mbr_i^{(k)}$, one can
rewrite the partition functions $Z_N$ of Eq.~(\ref{eq:ZN_PI}) in the form
\begin{eqnarray}
  Z_2 &=& V ~ 4\pi \int  \left\langle \e^{-\beta \overline{U_2}} \right\rangle ~
  r_{12}^2 \D r_{12} \label{eq:Z2_int}\\
  Z_N (N \geq 3) &=& V ~ 8 \pi^2 ~ \int \left\langle \e^{-\beta \overline{U_N}} \right\rangle
  \D \Omega_N \label{eq:ZN_int} \\
  \D \Omega_N (N \geq 4)  &=& \D \Omega_3   \prod_{i=4}^N \D \mbx_i^{(1)}, \label{eq:dOmegaN}
\end{eqnarray}  
where
\begin{eqnarray}
  \left\langle \e^{-\beta \overline{U_N} } \right\rangle =
  \int \e^{-\beta \overline{U_N} } \prod_{i=1}^N F_i \prod_{k=1}^{P-1} \D \Delta \mbr_i^{(k)}
\label{eq:ang_avg},
\end{eqnarray}
denotes the average of the Boltzmann factor of the potential energy over the
internal configurations of the ring polymers. Finally, using
Eqs.~(\ref{eq:Z2_int})--(\ref{eq:dOmegaN}) and the definition of the virial
coefficients (\ref{eq:B}) and (\ref{eq:C}), one obtains
\begin{widetext}
  \begin{eqnarray}
    B &=& -2\pi \NA \int \left \langle
    \e^{-\beta \overline{U_2}(r_{12})} - 1
    \right\rangle r_{12}^2 ~ \D r_{12} \label{eq:B_PI} \\
    B_\varepsilon &=&
    \frac{8 \pi^2}{3} \NA^2 \int \left\langle 
    \overline{\Delta \alpha_2}(r)
    \e^{-\beta \overline{U_2}(r_{12})}
    \right\rangle r_{12}^2 \D r_{12}
    \label{eq:Beps_PI} \\
    C &=& -\frac{8 \pi^2}{3} \NA^2 \int \left\langle \left[
      \e^{-\beta \overline{U_3}(r_{12}, r_{13}, r_{23})} - \sum_{i<j} \e^{-\beta \overline{U_2}(r_{ij})}
       + 2 - 3(\e^{-\beta \overline{U_2}(r_{12})} - 1) (\e^{-\beta \overline{U_2}(r_{13})} - 1) 
       \right]\right\rangle ~ \D \Omega_3 \label{eq:C_PI} \\
  C_\varepsilon &=& \frac{16\pi^2}{9} \NA^3 \int \left\langle \left[
    \left( \frac{\beta |\overline{\mbm_3}|^2}{3} + \overline{\Delta\alpha_3} \right)
    \e^{-\beta \overline{U_3}} - \sum_{i<j} \overline{\Delta\alpha_2}(r_{ij})
    \e^{-\beta \overline{U_2}(r_{ij})}
    - \right. \right. \nonumber \\
    & & \left. \left.
    6 \left(\e^{-\beta \overline{U_2}(r_{12})}-1\right)
    \overline{\Delta\alpha_2}(r_{13}) \e^{-\beta \overline{U_2}(r_{13})} 
    \right] \right\rangle~ \D \Omega_3, \label{eq:Ceps_PI} 
  \end{eqnarray}
\end{widetext}
which are very similar to the classical expressions reported in Sec.~\ref{sec:classical}. The
path-integral expressions are obtained by the classical expressions substituting the evaluation 
of potentials and polarizabilities as averages over the ring-polymer beads (see
Eq.~(\ref{eq:UNbar})) and averaging the resulting expressions over the configurations of the ring
polymers, as evidenced by the angular brackets in Eqs.~(\ref{eq:B_PI})--(\ref{eq:Ceps_PI}).  
The path-integral expression for $B_\mathrm{R}^{(2)}$ is obtained from Eq.~(\ref{eq:Beps_PI})
by the substitution of $\overline{\Delta\alpha_2}$ with $\overline{\Delta
  S(-4)}$.~\cite{Garberoglio2020}
Explicit expressions for the third acoustic virial coefficient in
the path-integral formulation are quite cumbersome, for reasons discussed in Appendix~\ref{app:RTg};
they can be found in Ref.~\onlinecite{Binosi2023}.

It is important to notice that in the case of $C(T)$ the terms coming from
$Z_2^2$ in Eq.~(\ref{eq:C}) actually involve averages over {\em four} ring
polymers, since these two terms involve two particles each and have to be
treated as {\em independent}, lest spurious correlations be introduced in
the calculation of the $\langle \cdots \rangle$ average. In fact, in the
last term of Eq.~(\ref{eq:C_PI}) two of these polymers are used to compute
$\e^{-\beta \overline{U_{2}}(r_{12})} - 1$ and the other two to compute
$\e^{-\beta \overline{U_{2}}(r_{13})} - 1$.
Similar considerations also apply when calculating $\gammaa$ and $C_\varepsilon$ using path
integrals. 

Quantum effects are taken into account by averaging over the ring-polymers configurations, and at
the same time evaluating the interaction energy as an average over the monomers, as in
Eq.~(\ref{eq:UNbar}). We recall that in Eqs.~(\ref{eq:B_PI}) and (\ref{eq:C_PI}) the radial
variables $r_{ij} = |\mbx_i^{(1)} - \mbx_j^{(1)}|$ are the distances between the first monomer of
particles $i$ and $j$.  In the classical limit, the size of the ring polymers shrinks to zero so
that one recovers the results of Sec.~\ref{sec:classical}.

It is worth noting that one can find several semi-classical approximations of the exact
path-integral expressions of Eqs.~(\ref{eq:B_PI}) -- (\ref{eq:Ceps_PI}). In general, they can be
obtained by expanding the full quantum-mechanical results in powers of $\hbar^2$, where the first
term is the classical one. This approach was pioneered by Wigner and
Kirkwood,~\cite{Wigner32,Kirkwood33} and subsequently developed by Feynman and
Hibbs,~\cite{Feynman1965} who put forward the idea of estimating semiclassical values by using the
classical expressions with suitably modified (and temperature-dependent) potentials.
Although the Feynman--Hibbs approach considered systems with pair potentials only, a systematic
derivation of semiclassical expressions in the case of three-body interactions has been developed by
Yokota.~\cite{Yokota60} Even if semiclassical approaches introduce uncontrolled approximations,
they are quite effective in the case of heavier atoms such as argon at high
temperatures and provide a useful check for the fully quantum calculations.

\subsubsection{Exchange effects}

The bosonic or fermionic nature of the particles enters in those terms of
Eq.~(\ref{eq:Q_MC}) where the permutation operator is different from the
identity.  In the case of the equivalent classical system, the main effect
of the permutation operators is that the condition of closed ring polymers,
that is $\mbx_i^{(P+1)} = \mbx_i^{(1)}$, is no longer valid.  For
a general permutation, one would have $\mbx_i^{(P+1)} = \mbx_{{\cal
    P}(i)}^{(1)}$ where ${\cal P}(i)$ denotes the particle exchanged with
$i$ under the action of permutation ${\cal P}$. This is equivalent to saying
that some of the ring polymers would coalesce into larger polymers, depending on
the specific permutation that is being considered in the sum of
Eq.~(\ref{eq:Q_MC}).  These larger ring polymers are still described by
probability distributions similar to those of the Boltzmann case, that is
Eq.~(\ref{eq:F}). As an illustrative example, let us see how the probability
distribution for the internal coordinates of particles $1$ and $2$ is
modified in the presence of exchange for bosons of spin 0. Defining $\mbR_i
= \mbx_1^{(i)}$ and $\mbR_{i+P} = \mbx_2^{(i)}$ for $i=1,\ldots,P$ as well
as $\Delta \mbR_i = \mbR_{i+1} - \mbR_{i}$ (notice that $\mbR_{2P} =
\mbR_1$ because we are considering the permutation involving only particle
$1$ and $2$), and $\Lambda_\mu = \sqrt{2} \Lambda$, the kinetic energy terms
that would give rise to the probabilities $F_1 F_2$ can be written as
\begin{eqnarray}
  F_1 F_2 &\to& \Lambda^6 \left( \frac{P^{3/2}}{\Lambda^3} \right)^{2P}
  \exp\left(-\frac{\pi P}{\Lambda^2} \sum_{k=1}^{2P} {\Delta \mbR_i}^2
  \right) \\
  &=& \frac{\Lambda^3 \Lambda_\mu^3}{2^{3/2}} 
  \left( \frac{(2P)^{3/2}}{\Lambda_\mu^3} \right)^{2P}
  \exp\left(-\frac{\pi 2P}{\Lambda_\mu^2} \sum_{k=1}^{2P} {\Delta \mbR_i}^2
  \right)  \\
  &\equiv& \frac{\Lambda^3}{2^{3/2}} F_\mu,
\end{eqnarray}
where we recognize the probability distribution of a single ring polymer
of $2P$ monomers describing a particle of mass $\mu = m/2$ at the same
temperature (cfr. Eq.~(\ref{eq:F})).
In the case of the second virial coefficient, where this is the only
exchange term present, this contribution is just a
simple average over the larger polymer, and can then be written as~\cite{Garberoglio2011a}
\begin{equation}
  B_\mathrm{xc}(T) = -\frac{2 \pi \Lambda^3 \NA}{2^{3/2}}
  \left\langle
  \exp\left(
  -\frac{\beta}{P} \sum_{i=1}^P U_2(|\mbR_{i+P} - \mbR_i|^2)
  \right)
  \right\rangle_\mu.
\end{equation}

In addition to this, the various terms in the sum over permutation of
Eq.~(\ref{eq:Q_MC}) also acquire factors depending on the number of nuclear
spin states of the particles, that is factors of $1/(2I+1)$ for a nuclear
spin $I$. A detailed derivation of these factors is reported in
Refs.~\onlinecite{Garberoglio2020,Garberoglio2021a}.

\subsection{Uncertainty propagation}
\label{sec:uncertainty}

As is apparent from their definition, the calculation of virial
coefficients depends on the knowledge of few-body properties of atoms,
namely interaction potentials, polarizabilities, and dipole moments.
In a completely {\em ab initio} calculation of virial coefficients, these
quantities -- as seen in Sec.~\ref{sec:abinitio} -- are determined by
electronic-structure calculations and are provided with a full uncertainty
estimation. In this section, we will show how this uncertainty can be propagated to the
uncertainty in virial coefficients, using the third virial coefficient
$C(T)$ as an example.

The first approach consists of calculating values of $C(T)$ using perturbed
pair and three-body potentials, that is:
\begin{equation}
  C_{\pm}^{[u_i]} = C(T; u_i \pm \delta u_i),
\end{equation}
and
\begin{equation}
  \delta C^{[u_i]} = \frac{1}{4} \left|
  C_{+}^{[u_i]} - C_{-}^{[u_i]}
  \right |,
  \label{eq:dC}
\end{equation}
where we have assumed that the uncertainties in the potential -- $\delta u_i$
for $i=2$ or $i=3$ in the case of the pair and three-body potential,
respectively -- are given as expanded $(k=2)$ uncertainties while $\delta
C^{[u_i]}$ is a standard uncertainty.
The overall standard uncertainty in $C(T)$ due to the uncertainty in the potentials
is then obtained as a sum in quadrature
\begin{equation}
  \delta C = \sqrt{ (\delta C^{[u_2]})^2 + (\delta C^{[u_3]})^2 }.
\end{equation}
Although this approach was used in early calculations of the virial
coefficients,~\cite{Garberoglio2009,Garberoglio2011} it is unsatisfactory
for several reasons. First of all, it considers only rigid shifts of the
potentials, while in principle the actual potential can be closer to the
upper bound for some configurations and closer to the lower bound for
others.
Secondly, the uncertainty (\ref{eq:dC}) is obtained as a difference of
quantities which are themselves computed with some statistical uncertainty. This
requires very long runs to make sure that the difference in Eq.~(\ref{eq:dC})
is not influenced by the statistical error in the calculation of
$C_{\pm}^{[u_i]}$.

A more satisfactory approach is obtained by considering that the virial coefficients are functions
of the temperature $T$ as well as functionals of the potentials.~\cite{Garberoglio2021a} A variation
$\delta u_i$ in the potential will then produce a corresponding variation in the value of the virial
coefficient, given by
\begin{equation}
  \delta C^{[u_i]} = \int \left| \delta u_i  \frac{\delta C}{\delta u_i} \right| ~ \D \Omega_3,
  \label{eq:dC_fd}
\end{equation}
where we have used the definition of the functional derivative $\delta C / \delta u_i$. The absolute
value in Eq.~(\ref{eq:dC_fd}) comes from the conservative choice of assuming that all the variations
will contribute with the same (positive) sign to the final uncertainty.  We note in passing that in
the case of the second virial coefficient, $B(T)$, Eqs.~(\ref{eq:dC_fd}) and (\ref{eq:dC}) produce
the same result.  As is apparent from Eq.~(\ref{eq:C}), the evaluation of Eq.~(\ref{eq:dC_fd})
requires the functional derivative of $Z_3$ and $Z_2$ with respect to the pair and three-body
potential. As a first approximation, one can use the classical expression Eq.~(\ref{eq:ZN_class})
(possibly augmented by semiclassical corrections).~\cite{Garberoglio2021a} More accurate results
(especially at low temperatures) are obtained by functional differentiation of the path-integral
expressions (\ref{eq:Z2_int}) and (\ref{eq:ZN_int}) so that one has~\cite{Binosi2023}
\begin{eqnarray}
  \left. \delta Z_2 \right|_{u_2} &=& -4 \pi V \beta
  \int \left\langle
  \overline{\delta u_2} \e^{-\beta \overline{U_2}}
  \right \rangle r_{12}^2 \D r_{12} \\
  \left. \delta Z_3 \right|_{u_2} &=& -8\pi^2 V \beta 
  \int \left\langle
  \sum_{i<j} \overline{\delta u_2}(r_{ij})
  \e^{-\beta \overline{U_3}}
  \right\rangle \D \Omega_3\\
  \left. \delta Z_3 \right|_{u_3} &=& -8\pi^2 V \beta 
  \int \left \langle
  \overline{\delta u_3} \e^{-\beta \overline{U_3}}
  \right\rangle \D \Omega_3,
\end{eqnarray}
where we have defined
\begin{equation}
  \left. \delta Z_i \right|_{u_j} =
  \int \delta u_j \frac{\delta Z_i}{\delta u_j} \D \Omega_i.
\end{equation}
The same approach can be used in the calculation of the propagated uncertainties for dielectric
virial coefficients.~\cite{Garberoglio2021,m3_2023}

In actual practice, these expressions enable rigorous estimation of the uncertainty
propagated from the potentials with a much smaller computational effort than that needed to compute
virial coefficients. Additionally, the {\em a priori} knowledge of a lower bound on the uncertainty
and its temperature dependence facilitates the process of finding the optimal set of parameters for
the path-integral simulations (cutoff distance, number of beads $P$, number of Monte Carlo
integration points) in order to make the statistical uncertainty of the calculation a minor
contributor to the final results.

\subsection{Mayer sampling and the virial equation of state}

Equations~(\ref{eq:B})--(\ref{eq:D}) show that the expressions for the virial coefficients become
more involved when the order is increased. Although these expressions can be systematically derived
using computer-algebra systems, their subsequent implementation in classical or quantum frameworks
becomes more and more time-consuming. Taking also into account the limited availability of {\em ab
  initio} many-body potentials (at the time of this writing, these are limited to three bodies and
have been developed only for a small set of atoms and molecules), it might seem that a fully {\em ab
  initio} calculation of the equation of state using virial expansions could not be feasible.
Nevertheless, it is observed that the largest contributions to the value of the virial coefficients
come from the many-body potentials of lower orders, as already discussed in Sec.~\ref{sec:highp}. As
a consequence, even if only pair and three-body potentials are available, a calculation of
higher-order virial coefficients can provide useful and reasonably accurate representations of the
equation of state.~\cite{Masters08,Schultz22}

A very efficient procedure to perform this task is based on the diagrammatic approach by
Ursell~\cite{Ursell27} and Mayer,~\cite{Mayer37,Mayer77} who showed how the various terms
contributing to the virial coefficients can be related to simpler cluster integrals that can be
catalogued using a diagrammatic form. The contributions from the various diagrams can be added very
efficiently using Monte Carlo sampling methods.~\cite{Singh04}
Although the number of diagrams increases exponentially with the order of the virial coefficient,
it has been shown that calculations can be kept within a manageable size up to virial coefficients
of order 16,~\cite{Wheatley13,Feng15,Wheatley20} resulting in equations of state with very good
accuracy up to the binodal (condensation) density.~\cite{Schultz22}

Mayer sampling methods, originally developed for monatomic systems, have been extended to
molecules~\cite{Shaul11} and therefore can also be used to perform path-integral calculations of
density~\cite{Shaul2012,Schultz19} and acoustic virial coefficients.~\cite{Gokul21} This approach
provides an independent validation of the framework outlined in this review. Virial coefficients
calculated using both approaches are found to be compatible within mutual
uncertainties.~\cite{Binosi2023}

\subsection{Numerical results for virial coefficients}

As seen in Sec.~\ref{sec:quantum_virials}, a fully first-principles calculation of virial
coefficients requires the knowledge of many-body potentials and, in the case of dielectric
properties, polarizabilities, which can be obtained by {\em ab initio} electronic structure
calculations.  Currently, as discussed in Sec.~\ref{sec:abinitio}, the only system for which these
calculations can be made without uncontrolled approximations is helium. Much effort has been devoted
to produce high-quality potentials from first principles. At the time of writing, the most accurate
pair potential is the one developed by Czachorowski {\em et al.},~\cite{Czachorowski2020} which
includes relativistic and QED effects. This potential was developed using exactly the same approach
as the potentiuial of Ref.~\onlinecite{Przybytek2017}, the only difference being that the
relativistic and QED corrections were computed using a larger basis set. As a consequence of
including the adiabatic corrections and recoil terms, slightly different pair potentials are
available for the ${}^4$He-${}^4$He, ${}^3$He-${}^3$He, and ${}^4$He-${}^3$He interaction.

Recently, a new three-body potential for ${}^4$He, including relativistic effects, has been
developed,~\cite{u3_2023} resulting in a significant increase of accuracy with respect to the
non-relativistic one that was previously available (see Sec.~\ref{sec:he_u3}).~\cite{Cencek2009} In
the case of dielectric properties, the single-atom polarizability has been calculated with
outstanding accuracy.~\cite{Puchalski:20} The most accurate pair-induced polarizability currently
available is that of Cencek {\em et al.}~\cite{Cencek2011} and, recently, fully {\em ab initio}
calculations of the three-body polarizability~\cite{a3_2023} and dipole moment~\cite{m3_2023} have
been performed, enabling a subsequent calculation of the third dielectric virial coefficient with
well-defined uncertainties completely from first principles.~\cite{m3_2023}

In the case of neon, the most recent pair potentials and polarizabilities have been computed by
Hellmann and coworkers.~\cite{Hellmann2021,Hellmann2022} Parametrizations of three-body potentials
have appeared in the literature,~\cite{Ne_3body} but no first-principles calculations have been
published so far.

Due to its easy accessibility and large measurement effects, argon has been the subject of many
theoretical studies. However, the large number of electrons prevents calculations of potentials and
polarizabilities with the same accuracy as the lighter noble gases, and some uncontrolled
approximations are still necessary. The most accurate pair potential so far has been developed by
Patkowski and Szalewicz,~\cite{Patkowski:10} while a three-body potential with well characterized
uncertainties was computed and characterized by Cencek and coworkers.~\cite{Cencek2013} Regarding
dielectric properties, the most accurate pair polarizability is the one developed by Vogel {\em et
  al.}.~\cite{Vogel10_b} In the case of neon and argon, no higher-order polarizabilities are
available beyond the second. Calculations have been performed using the {\em superposition
  approximation}~\cite{Alder80,Buck91} for the three-body polarizability. Although the results of
these calculations compare well with the available experimental data, their uncertainty is to a
large extent unknown.~\cite{Garberoglio2021}

We report in Table~\ref{tab:theoretical_virials} the most up-to-date references regarding {\em ab
  initio} calculations of virial coefficients. This table to some extent serves as an update to the
table of recommended data presented by Rourke.~\cite{Rourke2021a}

\begin{table}
  \caption{Bibliographic data for the most recent {\em ab initio} calculations of virial
    coefficients and transport properties. Notice that for neon and argon many of the values
    computed from first principles have a higher (and sometimes less rigorous) uncertainty than the
    best experimental determination.}
  \begin{tabular}{l|l|l|l}
    & Helium & Neon & Argon \\
    \hline
    \hline
    $B$ & Ref. \onlinecite{Czachorowski2020}       & Ref. \onlinecite{Hellmann2021} &
    Refs. \onlinecite{Wiebke12,Vogel10_b} \\
    $C$ & Ref. \onlinecite{u3_2023,Binosi2023} & Ref. \onlinecite{Wiebke12Ne} & Ref. \onlinecite{Cencek2013}\\
    $D$ & Ref. \onlinecite{Garberoglio2021a}       & --- & Ref. \onlinecite{Jaeger11} \\
    \hline
    $\betaa$  & Ref. \onlinecite{Czachorowski2020}        & Ref. \onlinecite{Hellmann2021} &
    Ref. \onlinecite{Wiebke12} \\
    $\gammaa$ & Refs. \onlinecite{u3_2023,Binosi2023} & --- & Ref. \onlinecite{Wiebke12} \\ 
    \hline
    $A_\varepsilon$ & Ref. \onlinecite{Puchalski:20}    & Ref. \onlinecite{Hellmann2022} &
    Ref. \onlinecite{Lesiuk2023} \\ 
    $B_\varepsilon$ & Ref. \onlinecite{Garberoglio2020}  & Ref. \onlinecite{Hellmann2021} & Ref. \onlinecite{Vogel10_b,Garberoglio2020}\\
    $C_\varepsilon$ & Refs. \onlinecite{Garberoglio2021,a3_2023,m3_2023} &
    Ref. \onlinecite{Garberoglio2021} & Ref. \onlinecite{Garberoglio2021} \\    
    \hline
    {$A_\mu$}${}^a$        & Ref. \onlinecite{Bruch02} & Ref. \onlinecite{Lesiuk2020} &
    Ref. \onlinecite{Lesiuk2023} \\
    $B_\mathrm{R}$ & Ref. \onlinecite{Garberoglio2020} & Ref. \onlinecite{Garberoglio2020,Hellmann2021}${}^b$ &
    Ref. \onlinecite{Garberoglio2020} \\
    \hline
    $\eta$    & Ref.~\onlinecite{Cencek2012} & Ref. \onlinecite{Hellmann2021} & Ref. \onlinecite{Vogel10_b}\\
    $\lambda$ & Ref.~\onlinecite{Cencek2012} & Ref. \onlinecite{Hellmann2021} & Ref. \onlinecite{Vogel10_b}\\
    \hline
    \hline
    \multicolumn{4}{l}{${}^a$ \footnotesize{Note that $A_\mathrm{R} = A_\varepsilon + A_\mu$.}}\\
    \multicolumn{4}{l}{${}^b$ \footnotesize{Best values can be
          obtained by applying the frequency 
          dependence}} \\
      \multicolumn{4}{l}{${}^{}$ \footnotesize{of Ref.~\onlinecite{Garberoglio2020} to $B_\varepsilon$ calculated from
      Ref.~\onlinecite{Hellmann2021}.}}
  \end{tabular}
  \label{tab:theoretical_virials}
\end{table}

\subsubsection{Density virial coefficients}

\begin{figure}[h]
  \center\includegraphics[width=0.8\linewidth]{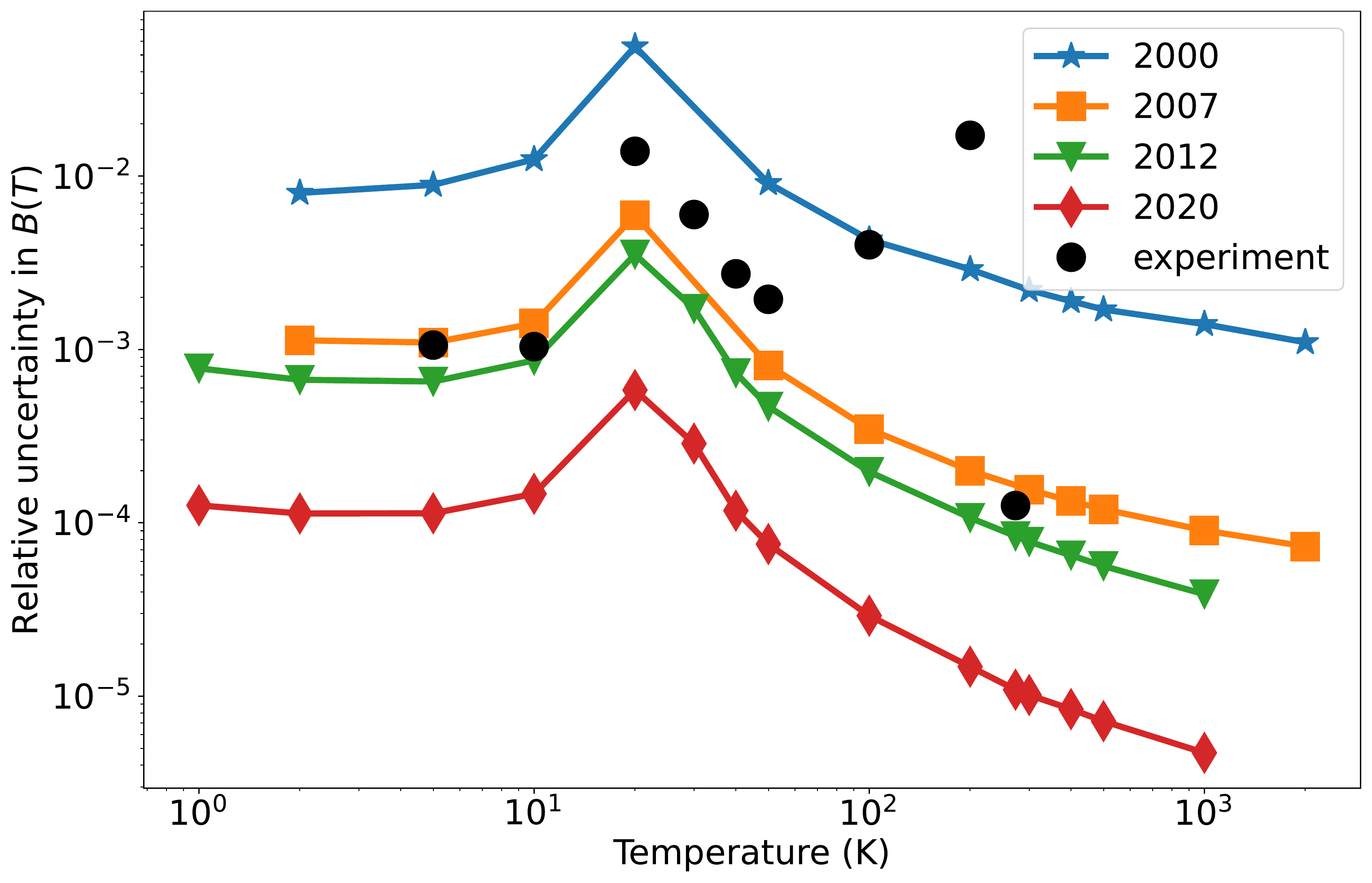}
  \caption{Values of the relative expanded ($k=2$) uncertainty in the calculated values of $B(T)$
    for ${}^4$He for 
    different pair potentials. Stars: the first calculation of thermophysical properties using {\em
      ab initio} potentials with well characterized uncertainties.~\cite{Hurly2000} Plus: the
    $\phi_{07}$ potential.~\cite{Hurly2007} Triangles: the first potential including a complete
    description of relativistic effects.~\cite{Przybytek2010,Cencek2012}  Diamonds: the latest pair
    potential.~\cite{Czachorowski2020}  The points are experimental data, reported in
    Ref.~\onlinecite{Czachorowski2020}. The peaks are due to the fact that $B(T)$
    crosses zero near $23$~K and hence relative uncertainties become large.} 
\label{fig:uB_He}  
\end{figure}

The most accurate {\em ab initio} values of the second virial coefficients of helium for both
isotopes are those computed by Czachorowski {\em et al.}.~\cite{Czachorowski2020}
In order to visualize the recent progress in this field, we report in Fig.~\ref{fig:uB_He} the
evolution of the theoretical uncertainty of $B(T)$ in the past 20 years. Theoretical and
computational improvements enabled a reduction of two orders of magnitude in the relative
uncertainty, which is presently on the order of $10^{-4}$ at low temperatures
($< 10$~K) and decreases to less than $10^{-5}$ at higher temperatures. In general, the current
theoretical uncertainties of $B(T)$ are more than one order of magnitude smaller than the best
experimental determinations.

\begin{figure}[h]
  \center\includegraphics[width=0.8\linewidth]{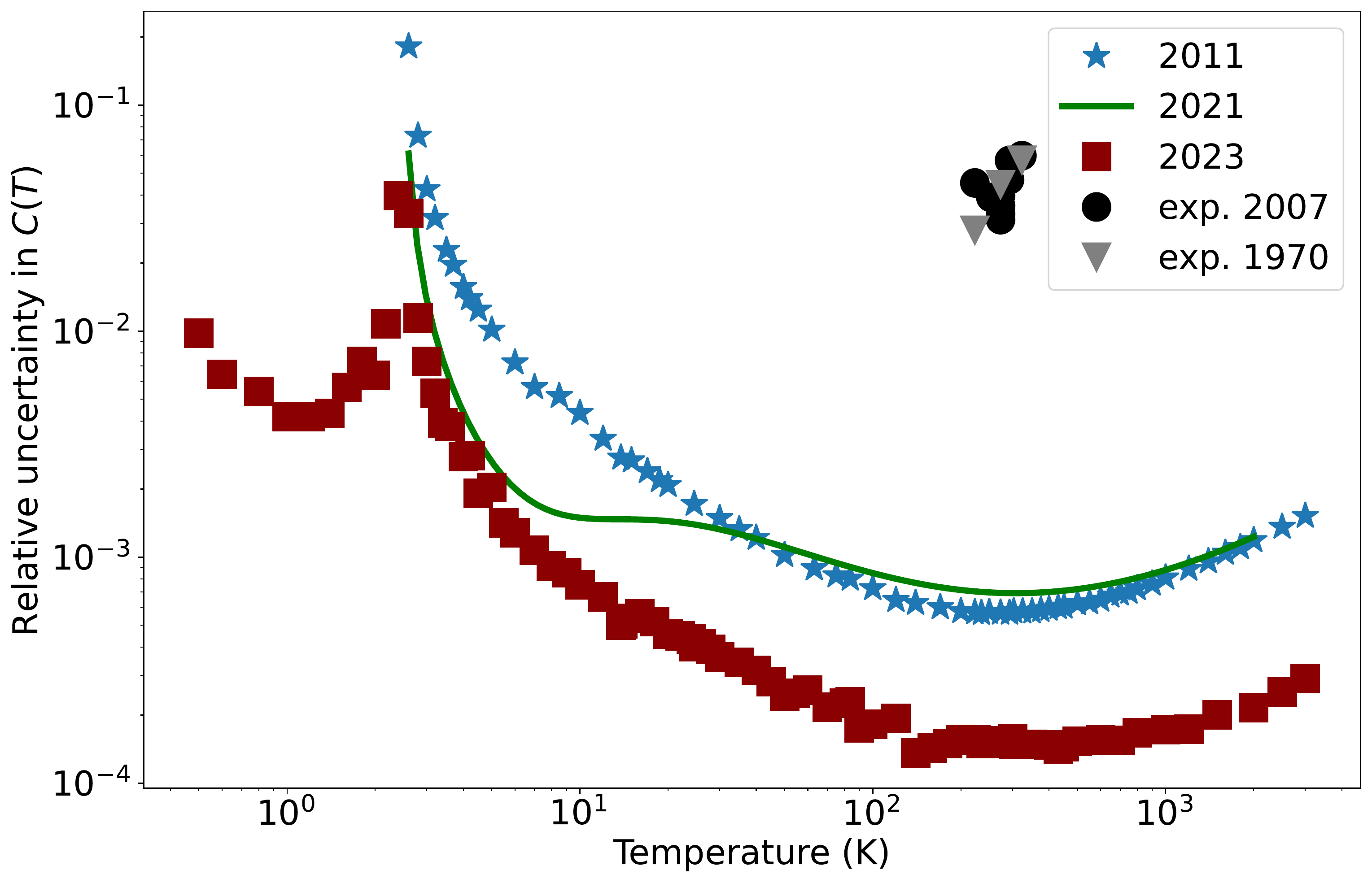}
  \caption{Values of the relative expanded ($k=2$) uncertainty in the calculated values of $C(T)$
    for ${}^4$He for different pair and three-body potentials. Stars: calculation from
    2011,~\cite{Garberoglio2011} using the pair potential of Ref.~\onlinecite{Przybytek2010} and the
    three-body potential of Ref.~\onlinecite{Cencek2009}. Solid line: calculation from 2021, using
    the latest pair potential~\cite{Czachorowski2020} and the three-body potential from
    Ref.~\onlinecite{Cencek2009}. Squares: latest calculation, using the pair potential from
    Ref.~\onlinecite{Czachorowski2020} and the latest three-body potential.~\cite{u3_2023}
    Experimental data are from McLinden and L{\"o}sch-Will~\cite{McLinden07} (circles) and from
    Blancett {\em et al.}~\cite{Blancett70} (triangles).
    The peaks around $T=3$~K are due to the fact that $C(T)$
    crosses zero and hence relative uncertainties become large.}
\label{fig:uC_He}  
\end{figure}

Figure~\ref{fig:uC_He} shows the development of the uncertainty in the calculations of $C(T)$ for
helium in the past 12 years, starting from the first calculation with fully characterized
uncertainties from 2011,~\cite{Garberoglio2011} whose results were independently confirmed a year
later using the Mayer sampling approach.~\cite{Shaul2012} One can clearly see that the subsequent
improvement of the pair potential resulted in a reduction of the uncertainty at the lowest
temperatures ($T \lesssim 50$~K), while the uncertainty at the highest temperatures is dominated by
the propagated uncertainty from the three-body potential. Recent improvements resulted in a further
reduction of the uncertainty by a factor of $\sim 5$ across the whole temperature range
$10$~K$-3000$~K. The current theoretical uncertainty in $C(T)$ is a few parts in $10^4$ at high
temperature, and increases to a few parts per $10^3$ below $50$~K. At temperatures below $\sim
10$~K, the theoretical uncertainty budget is dominated by the propagated uncertainty from the pair
potential.

Although no well-characterized four-body potential has yet been developed for helium, several
groups have performed calculations of the fourth virial coefficient, $D(T)$. Although initially the
effect of the four-body potential was neglected,~\cite{Shaul2012} more recent work tried to estimate
its contribution using known asymptotic values.~\cite{Garberoglio2021a} These results are in good
agreement with the limited experimental information.

In the case of neon, the most recent calculations for $B(T)$ with a pair potential having
well-characterized uncertainties~\cite{Hellmann2021} resulted in a relative
uncertainty at $T = 273.16$~K of $\ur(B) = 2 \times 10^{-3}$. As expected, this is larger than the
corresponding uncertainty for helium, due to the fact that electronic structure calculations for
the heavier atoms are much more computationally demanding.
Unfortunately, the three-body potential for neon is only approximately known at the moment. To the
best of our knowledge, no first-principles calculation is available in the literature, and
only a semi-empirical parametrization is currently known.~\cite{Ne_3body}
As a consequence, no {\em ab initio} calculation of higher-order coefficients has been performed to
date and only approximate values are known.~\cite{Wiebke12Ne}

The pair potential of argon is well characterized and has been calculated independently by two
groups,~\cite{Jaeger09,Patkowski:10} and hence thermophysical properties at the pair level are well
characterized.~\cite{Vogel10_b,Wiebke12} The relative uncertainty of $B(T)$ at $T=273.16$~K is $\ur \sim 0.6\%$.
The three-body potential for argon has also been computed independently by two
groups~\cite{Jaeger11,Cencek2013} and its uncertainty has been rigorously assessed. Therefore,
the third virial coefficient of argon is also known with rigorously propagated uncertainties. The
relative uncertainty is on the order of $\ur \sim 1\%$ at $T=273.16$~K and increases up to $\ur
\sim 6\%$ at $T=80$~K.
Analogously to the other noble gases, the four-body (and higher) non-additive contribution to the
potential energy of argon is not known from first principles. Nevertheless, higher-order virial
coefficients for argon, up to the seventh, have been computed based on pair and three-body
potentials.~\cite{Jaeger11} 

\subsubsection{Acoustic virial coefficients}

The situation regarding first-principles calculations of  acoustic virial coefficients closely
follows that of the density ones.
In the usual approach using phase shifts, the calculation of $B(T)$ also provides the temperature
derivatives needed to compute $\betaa(T)$, and therefore very accurate values for the second
acoustic virials for helium,~\cite{Czachorowski2020} neon,~\cite{Hellmann2021} and argon~\cite{Vogel10_b} can
be found in the papers where the pair potential and $B(T)$ calculations are reported.

In the case of the third acoustic virial coefficient, the situation is similar. The most accurate
values of $\gammaa$ for helium isotopes are reported in Refs.~\onlinecite{u3_2023} and
\onlinecite{Binosi2023}, which are in very good agreement with the values obtained independently
using the Mayer sampling approach.~\cite{Gokul21}
The current relative uncertainty in $\gammaa$ for helium from {\em ab initio} calculations is $\ur \sim
0.02 - 0.2\%$ across the temperature range from 10~K to 1000~K.~\cite{Binosi2023}

As already mentioned, the lack of an accurate three-body potential for neon has prevented a fully
first-principles calculation of the third virial coefficient, and hence no {\em ab initio} values of
$\gammaa$ are currently available for this substance.

Regarding argon, {\em ab initio} acoustic virial coefficients up to the fourth, together with a
thorough analysis of their associated uncertainties, have been reported by Wiebke {\em et
  al.}~\cite{Wiebke12} The uncertainty $\gammaa$ at $T=273.16$ is $\sim 1.4\%$.

\subsubsection{Dielectric and refractivity virial coefficients}
\label{sec:dielectric_theo}

The first dielectric virial coefficient $A_\varepsilon$ for helium has been computed with an
accuracy exceeding the best experimental determination in
Ref.~\onlinecite{Puchalski:20}.
In the case of neon and argon, the most accurate theoretical results are less accurate than the best
experimental determination.~\cite{Gaiser2018} The most accurate computed value for neon can be found in
Ref.~\onlinecite{Hellmann2022}, and a calculation for argon, including the frequency dependence
needed for refractivity estimates, has recently appeared.~\cite{Lesiuk2023}

Magnetic susceptibilities computed from first principles and the corresponding coefficients $A_\mu$ that
are used in RIGT are also available for helium,~\cite{Bruch02} neon,~\cite{Lesiuk2020}
and argon.~\cite{Lesiuk2023}
As noted by Rourke,~\cite{Rourke2021a} there are some discrepancies between the {\em ab initio}
calculations of this quantity and the experimental values often cited from Barter {\em et
  al.}~\cite{Barter60} This may be due to errors in the older argon data used in Barter's
calibration; a modern determination of $A_\mu$ for argon (or its ratio to that of helium) would be
highly desirable.

First-principles calculations of $B_\varepsilon(T)$ for helium have been available for a long
time.~\cite{Moszynski1995} Reference values from the latest pair potential and polarizability can be
found in Ref.~\onlinecite{Garberoglio2020}. These results have been independently confirmed (except
at the lowest temperatures) by semiclassical calculations.~\cite{Song2020}
Due to the recent development in three-body polarizabilities~\cite{a3_2023} and dipole-moment
surfaces,~\cite{m3_2023} {\em ab initio} values of $C_\varepsilon(T)$ with well-defined
uncertainties are also available for both helium isotopes.~\cite{m3_2023}
These values agree with the limited experimental data available, but have much smaller
uncertainties.

In the case of neon, the most accurate {\em ab initio} $B_\varepsilon(T)$ has been computed
by Hellmann and coworkers,~\cite{Hellmann2021} who also reported well-characterized
uncertainties. The results are in very good agreement with DCGT measurements.  The third dielectric
virial coefficient of neon is only approximately known from {\em ab initio} calculations, since the
contributions from the three-body polarizability and dipole-moment surfaces can only be estimated
with several uncontrolled approximations.~\cite{Garberoglio2021}

Regarding argon, the second dielectric virial coefficient has been computed using a fully {\em ab
  initio} procedure in Refs.~\onlinecite{Vogel10_b}, \onlinecite{Garberoglio2020}, and
\onlinecite{Song2020}.  Analogously to neon, the lack of {\em ab initio} three-body surfaces for
polarizability and dipole moment has prevented a fully first-principles calculation of
$C_\varepsilon(T)$ for argon. Approximate values were reported in
Ref.~\onlinecite{Garberoglio2021}.

Calculations of the second refractivity virial coefficient, $B_\mathrm{R}$, for helium, neon, and
argon were performed by Garberoglio and Harvey~\cite{Garberoglio2020} using the best
pair potentials and Cauchy moments available at the time, although in many cases a rigorous uncertainty
propagation was not possible. In the case of neon, the subsequent improved $B_\varepsilon$ from
Hellmann {\em et al.}~\cite{Hellmann2021} can be combined with the frequency-dependent correction
from Garberoglio and Harvey~\cite{Garberoglio2020} to provide improved values of $B_\mathrm{R}$. 

\subsection{Transport properties}
\label{sec:transport}

When the thermodynamic equilibrium of a gas is perturbed, dynamic processes will tend to restore
it. The actual response depends on the specific kind of induced non-homogeneity: density variations
will give rise to diffusive processes, relative motions will be damped by internal friction, and
temperature gradients will result in heat flowing through the system.

The kinetic theory of gases~\cite{Chapman90} provides a theoretical framework to analyze
non-equilibrium behavior and transport properties of gases, determining how the flux of matter,
momentum, or heat depends on the spatial variation of density, velocity, or temperature.
The most accurate description is based on the Boltzmann equation,
which describes the evolution of the state of a fluid where simultaneous interactions of three or
more particles are neglected; hence, it is valid in the low-density regime only. Despite this
limited scope, additional approximations are needed to make the kinetic equations manageable, for
example by limiting the strength of the inhomogeneities to the linear or quadratic regime, which are
situations that find widespread application.

In the following, we will briefly review the theory and the main computational results regarding
heat and momentum transport in monatomic fluids, and how the relevant quantities -- viscosity and
thermal conductivity -- can be calculated from first principles.  In the low-density and linear
regime, the shear viscosity ($\eta$) and thermal conductivity ($\lambda$) describe the linear
relation between momentum and temperature inhomogeneities, and the resulting internal friction and
heat
\begin{eqnarray}
  \pi_{ij} &=& p \delta_{ij} - \eta \left(\frac{\partial {\mathfrak u}_i}{\partial x_j} + \frac{\partial
    {\mathfrak u}_j}{\partial x_i} \right) \\
    q_i &=& -\lambda \frac{\partial T}{\partial x_i},
\end{eqnarray}
where $\pi_{ij}$ is the pressure tensor, $p$ the isotropic pressure, $\boldsymbol{\mathfrak u}$ the macroscopic
velocity, $\mathbf{q}$ the heat flux, and $T$ the temperature.
Kinetic theory shows how to compute $\eta$ and $\lambda$ from the details of the microscopic
interaction between atoms. To this end, it is useful to define
\begin{eqnarray}
  Q^{(l)}(E) &=& 2 \pi \int \left(1 - \cos^l\theta\right) \sigma(E,\theta) \sin\theta \D \theta
  \label{eq:Qk} \\
  \Omega^{(l,s)}(T) &=&  2  \int  \frac{\e^{-E/(\kB T)}}{(s+1)!} \left( \frac{E}{\kB T}
  \right)^{s+1} Q^{(l)}~ \frac{\D E}{\kB T},
  \label{eq:Omega}
\end{eqnarray}
where $\sigma(E,\theta)$ is the differential cross section for two particles with energy $E$ in the
scattering reference frame ($E = \mu v^2 / 2$, where $\mu = m/2$ is the reduced mass and $v$ the
modulus of the relative velocity). The quantities defined by Eq.~(\ref{eq:Omega}) are known as
collision integrals.  Equation~(\ref{eq:Qk}) is valid when the cross section is calculated either in
the classical or quantum regime; in the latter case one has to further consider the fermionic or
bosonic nature of the interacting atoms.~\cite{Hirschfelder1954}
The viscosity and thermal
conductivity are given by
\begin{eqnarray}
  \eta(T)    &=&  \frac{5}{16} \frac{\sqrt{2 \pi \mu \kB T}}{\Omega^{(2,2)}} f^{(k)}_\eta
  \label{eq:eta} \\
  \lambda(T) &=&  \frac{75}{64} \sqrt{\frac{\kB T}{2 \pi \mu}} \frac{1}{\Omega^{(2,2)}} f^{(k)}_\lambda,
  \label{eq:lambda}
\end{eqnarray}
where $f^{(k)}_\eta$ and $f^{(k)}_\lambda$ are factors of order 1 that depend on the specific order
$k$ of the approximations involved, which in turn involve collision integrals of higher
order.
In the quantum case, collision integrals cannot be computed using path-integral Monte Carlo methods,
but their value depends on the scattering phase shift (see Sec.~\ref{sec:quantum_virials}). For
example, on one has~\cite{Meeks94}
\begin{equation}
  Q^{(2)}(E) = \frac{4 \pi \hbar^2}{\mu E} \sum_{l=0}^\infty
  \frac{(l+1)(l+2)}{2l+3} \sin^2\left(\delta_l(E) - \delta_{l+2}(E) \right),
\end{equation}
and explicit expressions for $f^{(k)}_\eta$ and $f^{(k)}_\lambda$ can be found in
Refs.~\onlinecite{Hirschfelder1954} and \onlinecite{Viehland95} for $k=3$ and $k=5$, respectively.

\begin{figure}[h]
  \center\includegraphics[width=0.8\linewidth]{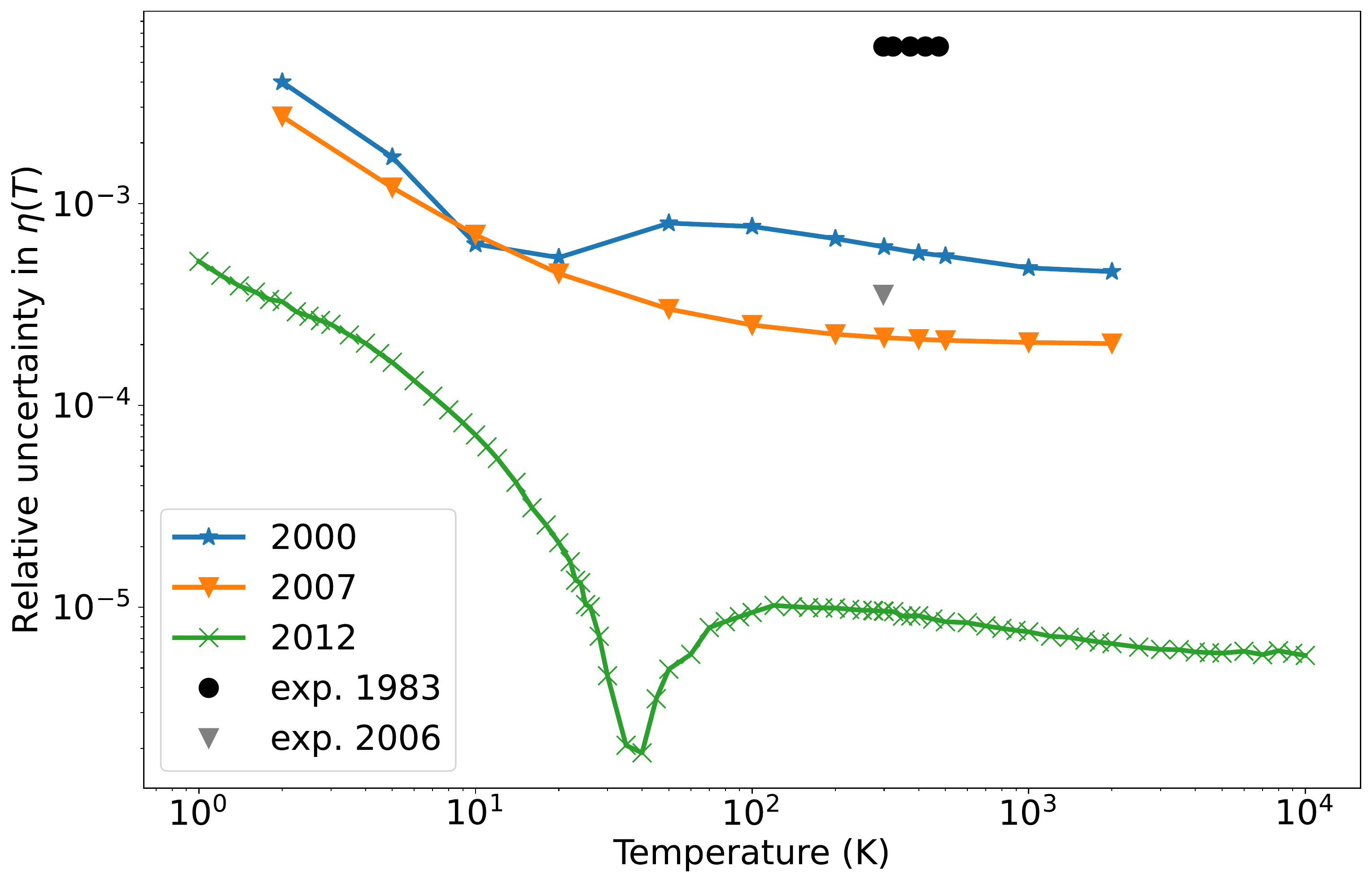}
  \caption{The evolution of the relative uncertainty in the {\em ab initio} calculation of viscosity
    for ${}^4$He. Stars: calculations from Ref.~\onlinecite{Hurly2000}. Triangles: calculations from
    Ref.~\onlinecite{Hurly2007}. Crosses: calculations from Ref.~\onlinecite{Cencek2012}.
    The minimum of the calculated uncertainty near $40$~K is unphysical; see
    Sec.~\ref{sec:future_uncertainty}.
    Black dots: experimental data from Ref.~\onlinecite{Kestin1983}. The gray triangle is the
    experimental value reported in Refs.~\onlinecite{Berg05,Berg05_err}.
    } 
\label{fig:viscosity}  
\end{figure}

As pointed out in Sec.~\ref{sec:flow}, the accuracy of {\em ab initio} calculations of transport
properties for helium vastly exceeds that of experiments.  We report in Fig.~\ref{fig:viscosity} the
evolution of the relative uncertainty in the theoretical calculation of $\eta_\mathrm{He}$ in the
past 20 years. The most recent theoretical values, which have an accuracy that is more than enough
for several metrological applications, can be found in Ref.~\onlinecite{Cencek2012}. It is worth
noting that a more accurate pair potential has been published in the
meantime,~\cite{Czachorowski2020} although no corresponding calculation of transport properties has
yet been published.

In the case of neon, the best theoretical estimates of transport coefficients are given in
Ref.~\onlinecite{Hellmann2021}, while for argon they can be found in Ref.~\onlinecite{Vogel10_b}.  In both
cases, the best experimental results are obtained from ratio measurements using the {\em ab initio}
value of the viscosity or thermal conductivity of helium.

\section{Molecular Systems}
\label{sec:molecules}

While the focus of this review is on noble gases, which are the fluids of choice for most {\em ab
  initio}-based primary temperature and pressure metrology, first-principles thermophysical
properties for molecular species can also be of interest and make significant contributions. Three
of the most promising areas are humidity metrology, low-pressure metrology, and atmospheric physics.

There are two main factors that make rigorous {\em ab initio} calculations
of properties much more difficult for molecules than for monatomic
species. The first is the increased dimensionality, where interactions
depend not only on distance but on the relative orientations of the
molecules. This not only complicates the development of potential-energy
surfaces between molecules, but also makes the calculation of properties
such as virial coefficients a sampling problem in many dimensions. Second,
for rigorous calculations the internal degrees of freedom of the molecule
must be considered, because properties of interest (such as the mean
polarizability) depend on the molecular geometry and a distribution of
geometries is sampled for each quantum state of the molecule. In some cases
it may be adequate to assume a rigid molecule, but at a minimum an estimate
of the uncertainty introduced by this assumption is needed, even though it might be difficult to
compute.

In this section, we will describe the calculation of single-molecule
quantities and quantities involving two or more molecules, along with the
use of these quantities to calculate properties of interest for
metrology. Particular attention will be given to methods for addressing
the challenges specific to molecular species. Finally, we will discuss some
metrological applications that use properties of molecular species.

\subsection{Single-molecule calculations}

\subsubsection{Intramolecular potentials}
In order to compute values of a property of a molecule averaged over nuclear motions, it is
necessary to have a potential-energy surface for the molecule. Such surfaces can be developed with
{\em ab initio} calculations, and they can often be refined if accurate spectroscopic measurements
are available. Development of the intramolecular potential is relatively straightforward for
diatomic molecules such as H${}_2$, N${}_2$, and CO because the potential is one-dimensional, but
the dimensionality and complexity increases quickly with the number of atoms. Surfaces of
sufficiently high quality for most purposes have been developed for the triatomic molecules
H${}_2$O~\cite{PES15K} and CO${}_2$.~\cite{Huang17} These intramolecular potential-energy surfaces
are also needed in order to sample configurations when considering molecular flexibility for pair
calculations as described in Sec.~\ref{sec:mol_virials}.
Except for few-electron diatomics and two-electron triatomics, pure {\em ab initio} surfaces are not
accurate enough to provide rovibrational spectra competitive with experiments, and the most accurate
molecular surfaces are always semiempirical.

\subsubsection{Electromagnetic properties}
\label{sec:mol_pol}

In contrast to noble gases, molecular species have multipole moments in the BO approximation
(dipole, quadrupole, {\em etc}.). The most significant for metrology is the
electric dipole moment.  Rigorous {\em ab initio} calculation of the dipole
moment for a molecule such as H${}_2$O requires the development of a
dipole-moment surface in which the dipole moment vector is given as a
function of atomic coordinates, along with the single-molecule PES. The
dipole moment for a given rovibrational state can then be computed as the
expectation value averaged over the wave function of that state. Because the
population of states changes with temperature, the average dipole moment
will also change (slightly) with temperature; this has been analyzed for
H${}_2$O and its isotopologues by Garberoglio {\em et al.}~\cite{Garberoglio23}

The polarizability is another important quantity, both in the static limit
for capacitance-based metrology and at higher frequencies for metrology
based on optical refractivity. Unlike a noble gas whose polarizability at a
given frequency is a single number, the polarizability of a molecule is a
tensor that reflects the variation with directionality of the applied field
and of the molecular axes. However, the quantity of interest for metrology
is the mean polarizability, defined as $1/3$ of the trace of the
polarizability tensor.

Polarizability reflects the response of the electrons to an electric field. It can be computed {\em
  ab initio} in a relatively straightforward way.  While for monatomic species (and homonuclear
diatomic species) the electronic polarizability is the only contribution, more complicated molecules
have an additional contribution in the static limit and at low frequencies; this is usually called
the vibrational polarizability. It can be thought of as the electric field distorting the molecule
(and therefore its charge distribution) by pushing the negatively and positively charged parts of
the molecule in opposite directions.

The molecular dipole moment and polarizability are defined as the first- and second-order response
to an externally applied electric field $E_0$, respectively. They can be computed by numerical
differentiation of the molecular energy computed in the BO approximation as a function of $E_0$, or
by perturbation theory. Although in principle these two approaches should give the same result, in
practice some differences are observed. For atomic systems, the results from perturbation theory are
found to be more accurate than numerical differentiation and are generally
preferred.~\cite{Cencek2011} In the case of water, numerical differentiation is considered more
accurate for dipole-moment calculations.~\cite{DMS2018}

Once intramolecular potential-energy surfaces, polarizability surfaces, and dipole-moment surfaces
are available, one can calculate the temperature-dependent electromagnetic response of a molecule,
that is the first dielectric virial coefficient $A_\varepsilon$ (see Eq.~(\ref{eq:cm})), which is
generally given by two contributions:~\cite{Garberoglio23} the first is proportional to the
rovibrational and thermal average of the electronic polarizability surface, while the second depends
on the squared modulus of the transition matrix element of the dipole-moment surface.  Additionally,
one can separate the contribution from the dipole-moment transition matrix elements into those
transitions where the vibrational state of the molecule changes and those for which the vibrational
state of the molecule does not change, but the rotational state does: these two components of the
dipole-moment contribution to the molecular polarizability are known as vibrational and rotational
polarizabilities, respectively.~\cite{Bishop90}

For small molecules (two or three atoms), one can solve directly the many-body Schr{\"o}dinger
equation for nuclear motion~\cite{DVR3D} ({\em e.g.}, using the efficient discrete-variable
representation~\cite{Baye15} of the few-body Hamiltonian~\cite{Sutcliffe87}) and then perform the
appropriate rovibrational and thermal averages to obtain $A_\varepsilon$.  It has recently been
shown that the path-integral approach outlined in Sec.~\ref{sec:thermo} can be successfully used to
compute the first dielectric virial coefficient of water.~\cite{Garberoglio23} It can possibly be
generalized to larger molecules, where the direct solution of the many-body Schr{\"o}dinger equation
becomes very demanding in terms of computational power.

In the case of water, computational results using the most accurate intramolecular potential-energy
surface,~\cite{PES15K} polarizability surface,~\cite{Lao2018} and dipole-moment surface for
water~\cite{DMS2018} are in good agreement with the experimental values of the dipole-moment
contribution to $A_\varepsilon$ (which has an uncertainty of $\sim 5\%$), and within 0.1\% of the
experimental values for the static dipole moment, although the theoretical surfaces for water do not
yet have rigorously assigned uncertainties.

\subsubsection{Spectroscopy}
\label{sec:mol_spectroscopy}

It is now possible, especially for molecules containing only two or three atoms, to compute the
positions and intensities of spectroscopic lines {\em ab initio}.  The calculation of line positions
requires only the single-molecule potential-energy surface. The more important quantity for
thermodynamic metrology, however, is the intensity of specific lines. This requires both the PES and
a surface for the dipole moment as a function of the coordinates.  Accurate {\em ab initio}
dipole-moment surfaces have been developed for H${}_2$O,~\cite{DMS2018}
CO${}_2$,~\cite{Fleisher21,Huang22} and CO.~\cite{Bielska22} The possible use in pressure metrology
of intensities calculated from the surfaces for CO and CO${}_2$ will be discussed in
Sec.~\ref{sec:mol_pressure}

\subsection{Calculations for molecular clusters}

\subsubsection{Interaction potentials}
\label{sec:mol_pot}

The development of interaction potentials for molecular gases is more difficult than for atomic ones
due to the additional degrees of freedom, but much of the description in Sec.~\ref{sec:abinitio} is
still applicable. A common approximation when developing intermolecular pair potentials is to treat
the molecules as rigid rotors, which reduces the dimensionality considerably. For example, the PES
of a pair of flexible water molecules has 12 degrees of freedom. By freezing the four OH bond
lengths and the two HOH bond angles, only six degrees of freedom, usually taken to be the
center-of-mass separation and five angles describing the mutual orientation, remain. To minimize the
consequences of freezing the intramolecular degrees of freedom, the zero-point vibrationally
averaged structures of the monomers are often used instead of the corresponding equilibrium
structures.~\cite{Mas96,Jeziorska:00}

However, even a six-dimensional dimer PES requires investigating thousands or even tens of thousands
of pair configurations with high-level {\em ab initio} methods. As discussed in
Sec.~\ref{sec:abinitio}, the most commonly applied level of theory is CCSD(T)~\cite{Raghavachari89}
for molecular monomers; this method is usually applied with the frozen-core (FC) approximation.
Such a level of theory was only the starting point in the schemes used to develop the most accurate
pair potentials for the noble gases beyond helium.
For the CCSD(T) method, the computational cost scales with the seventh power
of the size of the molecules, and the scaling becomes even steeper for post-CCSD(T)
methods.

In recent years, several intermolecular PESs have been developed that go beyond the CCSD(T)/FC level
of electronic structure theory.  The first step is to include all electrons in the calculations.
Examples of all-electron (AE) surfaces are the flexible-monomer water dimer PES of
Ref.~\onlinecite{Metz:20b} and the rigid-monomer ammonia dimer PES of Ref.~\onlinecite{Jing:22}.
Also, post-CCSD(T)/AE terms were used in the H$_2$-CO flexible-monomer PESs starting in
2012.~\cite{Jankowski:12,Jankowski:13}  The T(Q) contributions were shown to have surprisingly large
effects on the H$_2$-CO spectra.~\cite{Jankowski:21}

Intermolecular pair potentials can be accurately represented analytically by a number of different
base functional forms. Mimicking the anisotropy of the PES is most commonly achieved either by
using spherical harmonics expansions or by placing interaction sites at different positions in the
molecules, with each site in one molecule interacting with each site in the other molecule through
an isotropic function. The site-site potential form is also often used for the empirical effective
pair potentials commonly employed in molecular dynamics and Monte Carlo simulations of large
molecular systems. The analytical functions used to represent high-dimensional {\em ab initio} PESs
for pairs of small rigid molecules typically have a few tens up to a few hundred fit parameters.

Determination of these parameters, {\em i.e.}, fitting a PES to a set of grid points in a dimer
configurational space and the corresponding interaction energies, was until recently a major task
taking often several months of human effort.  This bottleneck has recently been removed by
computer codes that perform such fitting automatically.  In particular, the autoPES
program~\cite{Metz:16,Metz:20b} can develop both rigid- and flexible-monomer fits at arbitrary level
of electronic structure theory.  The automation is complete: a user just inputs specifications of
monomers and the program provides on output an analytic PES.  This means that the program determines
the set of grid points, runs electronic structure calculations for each point, and performs the fit.
In addition to developing automation, the autoPES project introduced several improvements in the
strategy of generating PESs.  In particular, the large-$R$ region of a PES is computed {\em ab
  initio} from the asymptotic expansion.  Such expansion predicts interaction energies well down to
$R$ about two times larger than the van der Waals minimum distance.  This means that no electronic
structure calculations are needed in this region and autoPES can develop accurate PESs for dimers of
few-atomic monomers using only about 1000 grid points, while most published work used dozens of
thousands of points.

Accurate analytical rigid-rotor PESs exist for a large number of both like-species and
unlike-species molecule pairs. For metrology, the most noteworthy of these are the
N${}_2$--N${}_2$,~\cite{Hellmann2013} 
CO${}_2$--CO${}_2$,~\cite{Hellmann_2014,Yue:22}
H${}_2$O--CO${}_2$,~\cite{Hellmann2019}
H${}_2$O--N${}_2$,~\cite{Hellmann2019a} and
H${}_2$O--O${}_2$.~\cite{Hellmann2020}
Other accurate PESs of this type are: N${}_2$--HF,~\cite{Jankowski:01}
H$_2$O--H$_2$O,~\cite{Jankowski15,Metz:20b} 
(HF)$_2$,~\cite{Ovsyannikov22}
(HCl)$_2$,~\cite{Jiang:05} CH$_4$--H$_2$O,~\cite{Akin-Ojo:05,Akin-Ojo:19} and 
H$_2$--CO.~\cite{Jankowski:12,Jankowski:21}

Many of these PESs ({\em e.g.}, those from Refs.~\onlinecite{Hellmann2013,Hellmann_2014,Hellmann2019,Hellmann2019a,Hellmann2020}) are based on nonrelativistic interaction energies corresponding to the frozen-core
CCSD(T) level of theory in the CBS limit and are represented analytically by site-site potential
functions, with each individual site-site interaction being modeled by a modified Tang--Toennies
type potential~\cite{Tang84} with an added Coulomb interaction term. In the case of the
N${}_2$--N${}_2$ PES,~\cite{Hellmann2013} corrections to the interaction energies for post-CCSD(T), relativistic, and
core-core and core-valence correlation effects were considered.  Motivated by the availability of
extremely accurate experimental data for the second virial coefficients of N${}_2$ and CO${}_2$, the
N${}_2$--N${}_2$~\cite{Hellmann2013} and CO${}_2$--CO${}_2$~\cite{Hellmann_2014} PESs were additionally fine-tuned such that these data are
almost perfectly matched by the values resulting from the PESs. The maximum well depths of the PESs
were changed by the fine-tuning by less than 1\%. Such fine-tuning does, however, mean that
properties such as virial coefficients calculated from these tuned potentials cannot be considered
to be truly from first principles for the purpose of metrology.

The second group of PESs listed above was also developed using either CCSD(T), with FC or AE, or
SAPT.  Post-CCSD(T) terms were considered in some cases, as already mentioned above.  A range of
different functional forms was used in the fitting, for larger monomers it was most often the
site-site form.

While the error introduced by approximating molecules as rigid rotors is
believed to be small for the molecules considered here, more rigorous
calculations should include the intramolecular degrees of freedom; this has
been done for example for the H${}_2$--H${}_2$, H$_2$--CO, and H${}_2$O--H${}_2$O
potentials.~\cite{Jankowski:12,Jankowski:13,Garberoglio2014,Garberoglio2018,Metz:20b}
There are several difficulties involved in the generation of fully flexible potentials. The first
one is the larger number of degrees of freedom. A system of $N$ molecules approximated as rigid
rotors can be described by $C_\mathrm{r} = 6N-6$ coordinates, while $C_\mathrm{f} = 3nN-6$
coordinates are necessary to fully describe a configuration of the same molecules if each of the
monomers has $n$ atoms. For sampling $c$ configurations per degree of freedom, the number of
calculations needed to explore the potential-energy surface grows exponentially as
$c^{C_\mathrm{r|f}}$. In the case of, say, the water trimer ($N=3$, $n=3$), even assuming $c=3$ one
goes from $3^{12} \approx 5 \times 10^5$ configurations for rigid models to $3^{21} \approx 10^{10}$
configurations for a fully flexible approach. The exponential increase of the number of
configurations as a function of the number of degrees of freedom to be consdiered is sometimes
called the dimensionality curse.
Not all of these configurations are equally important
and there is room for significant pruning and clever sampling strategies: one of the most useful
starts from potentials developed for rigid molecules and enables the development of fully flexible
versions optimizing the number of additional molecular configurations to be
evaluated.~\cite{Murdachaew01,Murdachaew02}
More generally, even for a few degrees of freedom, the product of
dimensions strategy leading to the $c^C$ is the worst strategy to
follow.  Instead, one uses various types of guided MC generation of grid
points.  In particular, the statistically guided grid generation method
of Ref.~\onlinecite{Metz:20a} reduces the number of points needed for a
6-dimensional PES to about 300 (assuming the use of {\em ab initio}
asymptotics).
Another important issue regards the choice of a suitable
form for the analytic potential and the fitting procedure. As in the case of rigid potentials,
site-site interaction models (based on exponential functions at short range, inverse powers at long
range, and Coulomb potentials) are commonly used also for intermolecular flexible potentials. For
the intramolecular interactions, Morse functions are often used but polynomial expansions work
sufficiently well for molecules in their low-energy rovibrational state.~\cite{Metz:20b}
Nevertheless, the dimensionality curse drastically limits the development of fully
flexible potentials and for the time being only pair and three-body potentials involving diatomic
and triatomic molecules (notably water~\cite{Wang11,Babin14,Jankowski15}) have been developed.

\subsubsection{Density virial coefficients}
\label{sec:mol_virials}

The calculation of density virial coefficients for molecular systems can be performed in a way very
similar to that for noble gases.  The main difference concerns the evaluation of the matrix elements of
the free-molecule kinetic energy operator, that is the generalization of Eq.~(\ref{eq:KE_PIMC})
which in turn depends on the specific degrees of freedom considered in the molecular model under
consideration.

In the most general case, one considers the translational degrees of freedom of all the atoms in the
molecule. Equation (\ref{eq:KE_PIMC}) remains the same (with the obvious modification of an
atom-dependent mass $m$), but one needs an intramolecular potential to keep the molecule bound and,
in general, a large number of beads, especially if light atoms (such as hydrogen or one of its
isotopes) are to be considered.
This approach allows flexibility effects to be fully accounted for
and has been applied to investigate the second virial
coefficient of hydrogen~\cite{Garberoglio2014} and water~\cite{Garberoglio2018} isotopologues.
As one might expect, flexibility is more important at higher temperatures.
On the other hand, this approach requires intramolecular and intermolecular potentials that depend
on all the degrees of freedom, which in turn call for very demanding {\em ab initio} electronic
structure calculations.

At sufficiently low temperatures, molecules occupy their vibrational ground state, and rigid monomer
models are expected to be quite (although not perfectly) accurate. In this case, a whole molecule is
described as a rigid rotor, that is by 3 translational and 3 rotational degrees of freedom (2 in the
case of linear molecules). The matrix elements of the kinetic energy operator are, in this case,
more complicated than that in Eq.~(\ref{eq:KE_PIMC}), but their expression has been worked out for
both linear~\cite{Marx99} and non-linear~\cite{Noya11_1,Noya11_2} rotors.

The rigid-rotor approximation of a molecular system is, in principle, an uncontrolled
approximation and, consequently, cannot directly provide useful data for metrological applications.
On the other hand, the associated uncertainties can be partially offset by the fact that
potential-energy surfaces can be generated higher accuracy than in the case of fully flexible
models.~\cite{Patkowski2008,Garberoglio2012}
Validation of the {\em ab initio} results with experimental data can be used to establish the
temperature range in which a rigid model is valid, and provide useful estimates of virial
coefficients where experimental data are lacking.
Additionally, rigid models can be a stepping stone towards the more accurate fully flexible
approaches.

Also, semiclassical approximations of density~\cite{Schenter2002} or dielectric virial
coefficients~\cite{Gray2011} for molecular systems are available. They are generally much easier to
evaluate than by path-integral calculations, and are quite accurate in many
cases.~\cite{Garberoglio2018,Garberoglio23}

\subsubsection{Dielectric and refractivity virial coefficients}
\label{sec:mol_dielectric}

The calculation of dielectric and refractivity virial coefficients for molecular species is much
more difficult than for the monatomic systems discussed in Sec.~\ref{sec:dielectric_theo}. In
addition to the increased dimensionality, the charge asymmetry creates additional polarization
effects in interacting molecules. A complete treatment must therefore include the effect of the
molecular interactions not only on the polarizability of the molecules, but also on their charge
distribution. Because of this complexity, it seems unlikely that coefficients beyond the second
virial will be calculated in the foreseeable future, and quantitatively accurate calculations with
realistic uncertainty estimates may be limited to diatomic molecules such as N${}_2$ or H${}_2$.

The only attempt at such calculations we are aware of for realistic
(polarizable) molecular models is the work of Stone {\em et
  al.},~\cite{Stone_2000} who calculated the second dielectric virial
coefficient for several small molecules, including CO and H${}_2$O. A
recent experimental determination of the second dielectric virial
coefficient for CO~\cite{Tsankova18} was in qualitative but not
quantitative agreement with the prediction of Stone {\em et al.}

For rigorous metrology, it would be necessary to characterize the
uncertainty of the surfaces describing the mutual polarization and pair
polarizability of the molecules.  The dimensionality, and therefore the
complexity, of these calculations for a diatomic molecule like N${}_2$ would be
similar to that for the three-body polarizability and dipole surfaces for
monatomic gases.

\subsubsection{Molecular collisions}
\label{sec:mol_coll}

In some pressure-metrology applications near vacuum conditions, collision rates, which are related
to collision integrals, are required. We already introduced collision integrals for atom-atom
collisions in Sec.~\ref{sec:transport}, but the concept can be generalized to include atom-molecule
and molecule-molecule collisions, which enables the calculation of transport properties for
dilute molecular gases. While the collision integrals for atom-atom collisions result in a classical
treatment from the solution of the linearized Boltzmann equation and in the quantum-mechanical case
from the solution of the linearized Uehling--Uhlenbeck equation,~\cite{Uehling33} the corresponding
classical and quantum-mechanical equations for collisions involving molecules are the linearized
Curtiss--Kagan--Maksimov equation~\cite{Curtiss81,Curtiss92,Kagan61,Kagan66} and the linearized
Waldmann--Snider equation.~\cite{Waldmann57,Waldmann58,Waldmann58b,Snider60}

The formalism for the calculation of collision integrals involving molecules is much more complex
than in the case of atom-atom collisions. Relations for classical collision integrals were derived
by Curtiss for rigid linear molecules~\cite{Curtiss81c} and extended to rigid nonlinear molecules by
Dickinson {\em et al.}~\cite{Dickinson2007} The quantum-mechanical calculation of collision
integrals involving two molecules has rarely been attempted because of the mathematical complexity
and large computational requirements, whereas atom-molecule collisions have been studied
quantum-mechanically more often. For collisions between a helium atom and a nitrogen molecule,
collision integrals were calculated both classically and
quantum-mechanically.~\cite{McCourt91,Vesovic95} The comparison showed that quantum effects are
small except at low temperatures.  The degree to which the quantum nature of collisions can be
neglected for pairs with larger expected quantum effects, such as H${}_2$O--H${}_2$O, remains an
open question, but the agreement with experiment of classically calculated dilute-gas viscosities
for H${}_2$O~\cite{Hellmann15v} suggests that the classical approximation is adequate for most
purposes.

\subsection{Humidity metrology}

Much humidity metrology requires knowledge of humid air's departure from
ideal-gas behavior. Because the densities are low, this can be described by
the virial expansion. The second virial coefficient of pure water has been
calculated~\cite{Garberoglio2018} based on flexible {\em ab initio} pair
potentials computed at a high level of
theory.~\cite{Wang11,Babin14,Jankowski15} It is necessary to take the
flexibility of the water molecule into account to obtain quantitative
accuracy.~\cite{Garberoglio2018}

The most important contribution to the nonideality of humid air comes from
the interaction second virial coefficient of water with air. While fairly
accurate measurements of this quantity exist near ambient temperatures,
it can now be computed with similar or better uncertainty by
combining the cross second virial coefficients for water with the main
components of dry air.~\cite{Harvey07} Good quality pair potentials exist for water with
argon,~\cite{Hodges02} nitrogen,~\cite{Hellmann2019a} and
oxygen,~\cite{Hellmann2020} and these have been combined by
Hellmann~\cite{Hellmann2020} to produce accurate water-air second virial
coefficients between 150~K and 2000~K.

For humidity metrology at pressures significantly higher than atmospheric,
corrections at the third virial coefficient level become significant. Only
very limited data exist for the relevant third virial coefficients
(water-water-air and water-air-air),~\cite{Hyland83} so {\em ab initio}
calculation of these quantities would be useful. This requires development
of three-body potential-energy surfaces for systems such as
H${}_2$O--N${}_2$--N${}_2$ and H${}_2$O--H${}_2$O--O${}_2$. To our
knowledge, no high-accuracy surfaces exist for these three-molecule
systems, but their development should be feasible with current technology.

The same framework can be used for humidity metrology in other
gases. Hygrometers are typically calibrated with air or nitrogen as the
carrier gas, but some error will be introduced if the calibration is used
in the measurement of moisture in a different gas. Calibrations can be
adjusted if {\em ab initio} values of the cross second virial coefficient
are known for water with the gas of interest. Such values have been
developed for several important gases, such as carbon
dioxide,~\cite{Hellmann2019} methane,~\cite{Akin06}
helium,~\cite{Hodges02He} and hydrogen.~\cite{Hellmann2023}

Some emerging technologies for humidity metrology can be aided by {\em ab
  initio} property calculations. Instruments to measure humidity from the
change in dielectric constant with water content of a
gas~\cite{Cuccaro12,Gavioso14} require the first dielectric virial
coefficient of water, which depends on its molecular polarizability and
dipole moment. These quantities, and their temperature dependence, have
been a subject of recent theoretical study.~\cite{Garberoglio23}

Spectroscopic measurement of humidity has also been proposed;~\cite{Underwood2017a} this requires
the intensity of an absorption line for the water molecule. Thus far, work in this area has used
measured line intensities due to their smaller uncertainty compared to {\em ab initio} values. The
recent work of Rubin {\em et al.}~\cite{Rubin22} demonstrated mutually consistent sub-percent
accuracy for both experimental and theoretical intensities based on a semiempirical PES for an
H${}_2$O line, offering promise for the future use of calculated intensities to reduce the
uncertainty of humidity metrology.

\subsection{Pressure metrology}
\label{sec:mol_pressure}

Molecular calculations are also promising for pressure metrology at low
pressures.~\cite{Jousten2017} Refractivity-based pressure measurements using noble gases are
discussed in Sec.~\ref{sec:pressure}.  Some proposed approaches use ratios of the
refractivity of a more refractive gas (such as nitrogen or argon) to that
of helium. Use of nitrogen in these systems would be aided by good {\em ab
initio} results for the polarizability of the N${}_2$ molecule and its second
density and refractivity virial coefficients.

For low pressures, on the order of 1 Pa and below, absorption spectroscopy is a promising approach
for pressure measurement. The absorption of a gas such as CO or CO${}_2$ can be used to measure low
gas densities (from which the pressure is calculated by the ideal-gas law, perhaps with a second
virial correction); this can be a primary pressure standard if the line intensity is calculated from
semiempirical potential-energy and dipole-moment {\em ab initio} surfaces tuned to spectral
data. Even if measured intensities are used, theoretical results are valuable to check their
accuracy. For CO${}_2$, uncertainty of intensity measurements and agreement between theory and
experiment below 0.5\% has been obtained.~\cite{Huang22,Polyansky15} The simpler CO molecule is more
amenable to accurate theoretical calculations; consistency between experimental and theoretical line
intensities on the order of 0.1\% has recently been achieved.~\cite{Bielska22}
In these calculations, the potential energy curve was purely empirical, but the dipole-moment
surface was obtained {\em ab initio}.
An unresolved question in this work so far is the uncertainty of {\em ab initio} calculated line
intensities, which must depend in a complex way on the uncertainties in the intramolecular potential
and in the dipole-moment surface. Without reasonable estimates for the uncertainty of calculated
intensities, the utility of this spectroscopic method for primary pressure standards is diminished.

For ultrahigh vacuum and below, gas densities can be measured based on the collision rate between the
gas and a collection of trapped ultra-cold atoms. Both lithium and rubidium have been proposed as
the trapped species.~\cite{Scherschligt17,Eckel18,Shen2020,Shen2021,Ehinger2022,Zhang22,Barker22}
While in some implementations an apparatus constant is derived from
measurements,~\cite{Shen2020,Shen2021} it has recently been recognized~\cite{Shen2022} that the
proposed procedure introduces error when light species (such as Li and H${}_2$) are involved in the
collisions.

It is also possible to determine the relevant proportionality factor for
the collision rate from first principles using collision cross sections
calculated from {\em ab initio} pair potentials and quantum collision
theory. These calculations have been performed for lithium with H${}_2$
(the most common gas in metallic vacuum systems) and
He;~\cite{Makrides2019,Makrides2020} {\em ab initio} calculations with
rubidium are more challenging due to the large number of electrons. A
recent paper has reported first-principles collision rate coefficients for
both Rb and Li with noble gases, H${}_2$, and N${}_2$.~\cite{Klos22} It is
also possible to measure the ratio of two collision pairs (for example,
Rb--H${}_2$ versus Li--H${}_2$) to obtain the coefficient for a system that
is more difficult to calculate {\em ab
  initio};~\cite{Scherschligt17,Shen2022} in this approach a low
uncertainty for the simpler-to-calculate system (that with fewer electrons)
is essential.

\subsection{Atmospheric physics}

In atmospheric physics, the interaction of radiation with atmospheric
gases, particularly H${}_2$O and CO${}_2$, has received increasing
attention for climate studies; it is also important for Earth-based
astronomy where the atmosphere is in the optical path. Scientists in these
fields rely on line positions and intensities in the HITRAN
database.~\cite{Gordon2022}
Increasingly, {\em ab initio} calculations are being used to supplement
experimental measurements for these quantities, as has recently been
summarized for CO${}_2$.~\cite{HITRAN2020-CO2}

\subsection{Transport properties}

While transport properties of molecular gases are of little relevance in
precision metrology, for the sake of completeness we mention briefly the
current state of the art for pure molecular gases. Most of the transport
property calculations for such gases performed so far are based on
classically calculated collision integrals for rigid molecules using the
formalism of Curtiss~\cite{Curtiss81} for linear molecules and of Dickinson
{\em et al.}~\cite{Dickinson2007} for nonlinear molecules (see Sec.~\ref{sec:mol_coll}.)

A representative example of such calculations for gases consisting of small
molecules other than H${}_2$ are the classical shear viscosity and thermal
conductivity calculations of Hellmann and Vogel~\cite{Hellmann15v} and
Hellmann and Bich,~\cite{Hellmann15i} respectively, for pure H${}_2$O. The
agreement with the best experimental data is within a few tenths of a
percent for the viscosity and a few percent for the thermal
conductivity. For both properties, these deviations correspond to the
typical uncertainties of the best experimental data.  The significant
contribution to the thermal conductivity due to the transport of energy
``stored'' in the vibrational degrees of freedom, which is not directly
accounted for by the classical rigid-rotor calculations, was estimated
using a scheme that only requires knowledge of the ideal-gas heat capacity
in addition to the rigid-rotor collision integrals.~\cite{Hellmann15i} The main assumption
in this scheme is that collisions that change the vibrational energy levels
of the molecules are so rare that their effects on the collision integrals
are negligible.

For pure H${}_2$, classical calculations are not accurate enough even at ambient
temperature. Fully quantum-mechanical calculations were performed by Mehl
{\em et al.}~\cite{Mehl10} using a spherically-averaged modification of a
H${}_2$--H${}_2$ PES,~\cite{Patkowski2008} thus 
reducing the complexity of the collision calculations to that for
monatomic gases. Despite this approximation, the calculated shear
viscosity and thermal conductivity values for H${}_2$ agree very well with the
best experimental data, particularly in the case of the viscosity where
the agreement is within 0.1\%.

\section{Concluding Remarks and Future Perspectives}
\label{sec:future}

The outstanding progress achieved during the last three decades by the {\em ab initio} calculation
of the thermophysical properties of pure fluids and mixtures has drastically reduced the uncertainty
of the measurement of these properties and of the thermodynamic variables temperature, pressure, and
composition.

For example, consider primary thermometry. {\em Ab initio} calculations directly contributed to 
the acoustic and dielectric determination of the value of the Boltzmann constant that is used in the new
SI definition of the kelvin. The remarkably accurate theoretical calculations of the polarizability
and the non-ideality of thermometric gases have also facilitated simplified measurement strategies
and techniques.~\cite{Moldover2014,Benedetto2004,Pitre2006} Consequently, new paths directly
disseminating the thermodynamic temperature are now available at temperatures below $25$~K, where
the realization of ITS-90 is particularly complicated.  Various methods of gas thermometry have
determined $T$ with uncertainties that are comparable to or even lower than the uncertainty of
realizations of ITS-90.~\cite{Mehl2009,Gaiser21primary,Gaiser2018} Improved theory has also
suggested that primary CVGT could usefully be revisited, as discussed in Section \ref{sec:cvgt}.

In the near future, technical achievements will likely further reduce the uncertainty of
measurements of the thermodynamic temperature and the thermophysical properties of gases.  Efforts
are underway to improve: (1) the purity of the thermometric gases at their point of use, (2)
implementing two-gas methods to reduce the uncertainties from compressibility of the apparatus, and
(3) developing robust microphones (possibly based on optical interferometry) to facilitate cryogenic
AGT.
In the remainder of this section, we will summarize current limitations and describe some
prospects for future contributions.

\subsection{Current limitations of {\em ab initio} property calculations}

As described in Sec.~\ref{sec:abinitio}, {\em ab initio} calculations of properties for individual
helium atoms and pairs of atoms have achieved extraordinarily small uncertainties. Even for
three-body interactions, the potential energy is now known with small uncertainty, and good surfaces
are available for the three-body polarizability and dipole moment surfaces. This enables accurate
calculations, with no uncontrolled approximations, of the second and third density, acoustic, and
dielectric virial coefficients. This high level of accuracy is due to the small number of electrons
involved; electron correlation at the FCI level is still tractable for three helium atoms with a
total of six electrons.

For DCGT and RIGT, it would be desirable to have similarly accurate properties for neon and argon,
because their higher polarizability (and therefore stronger response) reduces the relative effect of
other sources of uncertainty such as imperfect knowledge of the compressibility of the apparatus or
the presence of impurities in the gas. Unfortunately, this level of accuracy for neon and argon is
unlikely to be obtained in the foreseeable future. The neon atom has 10 electrons, as many as five
helium atoms, and argon has 18. While recent efforts have (at large computational expense)
significantly reduced the uncertainty of single-atom and dimer quantities for neon and
argon,~\cite{Lesiuk2020,Hellmann2022,Lesiuk2023,Hellmann2021,Lesiukprivate} they do not approach the
levels of accuracy achieved for helium. For example, the relative uncertainty of the best
calculation of the static polarizability of a neon atom~\cite{Hellmann2022} is more than $100$ times
greater than that of a helium atom.~\cite{Puchalski:20} Similarly, the relative uncertainty of the
pair potential minimum energy is about $100$ times larger for neon~\cite{Hellmann2021} than for
helium.~\cite{Czachorowski2020} Therefore, the relative uncertainties of calculated gas-phase
thermophysical properties will be much higher for other gases than for helium. In such cases, the
most accurate values of properties will be obtained by measuring ratios of properties relative to
that of helium.  This has already been done for the static polarizability of neon and
argon~\cite{Gaiser2018} and for the low-density viscosity of several
gases.~\cite{Berg12,Vogel21,Xiao20}

Refractivity-based thermal metrology~\cite{Rourke19,Rourke2021a} requires $A_\mathrm{R}$, and
preferably also $B_\mathrm{R}$ and $C_\mathrm{R}$. At microwave frequencies, the static values
($A_\varepsilon$, $B_\varepsilon$, etc.) can be used. At optical frequencies, $A_\mathrm{R}$ and
$B_\mathrm{R}$ have been computed at a state-of-the-art level for
helium,~\cite{Garberoglio2020,Puchalski:16} but corresponding calculations for neon and argon rely on
values for the dynamic polarizability and for the Cauchy moment $\Delta S(-4)$ that could be
significantly improved. Even with state-of-the-art {\em ab initio} results, it seems likely that
ratio measurements using helium, such as those of Egan {\em et al.} for
$A_\mathrm{R}$,~\cite{Egan19} will produce lower uncertainties. To our knowledge, the theory for
calculating $C_\mathrm{R}$ at optical frequencies is not available. Therefore, at the moment, it is
necessary to take rather uncertain values from experiment or assume (based on the small difference
between $B_\mathrm{R}$ and $B_\varepsilon$) that it is equal to $C_\varepsilon$.

As mentioned in Sec.~\ref{sec:dielectric_theo} and also noted by Rourke,~\cite{Rourke2021a} another
issue for refractivity methods is the unclear situation surrounding the $A_\mu$ contribution. The
best calculations of the magnetic susceptibility for helium,~\cite{Bruch02} neon,~\cite{Lesiuk2020}
and argon~\cite{Lesiuk2023} disagree with the old, sparse measurements of these
quantities~\cite{Barter60} by amounts much larger than their stated uncertainties. Independent
calculations of the magnetic susceptibility for one or more of these species would be helpful in
assessing this discrepancy, but what is most needed is a modern measurement of the magnetic
susceptibility of a noble gas (probably argon), either as an absolute measurement or as a ratio to a
substance with a better-known magnetic susceptibility, such as liquid water.

To reach higher pressures with helium-based apparatus, it would be desirable to have reliable
values, with uncertainties, for the fourth virial coefficient $D(T)$. The most complete
first-principles estimate so far~\cite{Garberoglio2021a} used high-accuracy two-body and three-body
potentials, but had a significant uncertainty component due to the unknown four-body
potential. Accurate calculations of the nonadditive four-body potential for helium are feasible with
modern methods.
A four-body PES for helium, even if its relative uncertainty was as large as 10\%, would allow
reference-quality calculation of $D(T)$ and enable improved metrology.

\subsection{Molecular gases}

Nitrogen is an attractive option for gas-based metrology due to its availability in high purity and
its longstanding use in traditional apparatus such as piston gauges, but its lack of spherical
symmetry and its internal degree of freedom add complication to {\em ab initio} calculation of its
properties.  The development of potential-energy surfaces for pair and three-body interactions for
rigid molecular molecules is certainly feasible. This is also possible for flexible models, pending
the difficulties already discussed in Sec.~\ref{sec:mol_pot}.  Once this data is available, the
methods for calculations of density virial coefficients have already been
proven.~\cite{Patkowski2008,Garberoglio2012,Garberoglio2018} (see Sec.~\ref{sec:mol_virials}.) To the
best of our knowledge, no fully {\em ab initio} calculation of dielectric virial coefficients for
molecular systems has been performed. This task will require the development of the molecular
interaction-induced polarizability function. The path-integral approach described in
Sec.~\ref{sec:thermo} can certainly be extended to compute these quantities as well as rigorously
propagate their uncertainties.

\subsection{Improved uncertainty estimations}
\label{sec:future_uncertainty}

As mentioned in Sec.~\ref{sec:uncertainty}, much progress has been made in estimating realistic
uncertainties for density and dielectric virial coefficients. The old method of simply displacing
the potentials in a “plus” and “minus” direction, while correct for one-dimensional integrations
such as $B$ and $B_\varepsilon$, is inefficient and can produce inaccurate results for higher
coefficients. The functional differentiation approach discussed in Sec.~\ref{sec:uncertainty}
provides more rigorous results.

However, it is not entirely clear how to obtain uncertainties for acoustic virial coefficients,
because they involve temperature derivatives of $B(T)$ and $C(T)$. The rigorous assignment of
uncertainty to a derivative of a function computed from uncertain input is an unsolved problem as
far as we are aware. Binosi {\em et al.}~\cite{Binosi2023} recently applied a statistical method
(the Schlessinger Point Method) to the estimation of uncertainties for acoustic virial coefficients;
this may provide a way forward.

A similar issue exists for the low-density transport properties. The very low uncertainty of the
viscosity of helium shown in Fig.~\ref{fig:viscosity} near $40$~K, obtained with the traditional
method of “plus” and “minus” perturbations to the pair potential, is an artifact of competing
effects on the collision integral of perturbations from different parts of the potential. While
$B(T)$, for example, exhibits monotonic behavior with respect to perturbations in the potential,
that is not the case for the collision integrals used to compute
transport properties, which can cause uncertainties to be artificially underestimated. This was
recognized by Hellmann and coworkers, who created potentials perturbed in additional ways to provide
a non-rigorous but reasonable estimation method for the uncertainty of low-density transport
properties for krypton,~\cite{Jaeger15} xenon,~\cite{Hellmann2017Xe} and neon.~\cite{Hellmann2021}
Further analysis would be welcome to improve the rigor of uncertainty estimates for transport
collision integrals.

\subsection{Transport properties}

In addition to the uncertainty issue just mentioned, we see two areas for improvement in the field
of transport properties. The first concerns the density dependence beyond the low-density limiting
values discussed in this work. As mentioned in Sec.~\ref{sec:flow}, for flow metrology it would be
desirable to know the viscosity with small and rigorous uncertainties not only at zero density, but
at the real densities at which instruments are calibrated. The first correction should be a
virial-like term linear in density, but the most successful theory so
far~\cite{Friend84,Rainwater87,Najafi98} 
relies on some simplifying assumptions. A more rigorous theory would be a significant advance. Even
if the initial density dependence were only known for helium, that would enable better metrology for
other gases because of the established methods for measuring viscosity ratios.

The second area is the transport properties of molecular species, such as N${}_2$ or H${}_2$O. As
mentioned in Sec.~\ref{sec:transport}, classical collision integrals can be calculated for these
species when they are modeled as rigid rotors. While it is believed that the errors introduced by
the assumptions of classical dynamics and rigid molecules are small, it would be desirable to have
verification from a more rigorous calculation. One might expect quantum effects to be significant
for the dynamics of H${}_2$O collisions, since they make a large contribution to $B(T)$ for
H${}_2$O.~\cite{Garberoglio2014} Since fully quantum calculation of collision integrals is currently
intractable for all but the simplest systems, the development of a viable “semiclassical” method for
transport properties would be desirable. No such formulation exists to our knowledge.

\subsection{Simulations of liquid helium}

While we have focused on the gaseous systems where {\em ab initio} properties are already making
major contributions to metrology, the thermophysical properties of condensed phases (particularly
for helium) are also important in temperature metrology. For example, the vapor pressures of liquid
${}^3$He and ${}^4$He are part of the definition of ITS-90.~\cite{PrestonThomas1990} With highly
accurate two- and three-body potentials for helium (perhaps eventually supplemented by a four-body
potential), high-accuracy simulation of thermodynamic properties of liquid helium may become
feasible.

In fact, path-integral simulations of liquid ${}^4$He can be performed without uncontrolled
approximations,~\cite{Ceperley1995} although, to the best of our knowledge, the most recent {\em ab
  initio} potentials have not been employed yet to compute any liquid helium property ({\em e.g.}, the
specific heat -- and hence the vapor pressure, via the Clapeyron equation -- or the
temperature of superfluid transition). Consequently, the accuracy of first-principles many-body
potentials in the case of fluid phases of ${}^4$He is largely unknown.  The use of three-body (or
higher, when available) potentials is expected to require considerable computational
resources, as has been recently observed in simulations of liquid para-H${}_2$,~\cite{Ibrahim2022}
but theoretical developments in efficient simulation methods for degenerate systems~\cite{Spada2022}
might pave the way for a fully {\em ab initio} calculation of the thermophysical properties of
condensed ${}^4$He.

In the case of fermionic systems such as ${}^3$He, the path-integral approach suffers in principle
from a ``sign problem''~\cite{Chakravarty1997} which generally requires some approximations, and
results in a large statistical uncertainty. However, two research groups have recently claimed to have 
overcome these limitations,~\cite{Filinov2022,Dornheim22} which might result in accurate calculations of
thermophysical properties in the liquid phase also for this isotope.

\subsection{Reproducibility and validation}

It is desirable for metrological standards to be based on multiple independent studies, so that they
will not be distorted by a single unrecognized error.  For example, for the recent redefinition of
the SI in which several fundamental physical constants were assigned exact values, it was required
that the value assigned to the Boltzmann constant be based on consistent results from at least two
independent experiments using different techniques and meeting a low uncertainty threshold.~\cite{Fischer2018}
Similarly, metrological application of the calculated results discussed in this Review would be on a
firmer basis if there was independent confirmation of the results.

The danger of an unrecognized error in calculated quantities is not merely hypothetical.  For
several years, the ``best'' calculated values of $C$ for $^3$He were in error below about 4.5~K
because the effects of nuclear spin on the quantum exchange contribution had been incorporated
incorrectly; this was eventually recognized and corrected in
Errata.~\cite{Garberoglio2011,Garberoglio2021a} An early quantum calculation of $B_\varepsilon$ of
argon~\cite{Rizzo2002} disagreed with a later study,~\cite{Garberoglio2020} apparently because of
inexact handling of resonance states in the earlier work.  Ideally, there would be independent
confirmation of all the results cited in Table~\ref{tab:theoretical_virials} so that any errors
could be detected. 

One helpful step in this direction would be more complete documentation of calculations, including
computer code, so that others can reproduce or check the work.  It is common to provide computer
code for potential-energy surfaces, but the calculation of virial coefficients has typically been
performed with specialized software that is not public.

More important for metrology, however, would be independent verification of the calculated results.
Conceptually, this has two parts: validation of the calculated quantities and surfaces described in
Sec.~\ref{sec:abinitio} (potential-energy, polarizability, and dipole surfaces; atomic and magnetic
polarizabilities) and validation of the calculation of virial coefficients from these quantities
(described in Sec.~\ref{sec:thermo}).

Validation of calculated virial coefficients is probably the easier of the two parts, because it is
typically less computationally demanding.  This has been done for a few quantities; for example, two
groups have performed fully quantum calculations (in one case neglecting exchange effects that
become important below 7~K) of $C$~\cite{Garberoglio2011,Shaul2012} and
$D$~\cite{Shaul2012,Garberoglio2021a} for $^4$He.  Consistency checks can also be made by comparing
different calculation methods, including classical and semiclassical approaches that should agree
with the quantum calculations at high temperatures.  The error in $B_\varepsilon$ for argon
mentioned above was detected by comparing phase-shift calculations to PIMC and semiclassical
calculations, showing the value of multi-method comparisons.

The independent validation of calculated atomic quantities and intermolecular surfaces is more
difficult, because these require large amounts of dedicated computer time.  There have been a few
cases where parallel efforts have produced independent, high-quality results; these include
$A_\varepsilon$ for neon~\cite{Lesiuk2020,Hellmann2022} and the three-body potential of
argon.~\cite{Jaeger11,Cencek2013} Some validation is also provided when the state of the art
advances and new potentials are produced that agree with previous potentials (but have smaller
uncertainties); this has been the case with the sequential development of pair potentials for helium
(Sec.~\ref{sec:He2}).  In some cases, however, these are not truly independent verifications because
they are developed by the same group and use many of the same methods.  While it may be difficult to
justify the extensive work required to independently confirm a state-of-the-art calculated surface,
there would be value in performing spot checks of a few points.  This would require developers of
surfaces to make their calculated points available (or at least a subset of them), and also the
multiple calculated quantities that typically contribute to each point.

We believe that more attention should be paid to the reproducibility and validation of the
calculated results that are increasingly important in precision metrology.  Work of this nature may
not be very attractive to funding agencies (or graduate students), but it is needed for more
confident use of gas-based metrology.

\medskip

\acknowledgements

We thank Mark McLinden, Patrick Egan, and Ian Bell of NIST for helpful comments, and Richard Rusby
of NPL for valuable discussion regarding the CVGT technique.
KS acknowledges support from the NSF grant CHE-2154908.

We acknowledge support from {\em Real-K} project 18SIB02, which has received funding from the EMPIR
programme co-financed by the Participating States and from the European Union’s Horizon 2020
research and innovation programme.

\section*{Author Declarations}
\subsection*{Conflict of Interest}
The authors have no conflicts to disclose.

\medskip

\appendix

\section{Formulae for the Third Acoustic Virial Coefficient, $\gammaa$}
\label{app:RTg}

As is apparent from Eqs.~(\ref{eq:beta_a}), (\ref{eq:gamma_a}), and
(\ref{eq:Q}), the explicit expression of the third acoustic virial coefficient as a
function of the pair and three-body potential is quite involved.
We found that it is most conveniently expressed by defining
\begin{eqnarray}
  b(r,T) &=& U_2(r) \e^{-\beta U_2(r)} -1 \\
  b_T(r,T) &=& \beta U_2(r) \e^{-\beta U_2(r)} \\
  b_{TT}(r,T) &=& \beta U_2(r) (\beta U_2(r) - 2) \e^{-\beta U_2(r)} 
\end{eqnarray}
and
\begin{eqnarray}
  c &=& \e^{-\beta U_3} - \sum_{i<j} \e^{-\beta U_2(r_{ij})} +2 \\
  c_T &=& \beta U_3 \e^{-\beta U_3} - \sum_{i<j} \beta U_2(r_{ij}) \e^{-\beta
    U_2(r_{ij})} \\
  c_{TT} &=& \beta U_3 (\beta U_3 - 2) \e^{-\beta U_3} - \nonumber \\
  & & \sum_{i<j} \beta U_3 (\beta U_3 - 2) \e^{-\beta U_3},
\end{eqnarray}
where $c$, $c_T$ and $c_{TT}$ are functions of the temperature $T$ as well as $r_{12}$,$r_{13}$, and
$r_{23}$ through their dependence on $U_3$.  Performing the substitution $\gamma_0 = 5/3$, we obtain
\begin{widetext}
  \begin{eqnarray}
    \RTg(T) &=& \frac{8 \pi^3}{3} \NA^2 \int \left[
    \frac{2}{15}b_{TT}(r_{12})b_{TT}(r_{13}) +
    \frac{14}{15} b_T(r_{12}) b_{TT}(r_{13}) +
     b(r_{12}) b_{TT}(r_{13}) + \right. \nonumber \\
    & & \frac{73}{30} b_T(r_{12}) b_T(r_{13}) +    
    \frac{34}{5} b(r_{12}) b_T(r_{13}) +
    \frac{33}{5} b(r_{12}) b(r_{13}) - \nonumber \\
    & & \left. \left( \frac{2}{15}c_{TT}(r_{12},r_{13},r_{23}) +
    \frac{16}{15}c_T(r_{12},r_{13},r_{23}) +
    \frac{13}{5}c(r_{12},r_{13},r_{23})
    \right) \right] \D \Omega_3.
    \label{eq:RTg_classical}
  \end{eqnarray}
\end{widetext}

The path-integral expression for $\gammaa$ is more complicated, due to the
fact that the ring-polymer distribution function $F$ of Eq.~(\ref{eq:F})
depends on temperature.
In particular, defining $\cal U$ so that
\begin{equation}
    F = \Lambda^3 \left( \frac{P^{3/2}}{\Lambda^3} \right)^P
  \exp\left(- \beta {\cal U} \right),
\end{equation}
one can show that
\begin{equation}
  \frac{\D F}{\D\beta} = \left({\cal U} - \frac{3(P-1)}{2\beta} \right) F,
\label{eq:dF}
\end{equation}
and derive path-integral expressions for $\gammaa$. However, this
straightforward approach is characterized by large variance in the Monte
Carlo simulations, since Eq.~(\ref{eq:dF}) has a form analogous to the {\em
  thermodynamic} estimator of the kinetic energy.~\cite{Tuckerman10}
It is possible to derive equivalent expressions with smaller variance,
using the same ideas that lead to the {\em virial} estimator.~\cite{Herman82,Tuckerman10} The resulting
formulae are very cumbersome, and can be found in Ref.~\onlinecite{Binosi2023}.

%


\begin{thebibliography}{420}
\bibitem [{\citenamefont {Andeen-Hagerling}()}]{Andeen}%
  \BibitemOpen
  \bibfield  {author} {\bibinfo {author} {\bibnamefont {Andeen-Hagerling}},\
  }\href@noop {} {}\bibinfo {note} {Model {AH} 2500{A}, {O}ption {E}.
  Andeen-Hagerling, Inc. Cleveland, Ohio, USA. In order to describe materials
  and procedures adequately, it is occasionally necessary to identify
  commercial products by manufacturer’s name or label. In no instance does
  such identification imply endorsement by the authors' institutions, nor does
  it imply that the particular product or equipment is necessarily the best
  available for the purpose}\BibitemShut {NoStop}%
\end{thebibliography}
\end{document}